\pdfoutput=1
\documentclass[fleqn,usenatbib]{mnras}

\usepackage{newtxtext,newtxmath}

\usepackage[T1]{fontenc}
\usepackage{ae,aecompl}

\usepackage{graphicx}	
\usepackage{amsmath}	
\usepackage{caption}
\usepackage{pdflscape}
\usepackage{subcaption}
\usepackage[normalem]{ulem}

\title[ALMA Survey of Lupus Class III stars]{ALMA Survey of Lupus Class III Stars: Early Planetesimal Belt Formation and Rapid Disk Dispersal}

\author[J. B. Lovell et al.]{J. B. Lovell$^{1}$\thanks{E-mail: jl638@cam.ac.uk}, M. C. Wyatt$^{1}$, M. Ansdell$^2$, M. Kama$^3,1$, G. M. Kennedy$^{4,5}$, C. F. Manara$^6$, \newauthor S. Marino$^{1,7}$, L. Matr\`a$^8$, G. Rosotti$^9$, M. Tazzari$^1$, L. Testi$^{6,10}$, J. P. Williams$^{11}$ \\
$^1$Institute of Astronomy, University of Cambridge, Madingley Road, Cambridge, CB3 0HA, UK\\
$^2$National Aeronautics and Space Administration Headquarters, 300 E Street SW, Washington DC 20546, USA\\
$^3$Tartu Observatory, University of Tartu, 61602 T\H{o}ravere, Estonia \\
$^4$Department of Physics, University of Warwick, Coventry, CV4 7AL, UK \\
$^5$Centre for Exoplanets and Habitability, University of Warwick, Gibbet Hill Road, Coventry CV4 7AL, UK\\
$^6$European Southern Observatory, Karl-Schwarzschild-Strasse 2, 85748, Garching bei M{\"u}nchen, Germany\\
$^7$Max Planck Institute for Astronomy, K\"onigstuhl 17, 69117 Heidelberg, Germany \\
$^8$School of Physics, National University of Ireland Galway, University Road, Galway, Ireland\\
$^9$Leiden Observatory, Leiden University, P.O. Box 9513, NL-2300 RA Leiden, the Netherlands\\
$^{10}$INAF - Osservatorio Astrofisico di Arcetri, L.go E. Fermi 5, 50125, Firenze, Italy\\
$^{11}$Institute for Astronomy, University of Hawai'i at Mänoa, Honolulu, HI, USA\\}

\date{Accepted XXX. Received YYY; in original form ZZZ}

\pubyear{2020}

\begin{document}
\label{firstpage}
\pagerange{\pageref{firstpage}--\pageref{lastpage}}
\maketitle

\begin{abstract}
Class III stars are those in star forming regions without large non-photospheric infrared emission, suggesting recent dispersal of their protoplanetary disks. 
We observed 30 class III stars in the 1-3\,Myr Lupus region with ALMA at ${\sim}856\mu$m, resulting in 4 detections that we attribute to circumstellar dust.
Inferred dust masses are $0.036-0.093M_{\oplus}$, ${\sim}1$ order of magnitude lower than any previous measurements; one disk is resolved with radius ${\sim}80$\,au.
Two class II sources in the field of view were also detected, and 11 other sources, consistent with sub-mm galaxy number counts. 
Stacking non-detections yields a marginal detection with mean dust mass ${\sim}0.0048M_{\oplus}$. 
We searched for gas emission from the CO~J=3-2 line, and present its detection to NO~Lup inferring a gas mass $(4.9\pm1.1)~\times~10^{-5}~M_{\oplus}$ and gas-to-dust ratio $1.0\pm0.4$.
Combining our survey with class II sources shows a gap in the disk mass distribution from $0.09-2M_\oplus$ for $>0.7M_\odot$ Lupus stars, evidence of rapid dispersal of mm-sized dust from protoplanetary disks. 
The class III disk mass distribution is consistent with a population model of planetesimal belts that go on to replenish the debris disks seen around main sequence stars.
This suggests that planetesimal belt formation does not require long-lived protoplanetary disks, i.e., planetesimals form within ${\sim}2$\,Myr.
While all 4 class III disks are consistent with collisional replenishment, for two the gas and/or mid-IR emission could indicate primordial circumstellar material in the final stages of protoplanetary disk dispersal. 
Two class III stars without sub-mm detections exhibit hot emission that could arise from ongoing planet formation processes inside ${\sim}1$\,au.
\end{abstract}



\section{Introduction}
\label{sec:intro}
Significant progress over the past ${\sim}$decade has been made in our understanding of the statistics of the planetary systems of nearby stars. We now know ${\sim}50\%$ of Sun-like stars have tightly packed systems of low mass planets at <1\,au, ${\sim}5\%$ have Jupiter-like planets, and ${\sim}1\%$ have Hot Jupiters \citep{Winn15}. We also know ${\sim}20\%$ of stars have detectable debris disks (i.e., belts of planetesimals at 20-100au) that are analogous to, but much more massive than, the Solar System’s Kuiper belt \citep{Wyatt08, Sibthorpe18, Hughes18}. The formation of these planets and planetesimals must occur in a star’s protoplanetary disk, which survives up to ${\sim}10$\,Myr, however these can disperse on timescales $<1\rm$\,Myr \citep{Williams11}. The finer details of planetesimal and planet formation processes and their timescales are uncertain.\\
\newline
The growth of initially $\rm{sub- \mu \rm{m}}$ dust into cm-sized pebbles, as well as their radial drift due to gas drag, is well understood \citep{Birnstiel10}. However, growth beyond cm-sized pebbles into planetesimals (that then either grow into planets, or feed the debris disks seen later on) is not. Growth beyond these pebbles is problematic due to radial drift and the bouncing barrier \citep{Windmark12, Pinte14, BoothR18}. Additionally such large objects cannot be observed directly. The way to form planetesimals may be to concentrate pebbles in regions sufficiently dense for gravitational instabilities to form them directly \citep{Johansen07}. Such concentrations could arise in the earliest phases of evolution ($<0.1\rm{Myr}$) while disks are gravitationally unstable \citep{BoothR16, Nixon18}, or at the end of a protoplanetary disk’s life as it disperses \citep[e.g., between 3-10\,Myr, see][]{Carrera17}, or indeed any time between these (i.e., $0.1-10\rm{Myr}$) if the disk is radially stratified \citep{Pinilla16, Carrera20}. Theoretically therefore, planetesimal formation could occur at any point within this broad range of ages.  \\
\newline
The collisional growth of planetesimals into planets is however well understood, it just takes time. This is not a problem close to the star where orbital timescales are short. Indeed the formation of close-in super-Earths must be largely complete by ${\sim}3\rm{Myr}$ as ${\sim}50\%$ of stars have such planets and ${\sim}50\%$ of stars have lost their protoplanetary disks by this age \citep[e.g.,][]{Haisch01}. However, at 10s of au, the long timescales required for planet growth may be more prohibitive \citep[e.g., ][]{Kenyon10}. Whilst pebble accretion may account for the growth of planets at these distances \citep{Bitsch15}, these require planetary seeds that are $\sim$\,Ceres sized, so this does not fully resolve this time scale issue. Nevertheless, annular gaps are seen at 10s of au in $0.5-1$\,Myr protoplanetary disks \citep{ALMAPart15, Sheehan18, Andrews18}, and these could be carved by planets \citep{Dipierro15, Rosotti16}, suggesting planet formation processes in outer disks could be well advanced by ${\sim}1\rm{Myr}$.  \\
\newline
While planetesimals are not detectable, the dust that is created in their collisions is. Such “debris disks” are seen around ${\sim}20\%$ of ($>10\rm{Myr}$) main sequence stars \citep{Wyatt08} showing planetesimal formation is common at 10s of au. It is not possible however to say by looking at main sequence stars when their planetesimals formed; i.e., these could have had either short-lived or long-lived protoplanetary disks. Neither can observations of protoplanetary disks answer this question, as second generation debris disk dust would be fainter than the primordial protoplanetary disk it is embedded within \citep[e.g., see Fig.3][]{Wyatt08}. Observations of young stars $<10$\,Myr that have recently dispersed their protoplanetary disks would shed light on disk dispersal processes, and provide valuable constraints on where and when planetesimals form. \\
\newline
Young Stellar Objects (YSOs) in star forming regions can be classified based on their mid-infrared emission (e.g., $24$\,$\mu \rm{m}$). Such emission originates from dust inside 10s of au of the star and is indicative of the presence of a protoplanetary disk \citep{Adams87}. The different classes can be considered as an evolutionary sequence, with newly born stars thought to evolve from class I for which a protoplanetary disk and envelope is present, to class II for which the protoplanetary disk emission dominates, and on to class III which are often considered to be diskless since no (or relatively low levels of) mid-infrared emission is present above the stellar photosphere. Such evolution can be readily seen in the spectral energy distributions of YSOs in these stages, although we note in between the class II and III stage also exist transition disks, with SEDs that can resemble a further intermediary evolutionary stage between these \citep{Espaillat14}. Based on the amount of dust available and analysis of grain growth, evidence exists that planet core formation may have begun prior to the class II stage \citep{Andrews18, Manara18b}, and potentially as early as the class 0/I phases \citep{Greaves11, Tychoniec20}. \\
\newline
A related classification scheme is based on evidence for ongoing accretion, placing YSOs into classical T-Tauri stars (CTTS) and Weak-line T-Tauri stars (WTTS), which are broadly similar to the class II and III except that some WTTS have significant infrared emission and so massive disks \citep{Manara13, Manara17, Alcala19}. Thus it is studies of class III stars, which is those with a mid-IR spectral index $\alpha_{\rm{IR}}<-1.6$ \citep{Lada87, Williams11}, which provide information on the nature of systems with recently dispersed protoplanetary disks \citep[although note that care must be taken when compiling YSO lists based on this classification, given that a range of wavelengths have been used to assess the mid-IR spectral index, see][]{Espaillat14}. \\
\newline
Despite this, class III stars have received little attention due to an inherent (apparent) lack of infrared emission. However, constraints on their circumstellar dust levels are usually weak, since star forming regions are $>100\rm{pc}$. Moreover, the absence of $24$\,$\mu\rm{m}$ emission (i.e., at the longest wavelength for which large scale surveys exist) does not rule out the presence of cold dust. Far-IR ($70-160\mu \rm{m}$) measurements are more sensitive to cold dust, but are not available for all stars, at least not at levels deep enough to detect dust emission except at protoplanetary disk levels. Nowadays, sub-mm observations with ALMA provide the best sensitivity to disks at 10s of au, and there have been several sub-mm surveys of class II stars \citep[e.g., Lupus, Taurus, Upper Sco, Ophiuchus;][]{Ansdell16, Andrews13, Barenfeld16, Cieza19}. However, only two sub-mm surveys included class III stars \citep{Hardy15, Cieza19}, and these were only sensitive enough to detect the most massive disks (at $M_{\rm{dust}}>0.2M_{\oplus}$). Thus, class III stars could host debris disks comparable in mass to those around nearby stars \citep[e.g., such as Fomalhaut with $\sim0.03M_\oplus$, see][]{Holland17}, a possibility also not ruled out by far-IR surveys of star forming regions.\\
\begin{figure*}
    \includegraphics[width=1.0\textwidth]{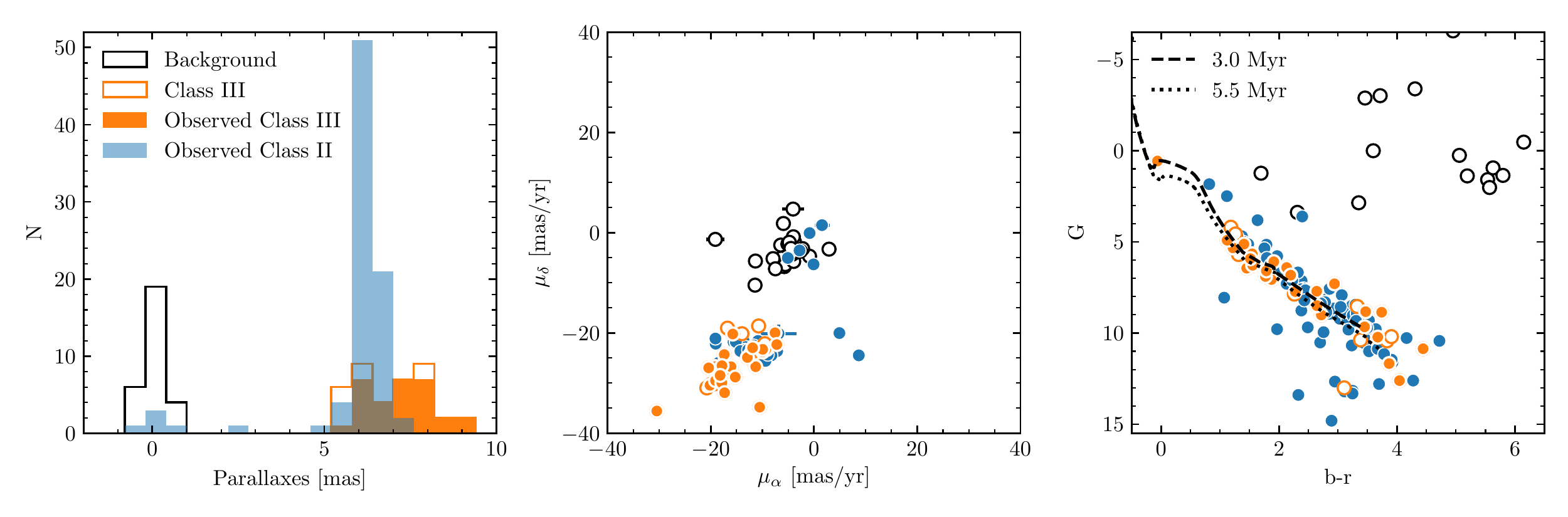}
    \caption{Gaia data for the 69 candidate class III stars and 85 class II stars (for which Gaia data is available):
    {\it (left)} histogram of parallaxes, {\it (middle)} proper motions, {\it (right)} G band absolute magnitude vs colour.
    The 30 class III sources deemed to be background objects are shown in black, while the final sample of 30 class IIIs are shown with filled orange markers, the remaining 9 class III sources (not observed) with empty orange markers, and the observed class II sources shown with blue markers. Isochrones were included on the magnitude - colour plot to show the sample consistency with the expected age of Lupus.}
    \label{fig:sourceSelection}
\end{figure*} 
\newline
In this work we report on the first dedicated ALMA survey of class III stars in the young ($1-3\rm{Myr}$) and nearby ($125-200\rm{pc}$) Lupus star forming region \citep{Comeron08}. Previous surveys have observed all class II stars in this region at both $890$\,$\mu \rm{m}$ and 1.3\,mm down to dust masses of ${\sim}0.2M_{\oplus}$ resulting in 71/96 detections in \citet{Ansdell16, Ansdell18} with a further 3/6 detections of class II sources in \citet{Sanchis20}. However, at this age ${\sim}25\%$ of stars have lost their protoplanetary disks \citep[e.g., Fig.1a of][]{Dunham15}, so these class II stars are the ${\sim}75\%$ with mid-IR indications of a protoplanetary disk. The class III stars we present here are those without protoplanetary disks, and therefore correspond to those within the population that for whatever reason had the shortest lived protoplanetary disks. The ALMA survey discussed here is deeper than any previous observations of class III YSOs by a factor of 5-10. At such levels we are able to detect emission down to debris disk levels (i.e., $0.024M_{\oplus}$). \\
\newline
This work is the first of two papers reporting on these class III observations. Here we discuss our survey of class III sources alongside a wider sample of Lupus Young Stellar Objects, while \citet{Lovell20} presents a more detailed analysis of NO~Lup, a class III source found in this survey to have a CO gas counterpart. In $\S$\ref{sec:sources} we discuss how we selected the sources for this survey, while further context for the photometric classification of the full sample of Lupus YSOs considered in this work is given in $\S$\ref{sec:photometry}. The observational set up of the ALMA measurements is described in $\S$\ref{sec:ALMAObsFull}. We analyse the continuum measurements in $\S$\ref{sec:contAnalysis}, the CO measurements in $\S$\ref{sec:COSurvey}, provide a detailed discussion of our results and their possible interpretations in $\S$\ref{sec:discussion}, and summarise our key findings and conclusions in $\S$\ref{sec:conclusions}.

\section{Source Selection}
\label{sec:sources}
We compiled a sample of 82 candidate class III sources from \citet{Hughes94, Padgett06, Merin08, Mortier11, Alcala14}.
We used Gaia astrometry to distinguish candidate class III young stellar objects from contaminant background objects from this sample. Indeed, it has been shown that many candidate class III young stellar objects in the Lupus star forming region are in reality background objects \citep{Manara18}.
Of the 82 class III candidates, 13 did not have Gaia astrometric solutions \citep{Gaia18} and were removed from the sample (although we note may still be class III Lupus YSOs). Of the remaining 69, 30 were deemed to be background sources due to Gaia parallax smaller than 2\,mas, corresponding to distances much greater than $200$\,pc whereas the distance to the majority of Lupus targets is ${\sim}160$\,pc \citep{Manara18, Alcala19}. In Fig.~\ref{fig:sourceSelection} the 30 background sources are seen with empty black markers, where it is also shown that the 39 class III sources (orange, either filled/unfilled) have proper motions consistent with Lupus YSOs, $\mu_{\alpha}$ and $\mu_{\delta} = (-16\pm2,-21\pm2)$\,$\rm{mas}\,\rm{yr}^{-1}$ \citep{Galli13} and lie close to the 3\,Myr isochrone in magnitude-colour space (see right plot, except for one class III faint G-band magnitude source). The final sample of 30 was then chosen from these sources (excluding the faint G-band source, and preferentially choosing the closer sources whilst maintaining a spread of distances), shown as filled orange markers in Fig.~\ref{fig:sourceSelection}. \\
\newline
The final sample of 30 Lupus class III's only included sources from Lupus Clouds I and III, and include 1 A-type, 14 K-type, and 15 M-types stars. 
Six of the sample of 30 are in known binary systems, with the survey star in each of these binaries being either an K-type or M-type star \citep[][]{Zurlo20}, see Table~\ref{tab:class3Objs} and $\S$\ref{sec:discBinary} for more details. Stellar masses were available from the literature for 16/30 of the sources \citep{Manara13, Galli15, Pericaud17}. We estimate the stellar masses of the remaining 14 class III's in $\S$\ref{sec:sed}. We note that the average distance to sources in the sample is $141$\,${\rm{pc}}$, slightly closer than the average distance for the class II sources of ${\sim}160\rm{pc}$. \\
\newline
For the class II's also being considered as part of a larger sample of Lupus YSOs, we include the 95 ALMA observed sources discussed in \citet{Ansdell16, Ansdell18}, the two class II sources J16104536-3854547 and J16121120-3832197 \citep[both of which were ruled out by][]{Ansdell18}, and the additional 5 brown dwarf class II sources discussed in \citet{Sanchis20} in Fig.~\ref{fig:sourceSelection} with blue markers. Although 17 of these 102 class II sources were not found to have Gaia astrometric solutions (and are not plotted) we still include these in our sample for it to be as complete as possible. In the left plot of Fig~\ref{fig:sourceSelection}, 6 sources have Gaia parallaxes smaller than 3\,mas, and 5 of those 6 also have low proper motions inconsistent with Lupus \citep{Galli13}. Although we cannot rule these out as having erroneous Gaia data, we excluded these 6 sources from our analysis; these are J15430227-3444059, J16104536-3854547, Sz~123B, J16120445-3809589 and J16121120-3832197 from \citet{Ansdell16} and Sz~102 from \citet{Ansdell18, Sanchis20}, i.e., we also rule out 2 of the same sources as \citet{Ansdell18}. We therefore consider a total of 96 class II and 30 class III YSOs to form a representative sample of 126 Lupus class II and III stars. Further detailed discussion of the class II sources is included in \citet{Ansdell16, Ansdell18, Sanchis20}, which we refer to throughout.

\section{Photometric Classification}
\label{sec:photometry}
\subsection{Mid-IR Photometry}
\begin{figure}
    \includegraphics[width=1.0\columnwidth]{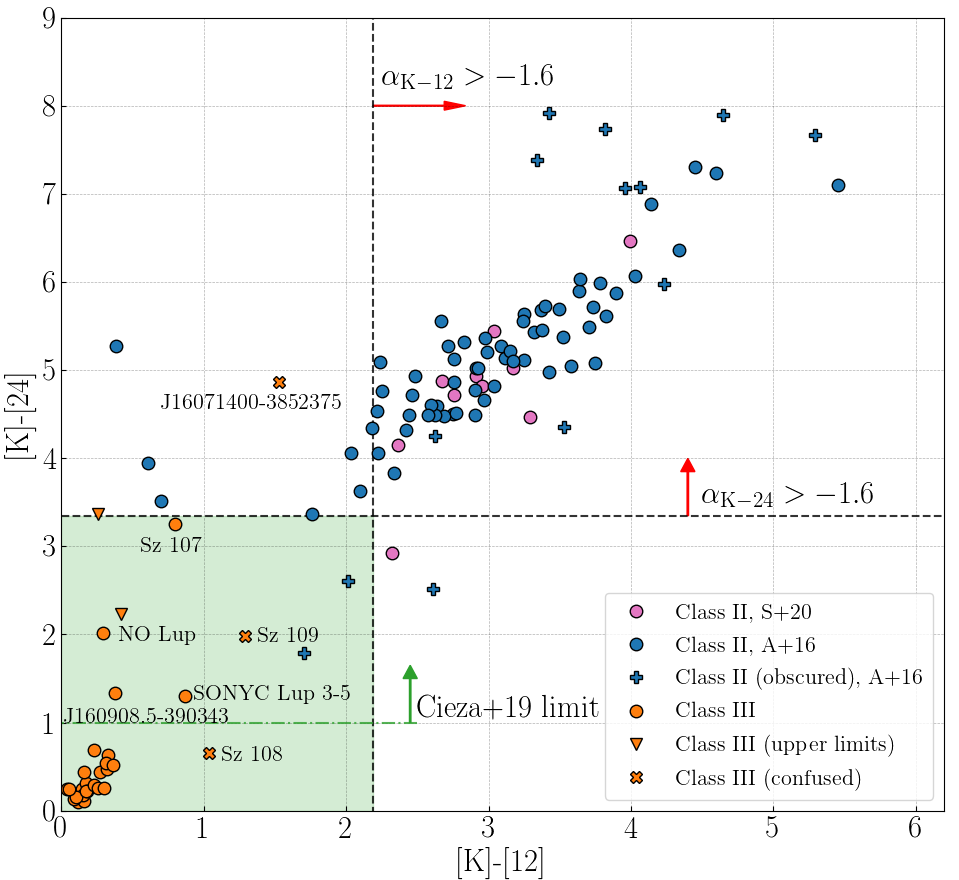}
    \caption{Lupus stars in our sample are shown in orange if class III, or else if they are class II, in blue or purple (if these were sourced from \citet{Ansdell16} or \citet{Sanchis20} respectively). Upper limits for those without 24\,$\mu$m detections are shown with downward pointing triangles, those marked as blue "+" crosses were deemed as obscured \citep[see][]{Ansdell16} for the class II sources, and those marked as orange "x" crosses are deemed confused class III sources in this work. The grey dashed lines show that corresponding to $\alpha_{\rm{K-12}}$ and $\alpha_{\rm{K-24}}$ = -1.6 (with the class III region shaded in green), and the green dot-dashed line represents the cut to which all YSOs in the Ophiuchus survey \citep{Cieza19} conformed (discussed further in $\S$\ref{sec:discussion}). We note that 24$\mu$m photometry for J16071400-385238 was not available, therefore this data point is plotted using its 22$\mu$m WISE 4 upper limit flux instead.}
    \label{fig:photometryC2C3}
\end{figure}

The sample of class III stars described in \S \ref{sec:sources} used the classification given in the literature \citep{Merin08, Evans09, Cutri13, Alcala14}. Fig.~\ref{fig:photometryC2C3} considers how the infrared properties of the class III sub-sample compare with those of the class II sub-sample, by showing their [K]-[12] and [K]-[24] colours. 
The grey dashed lines indicate the regions for which $d\log{\nu F_{\nu}} / d \log{\lambda}>-1.6$ between K-band (\textit{2MASS}) and either 12$\mu$m (\textit{WISE}) or 24$\mu$m (\textit{Spitzer}) fluxes (i.e., the photometrically defined boundary between class II and III, for which all sources below both are defined as being in the lower left quadrant).\\
\newline
It can be seen that 29/30 of the class III stars are inside this lower-left quadrant, alongside two class II sources. We note that both of these class II sources are defined in \citet{Ansdell16} as being obscured, and although form part of the sub-sample of class II stars, are removed from subsequent analysis for this reason. One class III source J16071400-385238 is outside of this region (in the upper left quadrant), however further inspection of this finds this source to be confused at 12$\mu$m (i.e., it does not appear to be a point source in WISE W3). This source also lacks MIPS 24$\mu$m photometry (i.e., it was too faint to be detected by Spitzer), and has its [24] magnitude estimated from its WISE W4 (22$\mu$m) photometry, which was an upper limit, and also happens to be the faintest in the sample (in the K band). One class II source identified in \citet{Sanchis20}, AKC2006-18, falls below the $\alpha_{\rm{K-24}}$ threshold, but not the $\alpha_{\rm{K-12}}$ threshold, placing it in the lower-right quadrant (i.e., although it may have 24$\mu$m photometry consistent with our class III definition, this is not the case for its 12$\mu$m photometry). Nevertheless, this source remains distinct from the population of class III YSOs. Therefore except for the three confused/obscured sources, the complete class II and III sample are seen to be well separated, and are shown to correspond as expected from $\alpha_{\rm{IR}}$ between the 2.2$\mu$m K band and 24$\mu$m mid-IR fluxes.\\
\newline
In addition, we find that the majority of our class III sample sit in a cluster in the lower left corner of Fig.~\ref{fig:photometryC2C3}, which is where stars exhibiting purely photospheric emission would be expected. However, falling outside or inside this cluster is no guarantee of a star having discernable excess emission or not. For that a more comprehensive analysis of the full SED is required, which we discuss in $\S$\ref{sec:sed}. For example, 4 of the sources in the lower-left class III cluster are found in such an analysis to have infrared excesses (MU~Lup, J155526-333823, MV~Lup, CD-31~12522), while 2 of the non-confused sources external to this cluster are found to have no significant excesses (J1609085-390343 and SONYC~Lup~3-5).\\
\newline
Considering the class III sources that lie outside the lower-left cluster in more detail, both NO~Lup and Sz~107 are later confirmed to have mid-IR excesses. Two class III sources are found outside the cluster but with 24$\mu$m upper limits that, while consistent with $\alpha_{\rm{K-24}}<-1.6$, could also be compatible with an excess being present (these are also broadly consistent with having no 12$\mu$m excess). There are two class III stars outside the cluster for which we consider the 12$\mu$m emission to be confused, and so their position on the figure should not be taken as evidence for excess emission (Sz~109 and Sz~108). Inspecting Sz~109 with the IRSA Finder Chart, it is found with significant levels of diffuse emission nearby and does not well resemble a point source in WISE W3. This is one of the two faintest YSOs in the sample (along with J16071400-385238) and is therefore deemed as having confused photometry. Sz~108 is nearby (${\sim}4\arcsec$) to the class II star, Sz~108B. This star could have contributed to Sz~108's $12$\,$\mu \rm{m}$ flux, and we note that the WISE photometric flag suggests this is variable. Accounting for these confused sources, we therefore note that whilst the existence of a photospheric excess is not ruled-out by being part of the lower-left cluster, there appears to be weak positive correlation between the sources outside/inside of this either having/not having a real mid-IR excess. We consider in more detail the significance of these $12$\,$\mu \rm{m}$ and $24$\,$\mu \rm{m}$ excesses in $\S$\ref{sec:sed}. \\
\newline
Six of the class II sources are found to have $\alpha_{\rm{IR}}<-1.6$ if 12\,$\mu$m is used instead of 24\,$\mu$m. This suggests that whilst these may have large 24\,$\mu$m fluxes, their warmer 12\,$\mu$m emission is less excessive in comparison to their photospheres. This might be expected if their disks have an inner cavity, i.e., the morphology of the class of protoplanetary disk known as a transition disk. From left-right (in [K]-[12]) these are J16083070-3828268, Sz~111, Sz~84, Sz~115, J16000060-4221567, and J16101857-3836125, and indeed the first three have previously been identified as transition disks \citep{Merin08, Merin10, Ansdell16, VDMarel18}. The left-hand side of this region (i.e., [K]-[12]<1.5, [K]-[24]>3.3) may be a reasonable probe for transition disk objects. We note that whilst Sz~107 sits below the [K]-[24] magnitude threshold that results in classification as a class III, it is closer to the class II's Sz~84 and Sz~111 than any of the other class III stars in our sample, which we will consider further in $\S$\ref{sec:discussion}.

\begin{figure*}
    \includegraphics[width=1.0\textwidth]{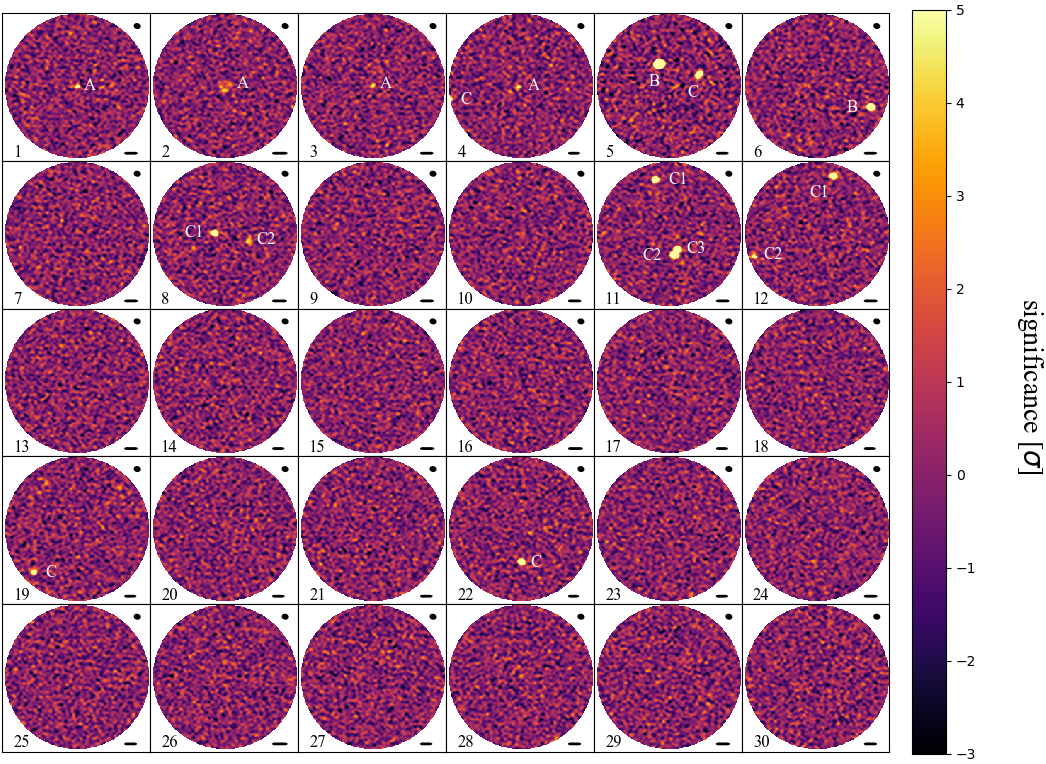}
    \caption{ALMA survey image showing all 30 observations in scale of significance (with respect to each image rms). Significant emission is noted with either an "A", "B" or "C", and numerically counted if more than one source type in a single image is detected. The source numbers in the bottom-left of each panel correspond to those in Table~\ref{tab:class3Objs}. In all images, North is up, East is left, the beam and 200\,au scale bar are shown in the upper and lower right respectively. }
    \label{fig:class3SurveyIm}
\end{figure*}

\subsection{H-alpha Equivalent Widths}
\label{sec:Halpha}
We briefly consider here whether or not the sample of 30 class III YSOs shows evidence for accretion based on their H$\alpha$ equivalent widths (EW). Using the classification scheme of \citet{White03} (i.e., based on the width of the H$\alpha$ line with respect to stellar type), we note here that 26/30 of our class III stars showed no evidence of accretion \citep{Krautter97, Cieza07, Wahhaj10, Manara13, Galli15}. In the case of 4/30 sources (J16071400-385238, J160758.9-392435, J160908.5-390343 and J160917.1-392710) data was not available in the literature to check this (i.e., these appear to be missing spectra). Therefore, although in the case of 4 YSOs evidence for ongoing accretion cannot be determined, none of the class III stars for which data was available showed evidence of accretion. We will revisit this in $\S$\ref{sec:discussion} when we consider the evidence for other processes in these YSOs.

\subsection{Summary of Photometric Classification}
In summary, the classification in the literature translates well into the cuts shown in colour-colour space. Accounting for confusion and flux upper limits, we have shown that our sample of class III YSOs is photometrically distinct to the class II sample in having low $12\mu \rm{m}$ and $24\mu \rm{m}$ spectral indices, defined as $[\rm{K}]-[24]<3.3$ and $[\rm{K}]-[12]<2.2$. We will discuss further the presence of mid-IR excesses in $\S$\ref{sec:sed}. Additionally, we showed that for all class III sources with data available, none showed evidence for accretion. 

\section{Observations}
\label{sec:ALMAObsFull}
ALMA was set up to observe the 30 class III YSOs in 21 separate scheduling blocks, each comprising of two scans of each star as part of programme 2018.1.00437.S. In total each source was observed 42 times (i.e., for a total of ${\sim}33.6$\,minutes on each). All measurements were made in ALMA cycle 8 over the course of the first two months of 2019, using 43 antennae and baselines ranging between 15-360m. The observing routine and sources used as bandpass, flux and phase calibrators are summarised in Table~\ref{tab:ObsRoutine}. The single phase calibrator J1610-3958 was used for all observations, and as both flux and bandpass calibrators, either J1517-2422 or J1427-4206 were used (as detailed in the table), and the precipitable water vapour (PWV) during these measurements ranged between 0.4-1.7\,mm.\\
\newline
ALMA was used in Band 7, having correlators set up with 3 spectral windows centred on 343.240\,GHz, 355.240\,GHz, and 357.135\,GHz each with a bandwidth of 2.0\,GHz and channel widths of 15.625\,MHz (for a total of 128 channels each) set up for continuum observations. A fourth spectral window was centred on 345.135\,GHz with a bandwidth of 1.875\,GHz and channel widths of 488.28\,kHz (for a total of 3840 channels) set up for $\rm{CO}$ J=3-2 spectral line observations. From these frequencies and bandwidths, we calculate a bandwidth weighted wavelength of $\lambda=856\mu$m and quote this as our ALMA Band~7 wavelength herein. \\
\newline
The Band 7 visibility data sets were calibrated using the CASA software version 5.1.1-5 with the standard pipeline provided by the ALMA Observatory. 
No data flagging was applied. 
Data for each source was concatenated using the CASA package \textit{concat}, and then the concatenated measurement sets (for the spectral windows conducting continuum measurements) were time averaged to 10-second widths and averaged into 4 channels (i.e., per spectral window) for each of the field observations. Separate measurement sets were produced for the analysis of J=3-2 CO emission, which were not channel or time-averaged, (i.e., these retained the full 3840 channels). The CASA \textit{statwt} task was run on all measurement sets with which we conducted our continuum analysis, as outlined in $\S$\ref{sec:contAnalysis}.

\begin{table}
    \centering
    \caption{Observation data for the ALMA observing routine for the 42 scans on each source. Note that the Scan Time column represents the amount of time spent for each number of scans in a given schedule block.}
    \begin{tabular}{c|c|c|c|c|c}
         \hline
         \hline
         D.M.Y & $N_{\rm{scans}}$ & Scan Time & Flux and & Phase \\
         &&[mins]&Bandpass&&\\
         \hline
         23.01.19 & 2 & 48 & J1517-2422 & J1610-3958 \\
         09.03.19 & 2 & 48 & J1517-2422 & J1610-3958 \\
         11.03.19 & 4 & 48 & J1517-2422 & J1610-3958 \\
         12.03.19 & 2 & 48 & J1427-4206 & J1610-3958 \\
         12.03.19 & 2 & 48 & J1517-2422 & J1610-3958 \\
         13.03.19 & 4 & 48 & J1517-2422 & J1610-3958 \\
         14.03.19 & 2 & 48 & J1517-2422 & J1610-3958 \\
         15.03.19 & 2 & 48 & J1517-2422 & J1610-3958 \\
         16.03.19 & 2 & 48 & J1517-2422 & J1610-3958 \\
         17.03.19 & 2 & 48 & J1427-4206 & J1610-3958 \\
         17.03.19 & 4 & 48 & J1517-2422 & J1610-3958 \\
         18.03.19 & 6 & 48 & J1517-2422 & J1610-3958 \\
         19.03.19 & 2 & 48 & J1427-4206 & J1610-3958 \\
         19.03.19 & 6 & 48 & J1517-2422 & J1610-3958 \\
    \hline
    \end{tabular}
    \label{tab:ObsRoutine}
\end{table}

\section{Survey Continuum Analysis}
\label{sec:contAnalysis}
This section presents analysis of the sub-mm continuum observations of the 30 class III stars. Detection types are described in \S \ref{sec:surveyanalysis}. In \S \ref{sec:sizemass} constraints are presented for the the sizes and masses of the detections, and upper limits are discussed in \S \ref{sec:stackedIm} for those where no such detection was made. SEDs of the class III sample are considered in \S \ref{sec:sed} to further constrain their dust distribution, and we consider the potential for extra-galactic confusion in \S \ref{sec:BkgnSources}.

\subsection{Continuum Detections}
\label{sec:surveyanalysis}
Non-interactive t-clean was used to image each measurement set, over 50,000 iterations with a threshold of ${\sim}122$\,$\mu$Jy (${\sim}$3 times the rms, see later) and \textit{auto-multithresh} masking applied, shown in Fig.~\ref{fig:class3SurveyIm}. The field of view of our images are set to show the entire primary beam (${\sim}12\arcsec$ radius). 
These images are scaled with respect to the image rms, which has a mean level across all images of $36.7\pm1.3$\,$\mu \rm{Jy}\,\rm{beam}^{-1}$. The average synthesised beam has a position angle $84.7^{\circ}$ and size ($0.75\times0.62$)$\arcsec$ when imaged with natural weighting. The range of major and minor axis extents, and position angles were $<0.005\arcsec$, $<0.017\arcsec$ and $<0.9^{\circ}$ respectively, demonstrating strong consistency in the synthesised beams for each source as expected given the observing sequence.\\ 
\newline
The Target ID and image number in Fig.~\ref{fig:class3SurveyIm} correspond to the "No." column in Table~\ref{tab:class3Objs}.
We identified potential sources as those where the images exhibited brightnesses above 3$\sigma$.
To confirm these as sources we required a $>5\sigma$ detection of emission within an aperture centred on the peak pixel of the source.
Three aperture sizes were used ${\sim}(1.1\times0.9)\arcsec$, ${\sim}(1.5\times1.2)\arcsec$, or ${\sim}(2.3\times1.9)\arcsec$, to account for the possibility that the sources are unresolved, partially resolved, or well resolved, respectively.
If the unresolved/partially resolved aperture sizes measured consistent flux values, then we report the measurement based on the smaller aperture (i.e., the one with the lower error).
As expected, the unresolved aperture flux measurement is consistent with that of the source's peak pixel value in units of Jy\,beam$^{-1}$.
In the case of the resolved source J155526.2-333823 (the only source requiring the largest aperture), the peak pixel is not well-centred on the source, and we centred this aperture on the ALMA phase centre (which is expected to correspond to the centre of the disk). \\
\newline
In total, we detected 17 sources in 11 separate images, which we classified as either type "A", "B" or "C" depending on their proximity to known class II and III stars.
Detections within $1.0\arcsec$ of the image centre were given an ID of "A" and considered associated with the class III star. 
Detections $>1.0\arcsec$ from the image centre consistent with the position of a known class II star were given an ID of "B". Significant detections $>1.0\arcsec$ from the image centre, and inconsistent with the position of known class II stars, were given an ID of "C". 
The separation between the image centre and the centroid of the detected sources is given for all significant source detections in the "Sep." column of Table~\ref{tab:class3Objs}. This analysis finds 4 A sources (1A, 2A, 3A, and 4A), 2 B sources (5B and 6B), and 11 C sources (4C, 5C, 8C1, 8C2, 11C1, 11C2, 11C3, 12C1, 12C2, 19C, and 22C), and Table~\ref{tab:class3Objs} is divided into two parts based on this classification (i.e., depending on whether the images contain a type A or B detection in the upper section, or contain either a type C detection or no detection at all in the lower section). The following apertures were used for the source fluxes given in Table~\ref{tab:class3Objs}: the large aperture (2A), the partially-resolved aperture (5B, 5C, 8C2, 12C1, 12C2, 19C and 22C) and the unresolved aperture (4C, 6B, 8C1, 11C1, 11C2 and 11C3). For the 26 class III stars where no type A detections were made, we determined the $3\sigma$ upper limit on their fluxes from the rms within apertures with an area equal to the size of the beam (i.e., $3\times36.7\mu\rm{Jy}$, which assumes disk radii $\leq \rm{FWHM_{beam}}\times d/2 {\sim}50$\,au). Given that the expected sub-mm stellar flux for class III and class II sources is significantly lower than the detection limits, we interpret these 4 A detections and 2 B detections as due to circumstellar dust. \\
\begin{landscape}
\begin{table}
    \small
    \centering
    \caption{Class III sample including stellar information and results from ALMA continuum imaging. Notes a: source partially lost at edge of image; total flux uncertain, b: two sources overlap; flux may be confused, c: \citet{Merin08}, d: \citet{Zurlo20}, e: \citet{Hardy15} f: \citet{Padgett06}, g: \citet{Koehler00}, h: \citet{Torres06}, i: \citet{Gaia18}, j: \citet{Manara17}, k: \citet{Krautter97}, l: \citet{Manara13}, m: \citet{Galli15}, n: this work.}
    \begin{tabular}{cccccccccccccccc}
         \hline
         \hline
         No. & Target & Other Name & RA & Dec &Lupus & SpT & $M_{\star}$ & d & Sep. & ID & $F_{\rm{856 \mu m}}$ & $M_{\rm{dust}}$ & Notes  \\
         & & & [J2000] & [J2000] & Cloud & & [$M_{\odot}$] & [pc] & [$\arcsec$] & & [mJy] & [$M_{\oplus}$] & \\
         \hline
         1&J154041.2-375619&MU~Lup&15:40:41.170$^{\rm{i}}$&-37:56:18.540$^{\rm{i}}$&I&K6.0$^{\rm{j}}$&$0.89\pm0.15^{\rm{m}}$& $135.3\pm0.8$&$0.12$&A& $0.26\pm0.05$&$0.045\pm 0.008$&-\\
         2&J155526.2-333823 &-&15:55:26.220 & -33:38:23.24&I&K5.0$^{\rm{g}}$&$0.91\pm0.15^{\rm{m}}$& $119.2\pm0.7$&$0.0$ &A& $0.68\pm0.07$& $0.093\pm 0.010$&-\\
         3& J160311.8-323920 & NO~Lup &16:03:11.812 & -32:39:20.31&I&K7.0$^{\rm{e,f}}$&0.7$^{\rm{e}}$ & $133.7\pm0.7$ & $0.0$& A & $0.21\pm0.04$ & $0.036\pm 0.007$&-\\
         4&J160841.8-390137 &Sz~107&16:08:41.79 & -39:01:37.0&III&M5.5$^{\rm{j,l}}$&$0.16\pm0.10^{\rm{l}}$& $152.3\pm2.5$&$0.38$&A& $0.22\pm0.04$& $0.050\pm 0.010$ &-\\
         &-&-&-&-&-&-&-&-&$12.3$&C&$1.45\pm0.20^{\rm{a}}$&-&-\\
         5&J160842.7-390618&Sz~108&16:08:42.73&-39:06:18.3&III&M1.0$^{\rm{c}}$&0.43$^{\rm{n}}$&$151.3\pm1.2$&-&-&<0.11&<0.024& Binary$^{\rm{d}}$\\
         &&Sz~108B&16:08:42.86&-39:06:15.04&III&&&$169\pm3$&$4.1$&B&$23.5\pm2.4$& $6.5\pm 0.6$ & - \\
         &-&-&-&-&-&-&-&-&$5.4$&C&$1.20\pm0.13$&-&-\\
         6&J160957.1-385948 &Sz~119&16:09:57.07&-38:59:47.9&III&M4.0$^{\rm{c}}$&$0.30\pm0.05^{\rm{m}}$& $110\pm4$&-&-&<0.11&<0.013&-\\
         &&Lup~818s&16:09:56.28&-38:59:51.90&III&&&$157.0\pm2.7$&$9.8$&B&$6.8\pm0.7$&$1.63\pm 0.21$&-\\
         \hline
         7& J153802.7-380723&MT Lup&15:38:02.724&-38:07:22.17&I&K5.0$^{\rm{k}}$ &0.90$^{\rm{n}}$&$133.0\pm0.7$&-&-&<0.11&<0.019&-\\
         8&J154038.3-342137 &Sz~67& 15:40:38.27&-34:21:36.7&I&M4.0$^{\rm{c}}$ &0.26$^{\rm{n}}$&$110.1\pm1.5$&-&-&<0.11&<0.013&-\\
         &-&-&-&-&-&-&-&-&$1.8$&C1& $1.00\pm0.11$&-&-\\
         &-&-&-&-&-&-&-&-&$4.1$&C2& $0.41\pm0.06$&-&-\\
         9&J154306.3-392020&-&15:43:06.245$^{\rm{i}}$&-39:20:19.482$^{\rm{i}}$&I&K6.0$^{\rm{j}}$&0.91$^{\rm{n}}$&$166\pm7$&-&-&<0.11&<0.030&Binary$^{\rm{d}}$\\
         10&J154641.2-361848&CD-35 10498& 15:46:41.199 & -36:18:47.44&I&K1.0$^{\rm{h}}$&$1.38\pm0.46^{\rm{m}}$&$149.5\pm1.7$&-&-&<0.11&<0.024&Binary$^{\rm{d}}$\\
         11&J154959.2-362958 &MV Lup& 15:49:59.219 & -36:29:56.59&I&K2.0$^{\rm{f}}$&$1.55\pm0.55^{\rm{m}}$&$139.3\pm0.7$&-&-&<0.11&<0.021&-\\
         &-&-&-&-&-&-&-&-&$9.5$&C1& $2.75\pm 0.29$&-&-\\
         &-&-&-&-&-&-&-&-&$3.6$&C2& $2.77\pm 0.28^{\rm{b}}$&-&-\\
         &-&-&-&-&-&-&-&-&$2.9$&C3& $1.48\pm0.15^{\rm{b}}$&-&-\\
         12&J155219.5-381932&MW Lup&15:52:19.522&-38:19:31.36&I&K7.0$^{\rm{f}}$&$0.80\pm0.12^{\rm{m}}$&$130.3\pm0.8$&-&-&<0.11&<0.018&-\\
         &-&-&-&-&-&-&-&-&$10.2$&C1& $2.73\pm0.29$&-&-\\
         &-&-&-&-&-&-&-&-&$11.2$&C2& $1.39\pm0.19$&-&-\\
         13&J155533.8-370941&MX Lup&15:55:33.795&-37:09:41.08&I&K6.0$^{\rm{f}}$&0.85$^{\rm{n}}$&$129.6\pm0.8$ &-&-&<0.11&<0.018&-\\
         14&J160159.2-361256&NN Lup& 16:01:59.272 & -36:12:53.89&I&K3.0$^{\rm{f}}$&$1.20\pm0.29^{\rm{m}}$&$136.3\pm0.7$&-&-&<0.11&<0.020&-\\
         15&J160352.5-393902&CD-39 10292&16:03:52.502&-39:39:01.28&III&K3.0$^{\rm{h}}$ &$1.01\pm0.25^{\rm{m}}$&$130\pm4$&-&-&<0.11&<0.019&Binary$^{\rm{d}}$\\
         16&J160430.6-320729&CD-31 12522&16:04:30.564&-32:07:28.81&III&K2.0$^{\rm{h}}$&1.17$^{\rm{n}}$&$130.6\pm0.8$&-&-&<0.11&<0.018&-\\
         17&J160713.7-383924&-&16:07:13.698&-38:39:23.88&III&K7.0$^{\rm{f}}$&$0.75\pm0.11^{\rm{m}}$&$158.9\pm2.3$&-&-&<0.11&<0.027&-\\
         18&J160714.0-385238&Lup~605& 16:07:14.00 & -38:52:37.9&III&M6.5$^{\rm{c}}$&0.03$^{\rm{n}}$&$124\pm4$&-&-&<0.11&<0.017&-\\
         19&J160749.6-390429&Sz~94&16:07:49.60&-39:04:29.0&III&M4.0$^{\rm{j,l}}$&0.28$^{\rm{l}}$ &$116.2\pm0.9$&-&-&<0.11&<0.015&-\\
         &-&-&-&-&-&-&-&-&$10.4$&C&$1.67\pm0.17$&-&-\\
         20&J160758.9-392435&Lup~714& 16:07:58.90 & -39:24:34.9&III&M5.0$^{\rm{c}}$ &0.05$^{\rm{n}}$&$162\pm3$&-&-&<0.11&<0.028&-\\
         21&J160816.0-390304&SONYC Lup3-5 & 16:08:16.03&-39:03:04.2&III&M6.5$^{\rm{j,l}}$&0.05$^{\rm{n}}$&$164\pm5$&-&-&<0.11&<0.029&-\\
         22&J160834.6-390534&V1027 Sco&16:08:34.60&-39:05:34.0&III&A3.0$^{\rm{c}}$&3.16$^{\rm{n}}$&$161.5\pm2.2$&-&-&<0.11&<0.028&-\\
         &-&-&-&-&-&-&-&-&$5.5$&C&$1.09\pm0.15$&-&-\\
         23&J160848.2-390419&Sz~109&16:08:48.16&-39:04:19.2&III&M5.0$^{\rm{c}}$&0.05$^{\rm{n}}$&$184\pm5$&-&-&<0.11&<0.036&-\\
         24&J160908.5-390343&Lup~608s&16:09:08.50 & -39:03:43.1&III&M5.0$^{\rm{c}}$&0.08$^{\rm{n}}$&$160.3\pm2.4$&-&-&<0.11&<0.028&-\\
         25&J160917.1-392710&Lup~710&16:09:17.13 & -39:27:09.7&III&M5.0$^{\rm{c}}$&0.03$^{\rm{n}}$&$131\pm3$&-&-&<0.11&<0.019&-\\
         26&J160942.6-391941&Sz~116& 16:09:42.57&-39:19:40.8&III&M1.5$^{\rm{c}}$&$0.43\pm0.10^{\rm{m}}$&$146.5\pm0.8$&-&-&<0.11&<0.023&Binary$^{\rm{d}}$\\
         27&J161012.2-392118&Sz~121&16:10:12.21&-39:21:18.3&III&M3.0$^{\rm{j,l}}$&$0.31\pm0.08^{\rm{m}}$&$158.0\pm1.7$&-&-&<0.11&<0.028&-\\
         28&J161016.4-390805&Sz~122&16:10:16.44&-39:08:05.4&III&M2.0$^{\rm{j,l}}$&$0.39\pm0.05^{\rm{m}}$&$138.9\pm0.6$&-&-&<0.11&<0.021&-\\
         29&J161153.4-390216 &Sz~124&16:11:53.35&-39:02:16.1&III&M0.0$^{\rm{c}}$&0.75$^{\rm{n}}$&$125.4\pm0.7$&-&-&<0.11&<0.017&-\\
         30&J161302.4-400433&V1097~Sco&16:13:02.416&-40:04:32.93&III&K7.0$^{\rm{f}}$& $0.80\pm0.13^{\rm{m}}$ & $139.9\pm2.5$ &-&-&<0.11& <0.021 &Binary$^{\rm{d}}$\\
         \hline
    \end{tabular}
    \label{tab:class3Objs}
\end{table}
\end{landscape}
\begin{table*}
    \centering
    \caption{Modelled parameters and constraints for the unresolved class III detections of MU~Lup, NO~Lup and Sz~107 in the upper section, and the modelled parameters for the resolved detection of J155526.2-333823 in the lower section.}
    \label{tab:2DFit}
    \begin{tabular}{c|c|c|c|c|c|c|c|c|c|c}
         \hline
         \hline
         Source & $I_0$ & $\sigma$ & $R_{95}$ & $R_{\rm{c}}$ & $\gamma_1$ & $\gamma_2$ & Inc & PA & $\Delta \rm{RA}$ & $\Delta \rm{Dec}$ \\
          & & [arcsec] & [au] & [au] & & & [deg] & [deg] & [mas] & [mas] \\
         \hline
         MU~Lup & $8.2\pm0.3$ & $0.14\pm0.07$ & $<53$ & - & - & - & $50\pm30$ & $90\pm50$ & $-50\pm50$ & $-90\pm30$ \\
         NO~Lup & $8.0\pm0.4$ & $0.19\pm0.07$ & $<56$ & - & - & - & $50\pm30$ & $120\pm50$ & $20\pm40$ & $30\pm40$ \\
         Sz~107 & - & - & $<60$ & - & - & - & - & - & - & - \\
         \hline
         J155526.2-333823 & $7.58\pm0.10$ & - & - & $80\pm15$ & $3.3\pm1.5$ & $3.8\pm0.8$ & $31\pm12$ & $138\pm21$ & $-30\pm60$ & $-90\pm50$ \\         
         \hline
    \end{tabular}
\end{table*}

In image 5 we did not detect emission from the class III star Sz~108, but emission consistent with Sz~108's class II binary companion, Sz~108B. Given the $>4.0\arcsec$ angular separation these sources are not confused. This is similarly true in image 6, in which the class III star Sz~119 was not detected, yet the nearby (${\sim}9.8\arcsec$ separated) class II source was \citep[i.e., J16095628-3859518, which we refer to as Lup~818s herein, see][]{Lopez05}. In the five other binary systems noted, neither the source nor its companion were detected. Further discussion of stellar multiplicity in Lupus for these 6 binaries is provided in \citet{Zurlo20}, and the effect of binarity on circumstellar disk evolution in our discussion section $\S$\ref{sec:discBinary}.

\subsection{Continuum Source Analysis}
\label{sec:sizemass}
\begin{figure*}
    \includegraphics[width=1.0\textwidth]{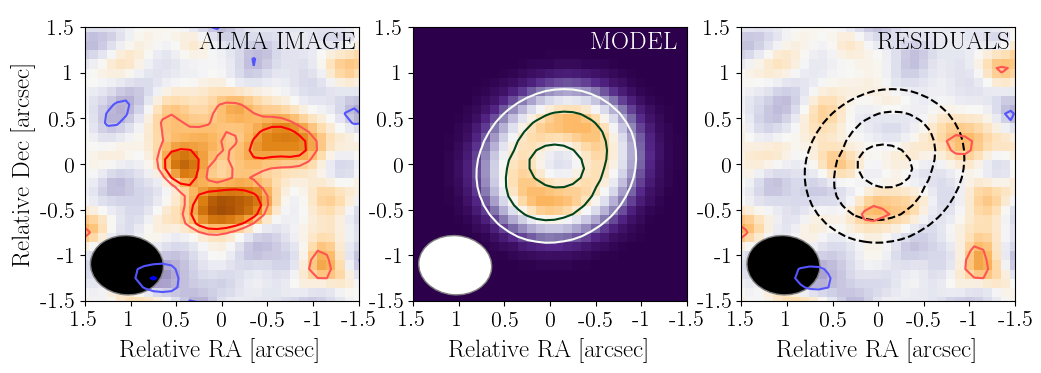}
    \caption{Model fits to the ALMA image of J155526.2-333823. {\it (Left)} ALMA cleaned image. The contours represent emission at $\pm2$ and $\pm3$ times the rms. {\it (Centre)} Best-fit model image based on values in Table~\ref{tab:2DFit} with contours set to demonstrate the ring-like nature of the model. {\it (Right)} Residual image (data minus model), with contours set at $\pm2$ and $\pm3$ times the rms, and the dashed lines showing the model contours for reference. In all images North is up and East is left, and the ellipse in the lower left corner represents the size of the synthesised beam (${\sim}90\times80\rm{au}$).}
    \label{fig:immodres_Source2}
\end{figure*}
\subsubsection{Continuum Source Modelling}
\label{sec:diskSizes}
The class III (type A) detections were unresolved for MU~Lup, NO~Lup and Sz~107, but resolved for J155526-333823.
To find limits on the radii of the unresolved sources, and disk parameters for the resolved source, we used GALARIO \citep{Tazzari18} to model the emission profiles of the 4 sub-mm detections.\\
\newline
For the unresolved observations the upper limit radii of the sources was determined using a simple parametrised Gaussian model defined as 
\begin{equation}
\label{eq:Gauss}
    I(r) = I_0 \exp { \Big[ - \Big( \frac{r}{2\sigma} \Big) ^2 \Big]},
\end{equation}
where $I_0$ is a normalisation factor, r is the deprojected sky radius, and $\sigma$ the Gaussian width of the profile. These parameters were fitted along with the positional offsets in RA and Dec, inclination (where $i=0^{\circ}$ corresponds to a face-on disk) and position angle (with positive values taken as anti-clockwise from North). By fitting with such a model, we find the average radius within which $95\%$ of the source flux is bounded, $R_{95}$, which we define as the radius limit for our unresolved sources \citep[in a manner consistent with][]{Sanchis20}.
For the resolved observation of J155526.2-333823, we followed the same procedure, but with a modified self-similar disk model with free parameters of the disk radius ($R_{\rm{c}}$) and inner and outer exponents ($\gamma_1$ and $\gamma_2$), defined as
\begin{equation}
\label{eq:modSelSim}
    I(r) = I_0 \Big( \frac{r}{R_{\rm{c}}} \Big)^{\gamma_1} \exp{ \Big[ - \Big( \frac{r}{R_{\rm{c}}} \Big)} ^{\gamma_2} \Big].
\end{equation}
\newline
All the values determined from our \textit{emcee} modelling \citep[for which 100,000 steps were used and either 36 or 48 walkers for the unresolved/resolved source modelling respectively][]{FM13} are shown in Table~\ref{tab:2DFit}. For Sz~107 the nearby bright type C source (likely background) means that the disk parameters found by GALARIO are much less accurate, so we use the results for NO~Lup to set an upper limit on the disk radius given their similar flux measurements (taking account of their different distances). \\
\newline
The constraints on the disk sizes for MU~Lup and NO~Lup (and estimate for Sz~107) show that this emission arises from dust within ${\sim} 60$\,au from their stars (and is entirely consistent with the size scales found by considering the FWHM of the source emission and the beam). Although the centroids of MU~Lup and NO~Lup are both well fitted by GALARIO, their inclinations and position angles are essentially unconstrained (as expected with these being unresolved). On the other hand, the disk of J155526.2-333823 is sufficiently large and bright that it was resolved and is well constrained by GALARIO, putting relatively tight bounds on its parameters (i.e., the radius, PA and inclination). This disk is the largest detected, with a radius ${\sim}80\rm{au}$, significantly greater than the blackbody radii obtained from SED fitting (see Table~\ref{tab:SEDBB}), where the positive $\gamma_1$ determines this to be a ring with an inner gap. Fig.~\ref{fig:immodres_Source2} shows the ALMA image, the disk model that was subtracted from this, and the resulting residual image, within which the lack of residuals $>3\sigma$ demonstrates the reasonable fit. \\
\newline
We note here that the aim of the survey was to maximise the S/N of the detections. As such, the ${\sim} 0.8\arcsec$ resolution was chosen that the majority should be unresolved, explaining why only 1 of our 4 detections was resolved. Follow-up high resolution measurements of the three unresolved sources (and additionally the resolved source) would be needed to gather much tighter constraints on the radii, widths, and other morphological features of these disks.

\subsubsection{Continuum Dust Masses}
\label{sec:Masses}
Based on the assumption that the type A and B detections are from optically thin, isothermal circumstellar dust, we estimated their dust masses (in $M_{\oplus}$) as
\begin{equation}
\label{eq:dustmass}
    M_{\rm{dust}} [M_{\oplus}] = \frac{F_{\nu}d^2}{\kappa_{\nu}B_{\nu}(T_{\rm{dust}})} \approx 0.22 \Big( \frac{d}{150} \Big)^2 F_{856\mu \rm{m}},
\end{equation}
where $d$ is distance in $\rm{pc}$ and $F_{856\mu \rm{m}}$ is the Band 7 flux in mJy, for an assumed dust temperature of $T_{\rm{dust}}{\sim}20\rm{K}$ \citep[e.g., the median for Taurus disks, see][]{Andrews05} and a dust grain opacity $\kappa_{\nu}$ of $10\rm{cm}^{2}\,\rm{g}^{-1}$ at 1000GHz, with an opacity power-law index $\beta=1$ \citep[e.g.,][]{Hildebrand83, Beckwith90}. These dust masses are noted in the Table~\ref{tab:class3Objs}. For the class III disks these masses range between $0.036 - 0.093 M_{\oplus}$, and for the class II disks they are $6.5$ and $1.63 M_{\oplus}$ respectively for sources Sz~108B (5B) and Lup~818s (6B). Given \citet{Ansdell16} assumed these 2 class II sources to be at $200\rm{pc}$, whereas here we have calculated their masses using their Gaia DR2 parallax distances, our dust mass measurements are smaller than those previously derived. After accounting for the different distances, these are consistent. For the type A sources with $3\sigma$ upper limits on their fluxes, eq.~\ref{eq:dustmass} is used to provide the $3\sigma$ upper limits on their masses. Given the type C detections do not coincide with known objects, these are most likely emission from background sub-mm extragalactic (SMG) sources, common in deep sub-mm observations (discussed further in $\S$\ref{sec:BkgnSources}), rather than stellar sources within Lupus. We therefore do not provide dust mass estimates for type C detections.

\begin{figure}
    \includegraphics[width=0.5\textwidth]{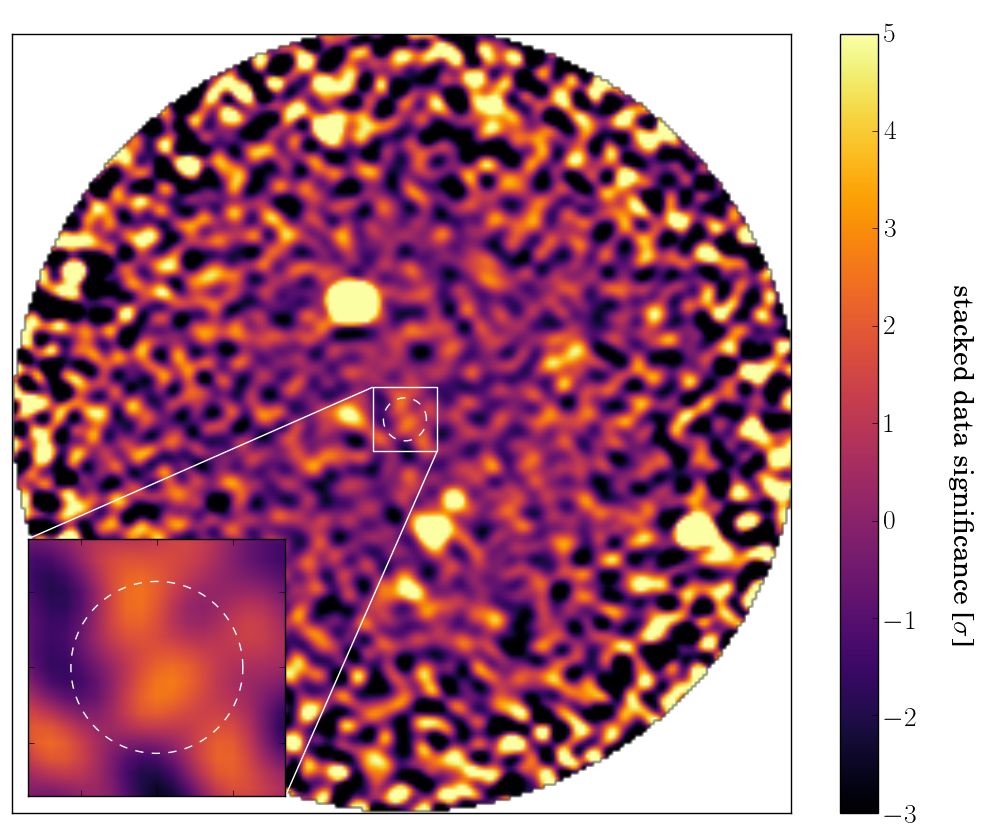}
    \caption{Stacked image of all 26 sources in scale of significance (with respect to the rms of the combined image stack, ${\sim}7.2\mu\rm{Jy}\,\rm{beam}^{-1}$). Within the inner ${\sim}0.68\arcsec$ radius circle there is a $2.8\sigma$ significant detection. Note that all other sources with significant emission are the non-type A detections as discussed in $\S$\ref{sec:contAnalysis}. In this stacked image, North is up, East is left.}
    \label{fig:class3Stack}
\end{figure}

\begin{figure*}
  \begin{center}
    \vspace{0.0in}
    \begin{tabular}{cc}
      \hspace{0.0in} \includegraphics[width=1.05\columnwidth]{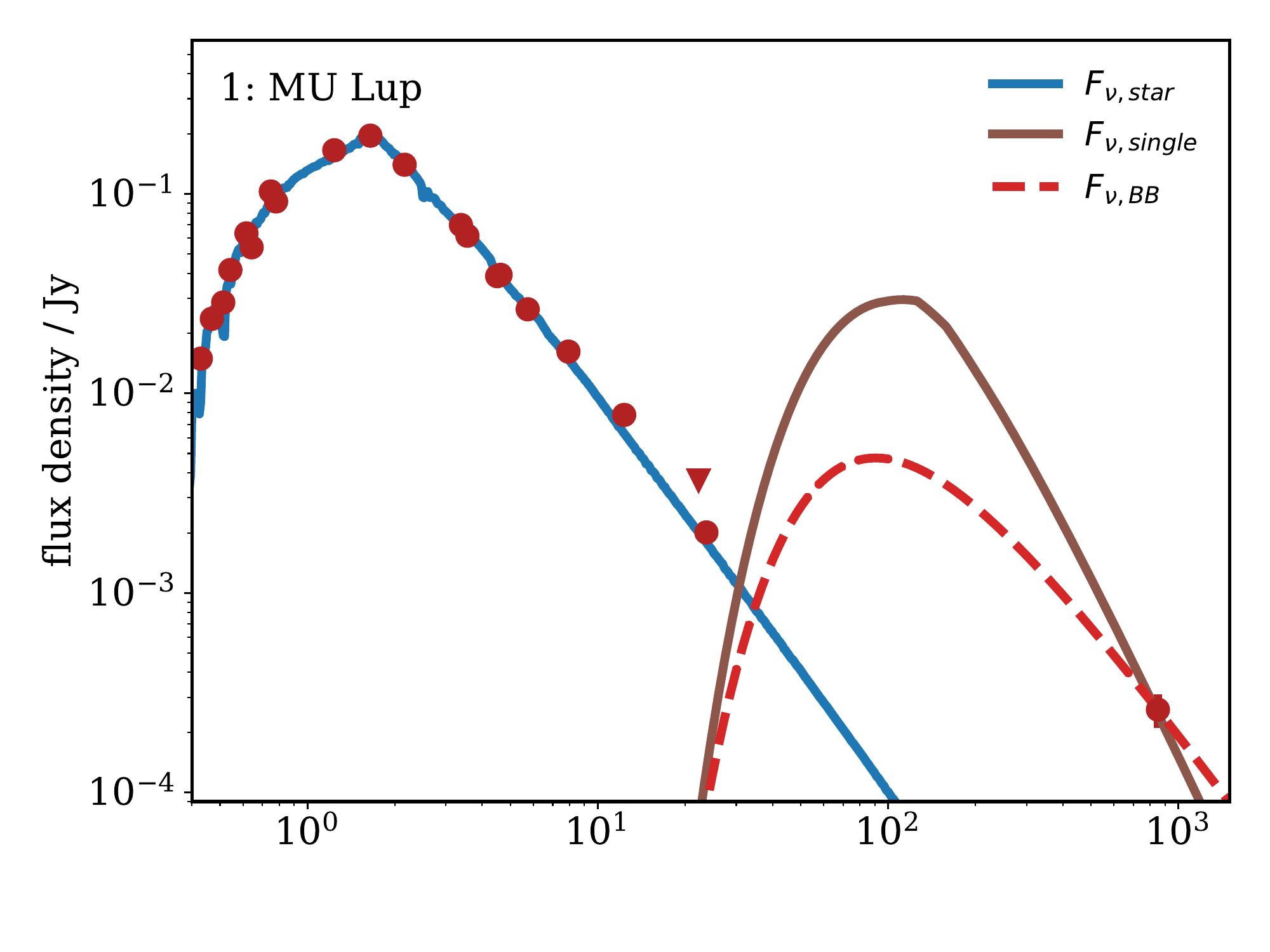} &
      \hspace{-0.22in} \includegraphics[width=1.05\columnwidth]{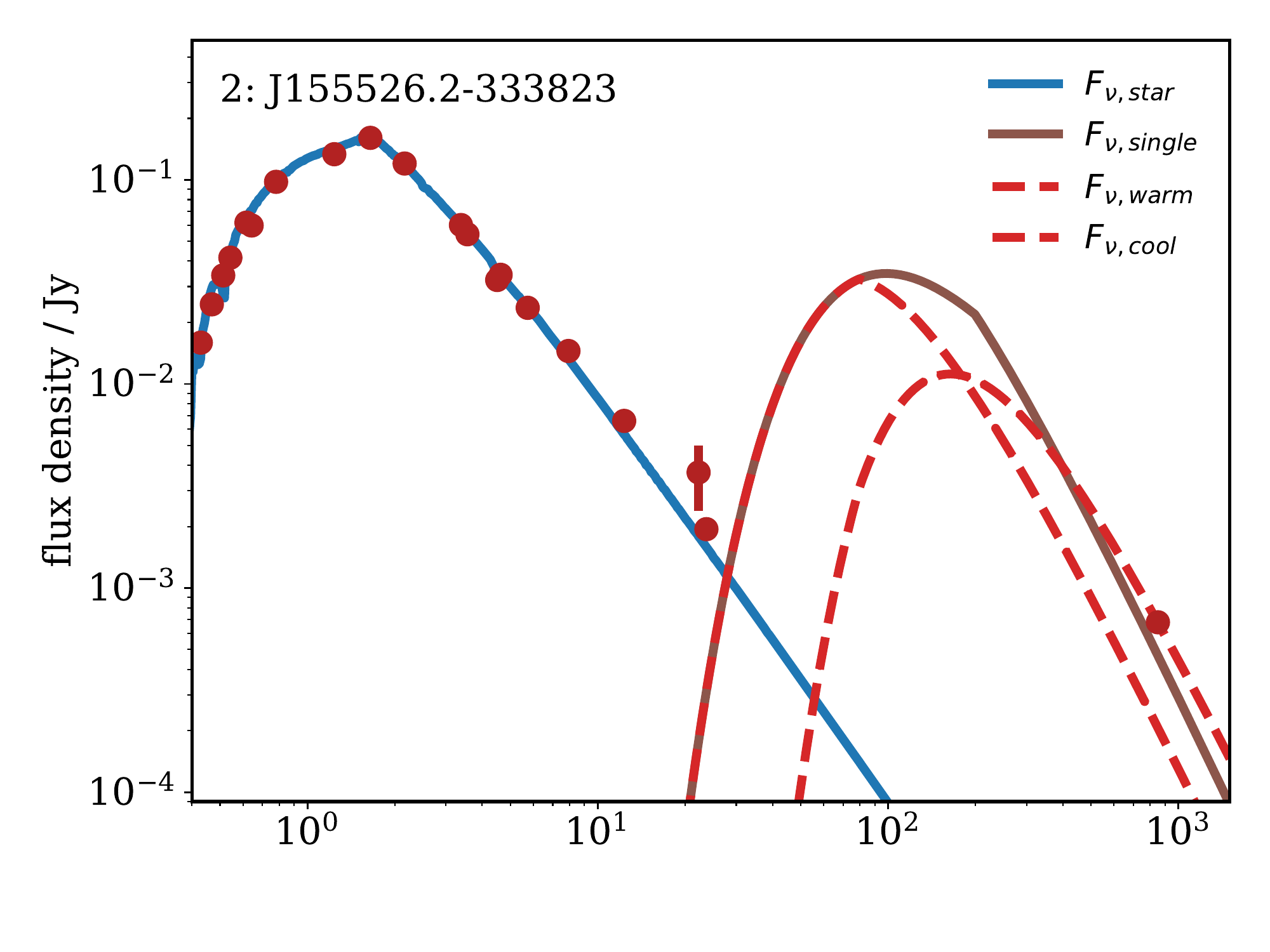} \\[-0.255in]
      \hspace{0.0in} \includegraphics[width=1.05\columnwidth]{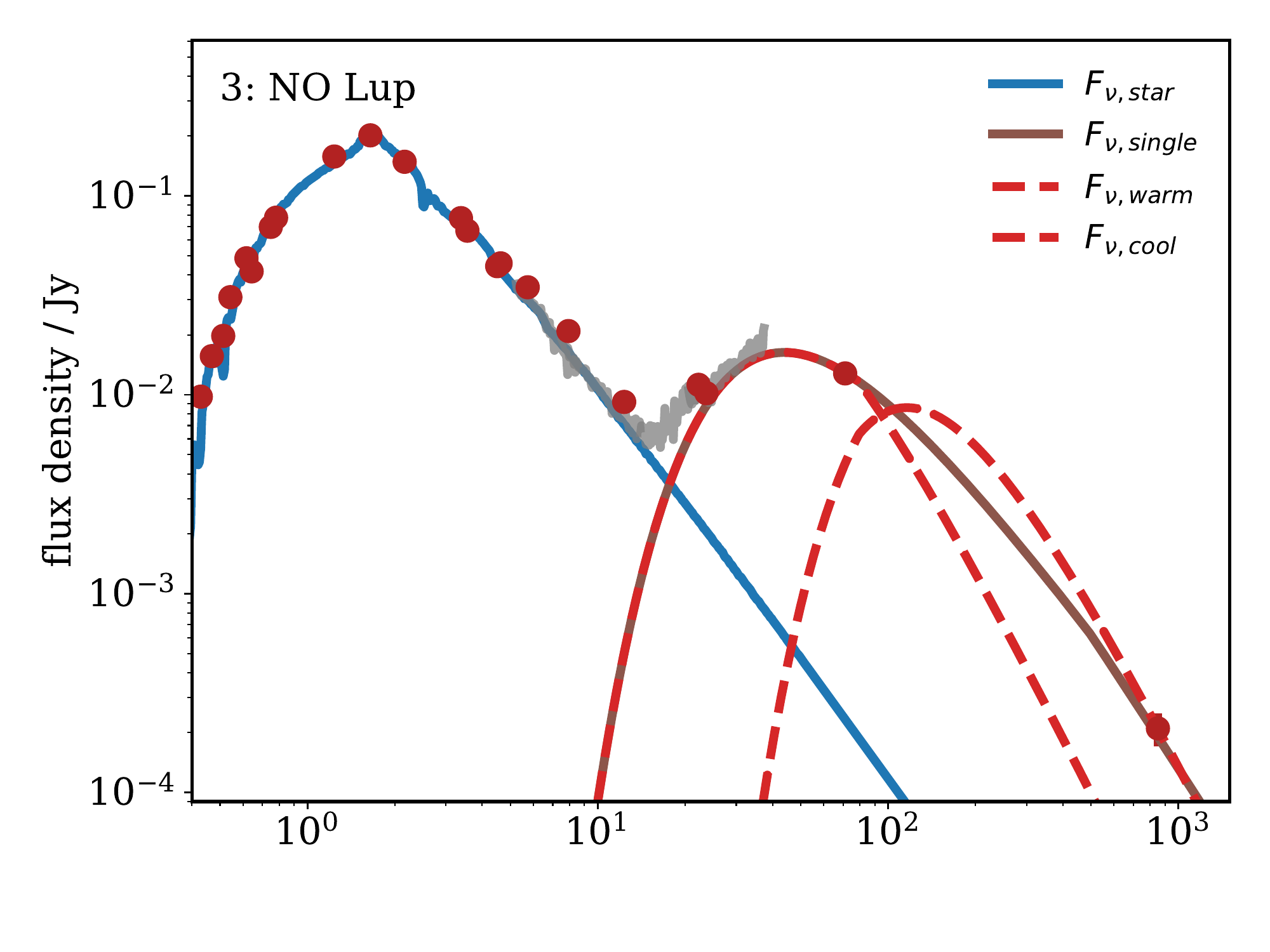} &
      \hspace{-0.22in} \includegraphics[width=1.05\columnwidth]{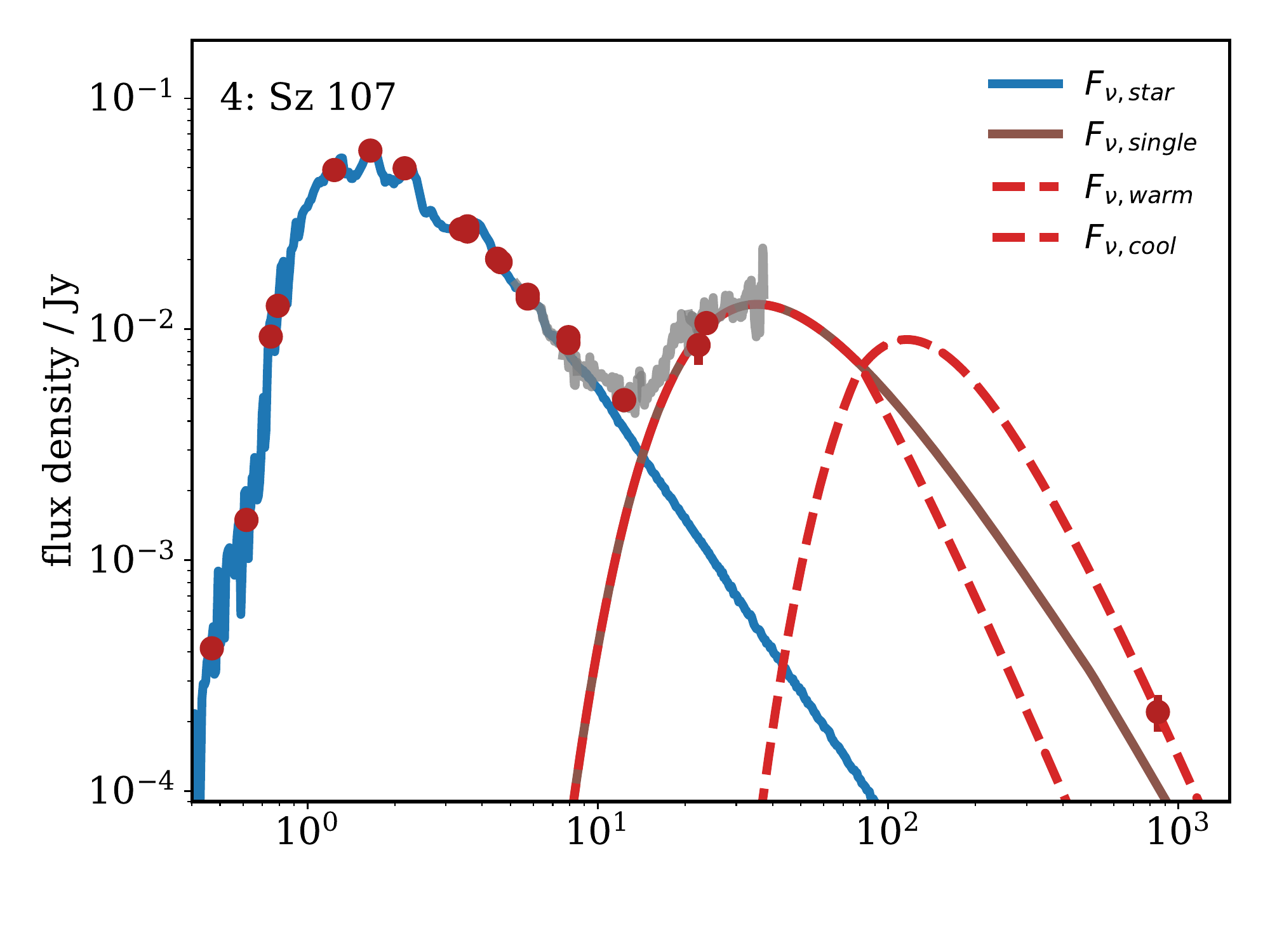} \\[-0.255in]
      \hspace{0.0in} \includegraphics[width=1.05\columnwidth]{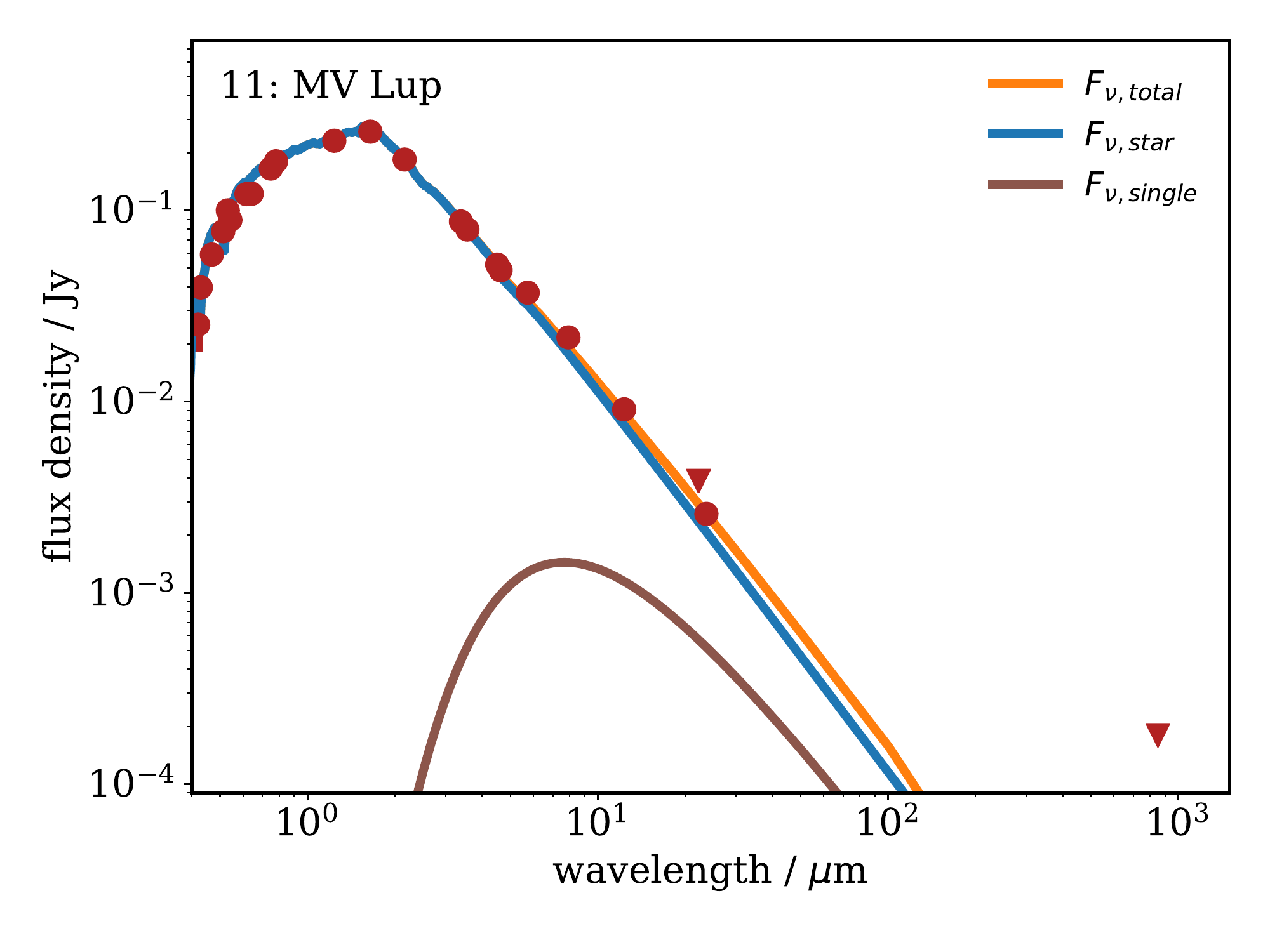} &
      \hspace{-0.22in} \includegraphics[width=1.05\columnwidth]{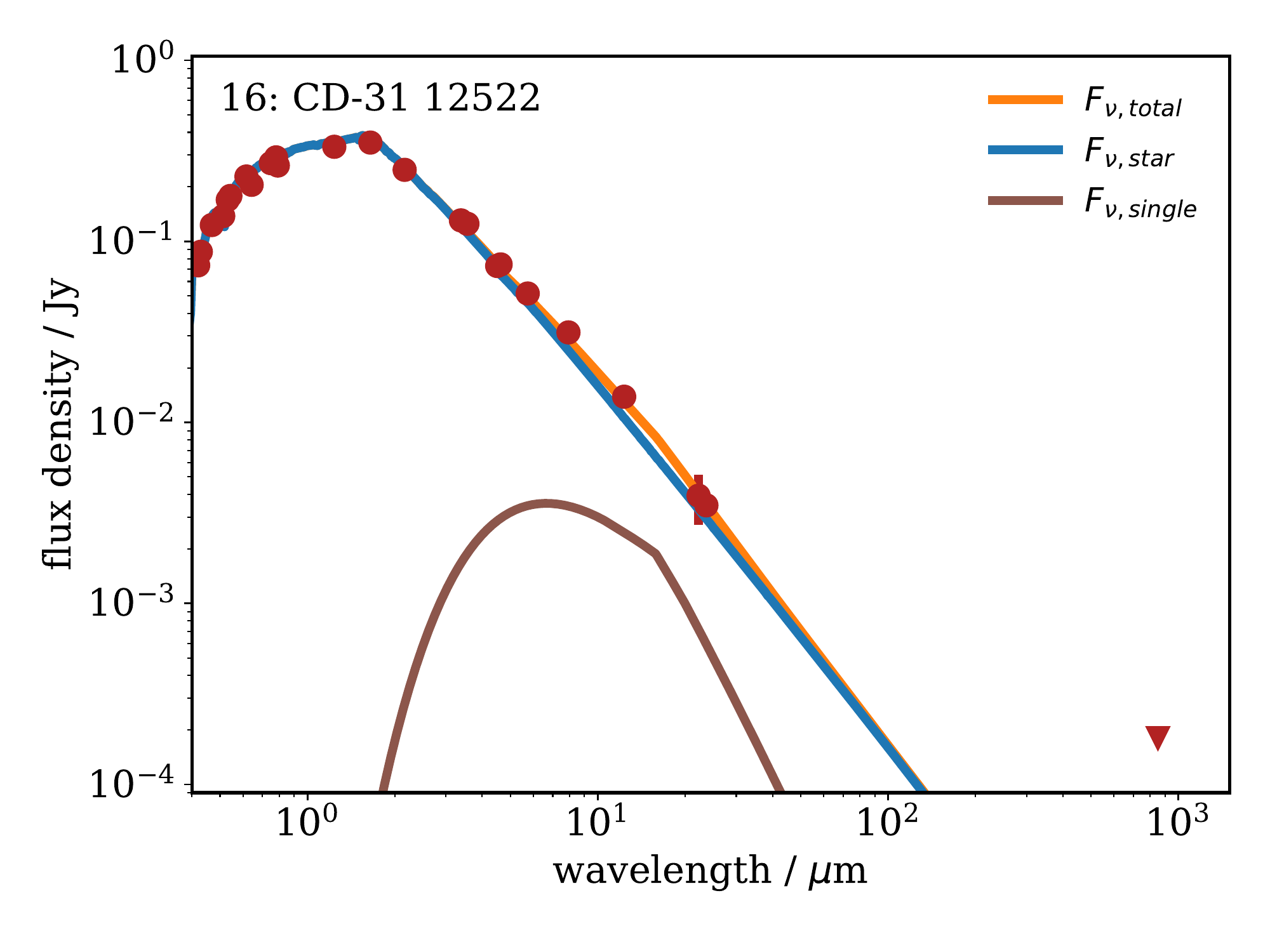}
    \end{tabular}
    \caption{SEDs for MU Lup (SpT: K6.0): top-left, J155526.2-333823 (SpT: K5.0): top-right, NO~Lup (SpT: K7.0): middle-left, Sz~107 (SpT: M5.5): middle-right, MV~Lup (SpT: K2.0): bottom-left, and CD-31~12522 (SpT: K2.0): bottom-right. Each plot shows the fit to the stellar photosphere and the circumstellar emission as a single modified blackbody (solid line). In the case of MU~Lup a single black body fit is plotted (red dashed line), and in the cases of J155526.2-333823, NO~Lup and Sz~107, two-component modified blackbody fits are plotted (red dashed lines). The dots and triangles show the available photometry and upper limits.}
   \label{fig:SEDSall6}
  \end{center}
\end{figure*}

\subsection{Survey Non-Detections}
\label{sec:stackedIm}
For the class III sources for which sub-mm emission was not detected, we produced a stacked image of their continuum images to constrain the average dust mass. The 26 images for which this was true combined to form the image shown in Fig.~\ref{fig:class3Stack}. Noticeable in this image are the B and C type detections that fall within the fields of view, however we are only concerned with the inner ${\sim}0.68\arcsec$ that is coincident with the class III stars (within one beam), as highlighted in the expanded central region. By integrating the emission within this inner region we found a total of $20.2\pm7.2 \mu \rm{Jy}$, i.e., a tentative ${\sim}2.8\sigma$ detection coincident with all of the class III stars which individually have non-detections. \\
\newline
The mean distance to the stars in our survey is $141{\rm{pc}}$, and so equation~\ref{eq:dustmass} shows that this tentative stacked detection (if from circumstellar dust) corresponds to a dust mass of $M_{\rm{dust}}=0.0048\pm0.0017M_{\oplus}$. This implies that whilst we only detected significant emission to 4/30 class III stars (i.e., type A sources), we may have detected further low intensity emission from the other sources. In the context of known dust masses, this would be a limit far below any previous measurements for stars of this age, and nearly an order of magnitude below the lowest mass type A detection in $\S$\ref{sec:Masses}.

\subsection{SED Analysis}
\label{sec:sed}
To investigate the nature of the infrared emission seen in the class III stars, Spectral Energy Distributions (SEDs) for all sources were produced using the methodology of \citet{Yelverton19}, i.e., we fit a stellar photosphere to determine $L_{\star}$ and $T_{\rm{eff}}$, plus a blackbody disk where merited, to the available photometry.
We include mid- and far-IR data from WISE, Spitzer and Herschel, but also data at shorter wavelengths that constrain the stellar photospheric emission. All 30 sources had photometry in J-, H- and K-bands \citep[2MASS;][]{Cutri13}, Gaia G, $B_P$ and $R_P$ bands \citep{Gaia18}, WISE bands W1, W2, W3 and W4 \citep{Wright10}, and the following number in each instrument and band: 28/30 in both IRAC 3.6$\mu$m and IRAC 4.5$\mu$m and 29/30 in both IRAC 5.8$\mu$m and IRAC 8.0$\mu$m, 29/30 in MIPS 24$\mu$m \citep{Werner04, Rieke04, Merin08B}, 2/30 in PACS 70$\mu$m \citep{Pilbratt10, Poglitsch10, Cieza13}, alongside the ALMA Band 7 (${\sim}856\mu$m) measurements presented in this work. \\
\newline
We found from the analysis of the full SEDs, that 24 of the sources are well explained with purely photospheric emission; i.e., there is no evidence for an additional circumstellar component in addition to their stellar photospheres. An additional emission component is required for the remaining 6 sources (MU~Lup, J155526.2-333823, NO~Lup, Sz~107, MV~Lup and CD-31~12522), which includes the 4 for which sub-mm continuum emission was detected, which we consider to be circumstellar emission. The SEDs of the 6 with excess emission are shown in Fig.~\ref{fig:SEDSall6}, and the 24 purely photospheric SEDs in Appendix~\ref{sec:AppendixA}. \\
\newline
The SED modelling process resulted in a fit to the stellar photospheric emission, predominantly constrained by the shorter wavelength data. In addition to being essential for predicting the stellar contribution at longer wavelengths, we used the resulting stellar parameters $L_{\star}$ and $T_{\rm{eff}}$ to estimate the stellar masses shown in Table~\ref{tab:class3Objs} for the 14/30 class III stars for which values were not available in the literature (i.e., those denoted with superscript "n"). To do so, we applied the models of \citet[][for V1027~Sco]{Siess00} and \citet{Baraffe15} for the 13 others, except Sz~109 and J160816.0-390304 for which estimates based on their spectral types were required (i.e., since these were poorly fitted by stellar models). In all cases where the stellar mass uncertainty is not stated in Table~\ref{tab:class3Objs} (i.e., in the literature or as derived here), we assume this to be 10\%. \\
\newline
We first consider the $12\mu \rm{m}$ and $24\mu \rm{m}$ excesses, given the discussion of [K]-[12] and [K]-[24] in $\S$\ref{sec:photometry} and the colour-colour diagram in Fig.~\ref{fig:photometryC2C3}. For the 6 sources with excess emission, these are all found to have significant excesses at 24$\mu$m. Significant excesses were found at 12$\mu$m for 4 sources (MU~Lup, NO~Lup, MV~Lup and CD-31~12522), while both J155526.2-333823 and Sz~107 were found to have tentative 12$\mu$m excesses ($2.6\sigma$ and $2.2\sigma$ respectively). Given that MU~Lup, J155526.2-333823, MV~Lup and CD-31~12522 were not deemed as outliers in Fig.~\ref{fig:photometryC2C3} (i.e., these sit with the majority of class III sources in the lower-left cluster), this illustrates that low [K]-[12] or [K]-[24] indices do not necessarily rule out the presence of excess emission. It is also notable that 4/6 of the class III sources with a 24\,$\mu$m excess were found to have detectable sub-mm emission in the ALMA survey. This suggests that the presence of a mid-IR excess may raise the sub-mm detectability for class III YSOs. We discuss the implications of this warm emission further in $\S$\ref{sec:discussion}. \\
\newline
The excess emission for these 6 sources was first fitted with a single-component modified blackbody, parameterised by a temperature, T, fractional luminosity, $f=L_{\rm{disk}}$/$L_{\star}$, $\lambda_0$ and $\beta$ (the latter two parameters describe the location and gradient beyond which the modification to the blackbody fit is implemented). This temperature is readily converted to a black body radius, R$_{\rm{bb}}$, by assuming this emission is from optically thin dust in thermal equilibrium with the stellar radiation of luminosity $L_\star$. Whilst this radius provides an indication of the size of the dust disk, it is expected that this radius will be systematically underestimated \citep[e.g., since the dust is likely to be small enough to emit inefficiently and so be hotter than a black body, see ][]{Pawellak14, Pawellek15}. These single-component blackbody SED fits are shown as brown solid lines ($F_{\rm{\nu,single}}$) on Fig.~\ref{fig:SEDSall6}, and the model fits for the temperature, associated blackbody radius and fractional luminosity are shown in the upper part of Table~\ref{tab:SEDBB}.
We have not included the $\lambda_0$ and $\beta$ values as these are poorly constrained and highly degenerate. In all 6 cases $\beta$ can be 0, for which $\lambda_0$ is then meaningless and unconstrained. However, the joint 2D parameter space is somewhat constrained for NO~Lup and Sz~107 because the dust temperatures are well constrained by their IRS spectra \citep{Lebouteiller11}.
For MU~Lup, J155526.2-333823, MV~Lup and CD-31~12522 the SED of the excess emission is poorly sampled which means that a wide range of single-component modified black body fits can model the data by choosing different $\lambda_0$ and $\beta$. In the case of MU~Lup, the solid line is plotted with $\lambda_0{\sim}120\mu$m and $\beta=1.2$, whereas the dashed line is a pure black body (for which the same degeneracy exists in the cases of J155526.2-333823, MV~Lup and CD-31~12522). 
However, despite this allowed range of model fits, a pure black body model (i.e., $\beta=0$) provides the warmest temperature fit (i.e., this corresponds to the lowest radius). For MU~Lup, such a pure black body radius is at 10\,au. Therefore, accounting for the range of allowed models, this suggests that the emission measured for MU~Lup is likely from a component at tens of au. \\
\begin{table}
    \centering
    \caption{SED blackbody model parameters for all class III stars determined to have excess emission above the stellar photosphere (note J155526.2-333823 is written here as J155526). The upper part of the table includes the 4 objects with sub-mm detections and 2 objects undetected in the sub-mm but with archival mid-IR observations found evidence for a warm component, and their respective modified blackbody fits. The lower part of the table includes the three sources determined as well-fitted by a two-component modified blackbody, though we note here the broad parameter space that can be fitted by these modified black bodies, and therefore whilst errors are defined, these are only weakly constrained.}

    Single Component Modelling \\
    \begin{tabular}{c|c|c|c}
         \hline
         Source & T & R$_{\rm{bb}}$ & $f$  \\
         & [K] & [au] & $\times\,0.01$ \\
         \hline
         MU~Lup & $50\pm10$ & $16\pm9$ & $0.1\pm0.5$  \\
         J155526 & $52\pm8$ & $14\pm5$ & $0.3\pm0.3$  \\
         NO~Lup & $117\pm2$ & $3.1\pm0.1$  & $0.34\pm0.01$  \\
         Sz~107 & $144\pm3$ & $1.16\pm0.05$ & $1.39\pm0.03$ \\
         MV~Lup & $600\pm200$ & $0.2\pm0.1$ & $0.09\pm0.06$  \\
         CD-31~12522 & $750\pm40$ & $0.13\pm0.01$ & $0.13\pm0.05$  \\
    \hline
    \end{tabular} \\
    Two Component Modelling \\
    \begin{tabular}{c|c|c|c|c|c|c}
         \hline
         Source & $T_1$ & $R_{\rm{bb1}}$ & $f_1$ & $T_2$ & $R_{\rm{bb2}}$ & $f_2$ \\
         & [K] & [au] & $\times\,0.01$ & [K] & [au] & $\times\,0.01$  \\
         \hline
         J155526 & >50 & <15 & 0.01-1 & 20-60 & 10-100 & 0.01-1  \\
         NO~Lup & 117 & 3.1 & 0.34 & 25-120 & 2-56 & 0.00002-0.4 \\
         Sz~107 & 144 & 1.16 & 1.39 & 20-150 & 1-60 & 0.00005-4  \\
    \hline
    \end{tabular}
    \label{tab:SEDBB}
\end{table}
\newline
Although their single-component black body radii are poorly constrained, the 16\,au radius found for MU~Lup is consistent with this source being unresolved in the sub-mm (i.e., <53\,au in $\S$\ref{sec:diskSizes}), and at $\leq0.2$\,au for both MV~Lup and CD-31~12522 it is not surprising that these were not detected in the sub-mm. However the single modified black body fits for J155526.2-333823, NO~Lup and Sz~107 are problematic. For J155526.2-333823, whilst this is found with a single black body radius of ${\sim}14$\,au, this source was resolved with a sub-mm radius of ${\sim}80$\,au (i.e., more than a factor of 5 larger), and both NO~Lup and Sz~107 \citep[as accurately measured with IRS spectra, see][]{Lebouteiller11} are found to have well-constrained temperatures that result in blackbody radii less than ${\sim}3$\,au. In the case of Sz~107, it can also be seen that the single component modified black body is not as well fitted to both the sub-mm measurement and the warm emission. 
For these reasons, we therefore modelled J155526.2-333823, NO~Lup and Sz~107 as having two separate modified black body components, representing both warm and cool emission. While such an additional component is not required to fit the SED, this is required to explain the excess emission seen towards many older main sequence stars \citep[e.g.,][]{Kennedy14} and so it is not unreasonable that these stars might exhibit two-component SEDs. Thus it is important to consider how such an additional component might affect our estimate of the radii of these disks. \\
\newline
The two-component models for J155526.2-333823, NO~Lup and Sz~107 are under-constrained in that their SEDs can be fitted by a range of two-component blackbody models, as illustrated by the shaded regions on the plots in Appendix~\ref{sec:AppendixAA}, where further discussion of their interpretation is provided. In the lower part of Table~\ref{tab:SEDBB} we provide ranges of allowed temperatures, associated blackbody radii and fractional luminosities. Fig.~\ref{fig:SEDSall6} shows examples for each system of one such possible two-component model with red dashed lines, with the warm component modelled to overlap with emission $\lambda<100\mu$m and the cool component modelled to overlap with the sub-mm detection. Both NO~Lup and Sz~107's warm components are tightly constrained, so match the results of the single modified blackbody fits, while the sub-mm emission originates in a cooler component that is consistent with being at 10s of au from the star (though must both still remain unresolved in our ALMA observations at $<56$\,au and $<60$\,au respectively). \\
\newline
In summary, despite the limitations of modelling the SEDs for the 6 sources shown in Fig.~\ref{fig:SEDSall6}, we have shown that reasonable single modified black body models are found for all sources, alongside alternative 2-component modified black body models for J155526.2-333823, NO~Lup and Sz~107. 
This modelling has shown that disks with black body radii of tens of au remain consistent with all four sub-mm detections (i.e., the circumstellar material detected by ALMA may be >10s of au away from the star).

\begin{figure}
    \includegraphics[width=0.475\textwidth]{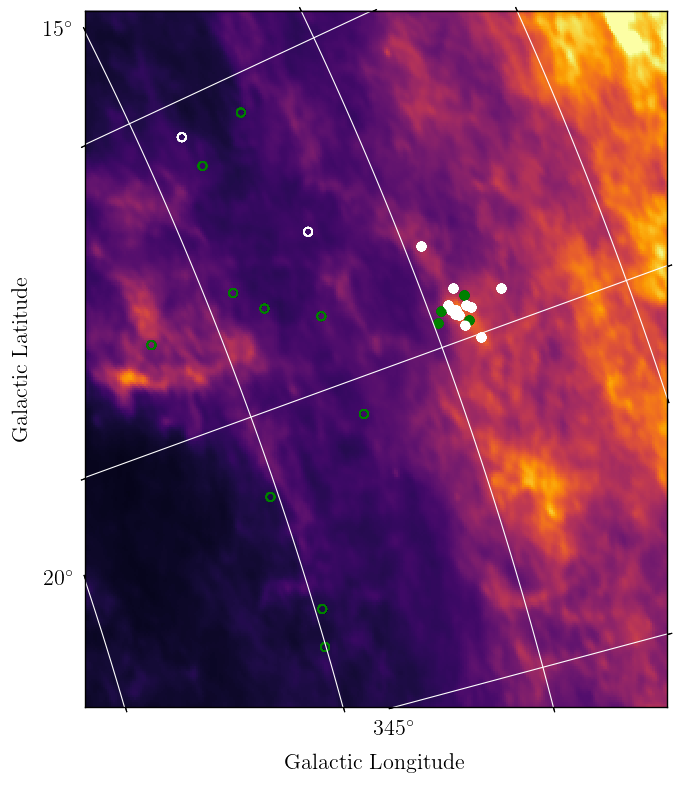}
    \caption{Sky map of the far-IR galactic dust extinction in the Lupus star forming region, where the locations of the class III stars in this survey have all been noted. Stars represented by "o" markers indicate that they are located in Lupus cloud I, whereas markers represented by a filled circle indicate the star is in Lupus cloud III. White markers indicate that background CO emission was detected in the channel maps, whereas green markers indicate that no background CO was detected, discussed further in $\S$~\ref{sec:COSurvey}.}
    \label{fig:class3SurveySkyIm}
\end{figure}

\begin{figure*}
    \includegraphics[width=1.0\textwidth]{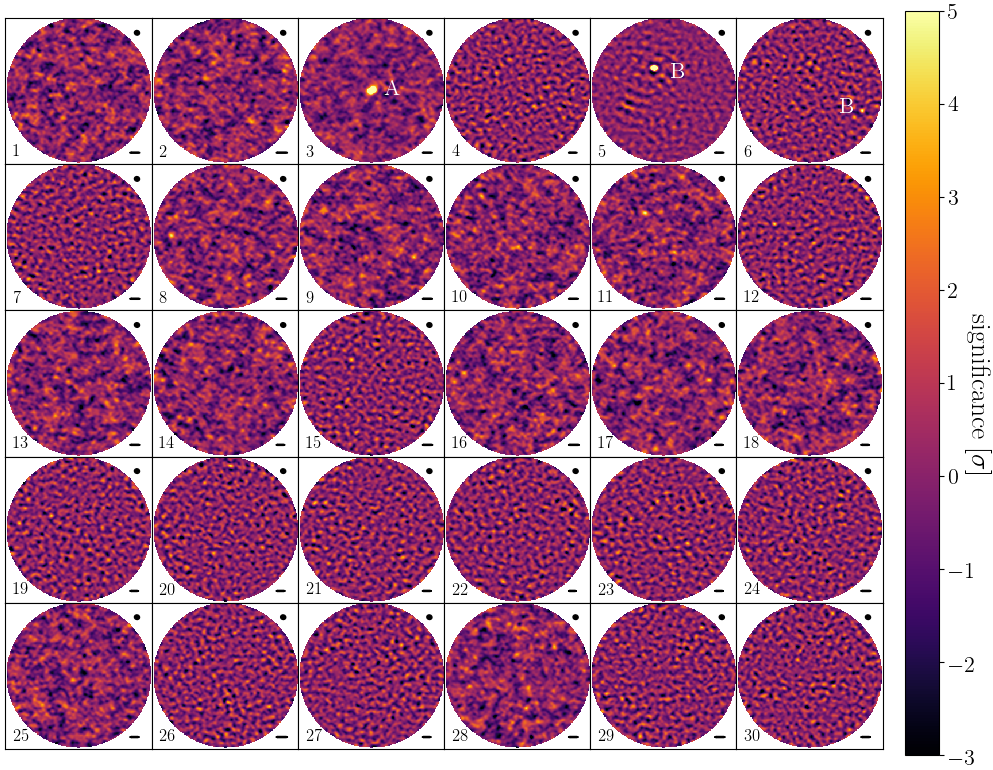}
    \caption{ALMA CO moment 0 images for all 30 class III stars. The colour scale is in units of significance with respect to each image's rms. The source numbers in the bottom-left of each panel correspond to those in Table~\ref{tab:class3Objs}. Note that the plots in which rms is higher were those which suffered from CO cloud contamination, further discussed in $\S$~\ref{sec:COSurvey}. In all images, North is up, East is left, and the beam and 200\,au scale bar are shown in the upper and lower right respectively.}
    \label{fig:class3SurveyMom0Map}
\end{figure*} 

\subsection{Background Sources}
\label{sec:BkgnSources}
In our sample we find 11 type~C detections, 10 of which have fluxes $>1.0$\,mJy, and 1 with a flux ${\sim}0.41$\,mJy. We previously inferred these to be sub-mm galaxies (SMGs) in $\S$\ref{sec:surveyanalysis}, given that their locations are inconsistent with other known stellar sources in Lupus. This detection rate can be compared with that expected from surveys designed to determine SMG number counts at ${\sim}850\mu$m \citep{Simpson15, Popping20}. To do so, we have to account for the primary beam correction, which means that while we could detect sources brighter than 5 times $36.7$\,$\mu$Jy (i.e., $\approx0.2$\,mJy) towards the centre of the image, sources would need to be ${\sim} 2$ times brighter than this at a radius of $8.4\arcsec$, and ${\sim} 5$ times brighter than this at $12.5\arcsec$ (i.e., at $\sim0.4$\,mJy and $\sim1.0$\,mJy respectively). \\ 
\newline
The \citet{Popping20} number counts predict that we should detect ${\sim} 8$ sources above 1\,mJy inside $12.5\arcsec$ and ${\sim}8$ sources above 0.4\,mJy inside $8.4\arcsec$ (all of which are scaled to cover the total survey area, i.e., inside an area equal to 30 circles with these three radii). Compared with the number of type C sources we measured in each of these regions, 10 and 6 respectively, these numbers therefore appear strongly consistent. At lower flux levels it may be notable that we only detected 1 SMG in the range $0.4-1$\,mJy inside $8.4\arcsec$, whereas the \citet{Popping20} number counts predict we should have detected ${\sim} 4$. Given that these numbers are both low, we do not ascribe great significance to this difference, but note that a significant difference might be evidence for cosmic variance or a turn-over in SMG number counts at low flux levels. Indeed, our survey could be considered as an independent measurement of 856$\mu$m SMG number counts in this region of sky down to 1\,mJy and indeed below, even if its interpretation is complicated by the possibility of Galactic sources in the field of view. \\
\newline
We further note that the 4 sub-mm detections towards class III stars (i.e., type A) were all detected within $0.4\arcsec$ of their corresponding ALMA phase centres, with fluxes exceeding 0.2\,mJy. Inside this region over the entire survey, the \citet{Popping20} number counts correspond to an SMG detection probability of just ${\sim}3\%$. The contamination rate may be even lower than this, given that our survey measured a lower number of $>0.4$\,mJy sources than predicted by \citet{Popping20}. Thus, whilst it is possible that one of these class III (type~A) detections is an SMG, this also appears highly unlikely.

\section{Survey CO Analysis}
In this section we use the ALMA observations to search for CO emission coincident with the class III stars, and with the two class II sources that fall in the field of view.

\label{sec:COSurvey}
\subsection{Survey Detections}
To produce image cubes for each measurement set, we first transformed this data to the barycentric reference frame, and subtracted a linear fit of the continuum that was fitted over line-free channels (i.e., those external to the region $\pm15$\,km\,s$^{-1}$ from the CO J=3-2 spectral line) using the CASA task $\rm{uvcontsub}$. These were then all imaged with the CASA task $\rm{tclean}$ non-interactively, for only the initial major cycle to produce data cubes, with a rest frequency set to the CO J=3-2 spectral line frequency, $f_{\rm{CO}}=345.79599$\,$\rm{GHz}$. 
The FOV was likewise chosen to show the entire primary beam.

\subsubsection{Cloud Contamination}
\label{sec:COSpur}
Channel maps were produced from the sub-cubes of all survey sources and in 16/30 of these high intensity, large scale CO emission was observed in the field of view in the channels in the range $f_{\rm{CO}} \pm 10$\,km\,$\rm{s}^{-1}$. In none of these cubes were emission signals found further away than a few channels from this frequency at this intensity. For all 30 sub-cubes, velocities from +13 to -16\,km\,s$^{-1}$ are shown in Appendix~\ref{sec:AppendixB} in Figures~\ref{fig:class3Spurious1} and~\ref{fig:class3Spurious2}, with 16 of these demonstrating large scale CO J=3-2 emission unrelated to any detected continuum emission (in images 4, 5, 6, 7, 12, 15, 19, 20, 21, 22, 23, 24, 26, 27, 29, 30). The full cubes for these 16 sources were investigated and the intensity and scale of the observed CO emission was found to be inconsistent with being due to flagged data, missing baselines, or visibility values, and found irrespective of the $\rm{uvcontsub}$ task being applied. \\
\begin{table}
    \small
    \centering
    \caption{Shown are the total CO fluxes, estimated CO gas masses for an optically thin, LTE (50K) assumption, and whether or not CO cloud emission was present in the data.}
    \begin{tabular}{ccccccc}
         \hline
         \hline
         No. & Name &  Flux (CO) & $M_{\rm{CO ,LTE }}$ & Cloud \\
         & & [$\rm{Jy}\,\rm{km}\,\rm{s}^{-1}$] & [$10^{-5}\times M_{\oplus}$] & Gas\\
         \hline
         1&MU~Lup&<0.12&<2.1&-\\
         2&J155526.2-333823 &<0.12&<1.6&-\\
         3&  NO~Lup &$0.29\pm0.07$& $4.9\pm1.1$ &-\\
         4&Sz~107&<0.19& <4.2 &Y\\
         5&Sz~108&<0.19&<4.2&Y\\
         &Sz~108B&$0.56\pm0.06$ & $12.5\pm1.4$& Y\\
         6&Sz~119&<0.19&<2.1&Y\\
         &Lup~818s&$0.037\pm0.012$ & $0.43\pm0.14$ & Y\\
         \hline
         7& MT Lup&<0.19&<3.2&Y\\
         8&Sz~67&<0.12&<1.4&-\\
         9&J154306.3-392020&<0.12&<3.2&-\\
         10&CD-35 10498&<0.12&<2.6&-\\
         11&MV Lup&<0.12&<2.3&-\\
         12&MW Lup&<0.19&<3.1&Y\\
         13&MX Lup&<0.12&<1.9&-\\
         14&NN Lup&<0.12&<2.2&-\\
         15&CD-39 10292&<0.19&<3.1&Y\\
         16&CD-31 12522&<0.12&<2.0&-\\
         17&J160713.7-383924&<0.12&<2.9&-\\
         18&J160714.0-385238&<0.12&<1.7&-\\
         19&Sz~94&<0.19&<2.4&Y\\
         20&J160758.9-392435&<0.19&<4.7&Y\\
         21&J160816.0-390304&<0.19&<4.7&Y\\
         22&V1027 Sco&<0.19&<4.7&Y\\
         23&Sz~109&<0.19&<6.0&Y\\
         24&J160908.5-390343&<0.19&<4.6&Y\\
         25&J160917.1-392710&<0.12&<2.0&-\\
         26&Sz~116&<0.19&<3.9&Y\\
         27&Sz~121&<0.19&<4.4&Y\\
         28&Sz~122&<0.12&<2.2&-\\
         29&Sz~124&<0.19&<2.9&Y\\
         30&V1097 Sco&<0.19&<3.5&Y\\
         \hline
    \end{tabular}
    \label{tab:class3ObjsGas}
\end{table}
\newline
We plotted the positions of all 30 class III stars on Fig.~\ref{fig:class3SurveySkyIm}, overlaid onto an image of far-IR Milky Way galactic dust extinction, obtained from the "SFD" dustmap as described in \citet{Schlegel98} and updated by \citet{Schlafy11}. This can be used to indicate whether galactic dust or CO gas might have contaminated our observations near the CO J=3-2 transition frequency (discussed later in $\S$\ref{sec:discussion}), and showed a correlation between the dust extinction coefficient and the presence of CO emission. \\
\newline
We therefore conclude that such large scale CO emission is highly likely from galactic CO gas, present in large abundance in the Lupus star forming region, and detected by our deep ALMA imaging (although we note that the intensity and extent of the gas emission may not be well traced by our images). All 16 of these sources are noted in Table~\ref{tab:class3ObjsGas} in the "Cloud Emission" column with a "Y" if the channel maps were affected. We note that cloud contamination was reported in the class II Lupus survey of \citet{Ansdell18}, however the depth of the observations in this class III survey has resulted in cloud emission being much brighter than in any previous ALMA imaging of Lupus. \\
\newline
Given the size scale of the emission, in the 16 measurement sets where CO cloud contamination was detected we implemented a uv-baseline cut during the imaging process to remove all visibility data with $\rm{uv-distance}<100\rm{k}\lambda$, significantly reducing the structure from emission on large scale sizes ($>2\arcsec$, i.e., larger than any CO emission that would be circumstellar). Moment-0 maps for all 30 sources were then produced from emission within 10\,km\,s$^{-1}$ of the CO spectral line, shown in Fig.~\ref{fig:class3SurveyMom0Map}. These CO moment-0 maps show significant source detections in positions coincident with continuum emission sources 3A (NO~Lup, i.e., one of the class III sources) and both class II sources, Sz~108B (5B) and Lup~818s (6B). No CO emission was detected within $1.0\arcsec$ of any other class III sources or type-C detections from our continuum maps.

\begin{figure*}
    \includegraphics[width=1.0\textwidth]{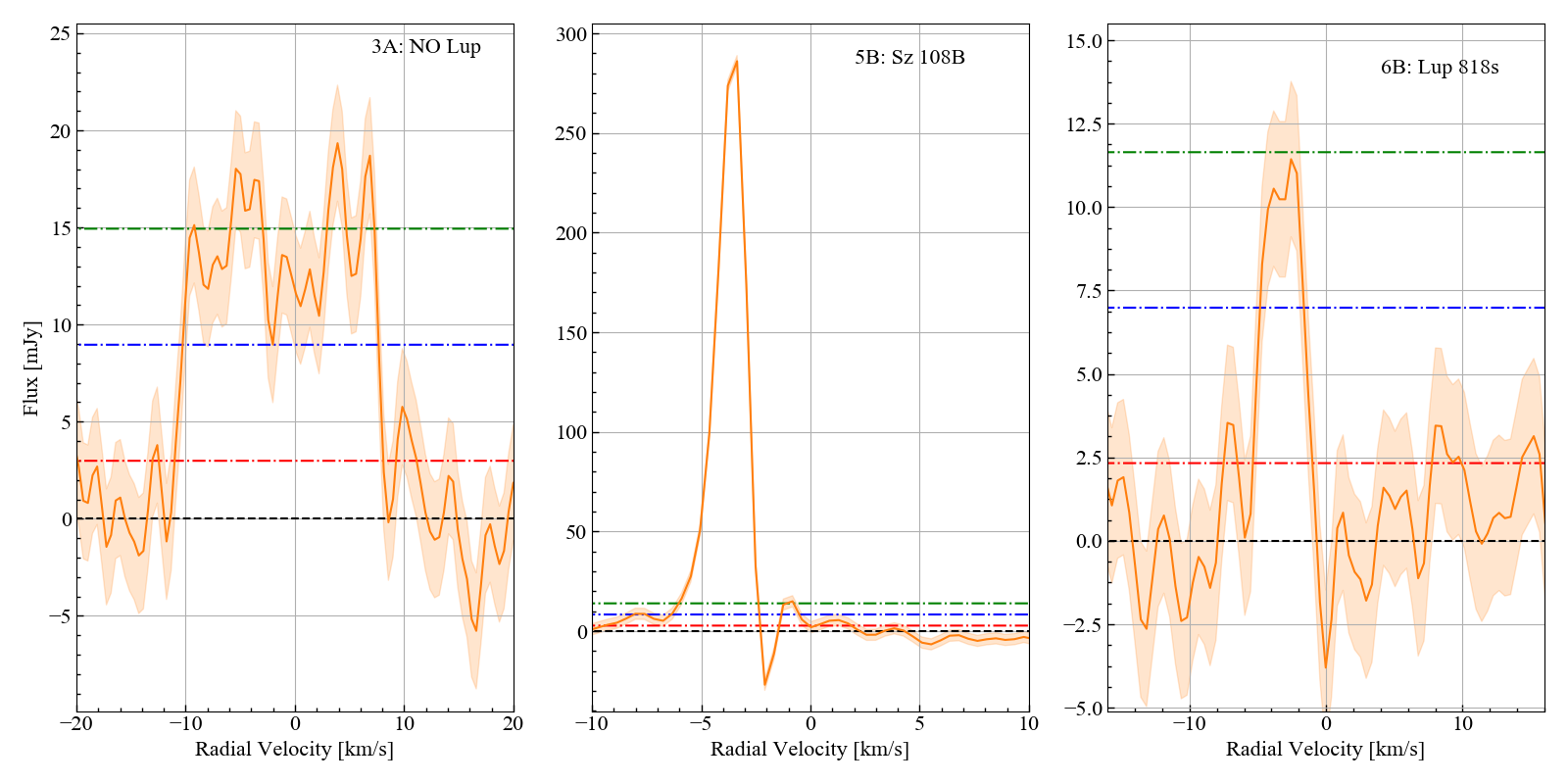}
    \caption{CO spectra of the class III star NO~Lup (left), and the class II stars Sz~108B (middle) and Lup~818s (right). The red, blue and green dashed lines show the per-channel $1\sigma$, $3\sigma$ and $5\sigma$ noise levels, respectively, calculated from the rms signal of the integrated flux in channels distant from the CO emission.}
    \label{fig:spectraClassIII8}
\end{figure*}

\subsection{Class II and III CO Analysis}
\label{sec:COII_III}
Of the class III stars for which sub-mm continuum emission was detected (i.e., those designated type~A), only 1 source was affected by the CO cloud contamination, Sz~107, for which we can only set an upper limit on the CO gas flux/mass (see later in this section). MU~Lup and J155526-333823 showed no significant CO emission coincident with the star, and we can only derive upper limit fluxes/masses for these two sources also. NO~Lup, however, demonstrates highly significant CO emission, in fact at higher significance than its continuum detection. Cloud contamination affected some channels for both of the class II sources (Sz~108B and Lup~818s), however the brightness of these two sources meant that their CO flux was still detectable above the background emission. We note here that it is also possible that cloud contamination has masked low level detections of CO gas in sources 19, 23, 24, and 30, given that these appear most affected at the class III stellar position. Further high resolution imaging of these sources may reveal if CO gas is present. \\
\newline
Spatially integrated spectra were produced for NO~Lup, Sz~108B and Lup~818s, and are shown in Fig.~\ref{fig:spectraClassIII8}. For the NO~Lup spectrum, emission was integrated within an elliptical mask with an aperture of $(2.5\times2.1)\arcsec$ (i.e., such that the mask fully enveloped all significant CO emission). For the class II spectra, emission was integrated within elliptical masks centred on the two class II sources with apertures of $(1.6\times1.4)\arcsec$ and $(1.2\times1.0)\arcsec$ for Sz~108B and Lup~818s, respectively. CO fluxes were determined by integrating the area under these spectra between the points at which these cross their respective $1\sigma$ thresholds; -11.0 to +8.1 $\rm{km} \, \rm{s}^{-1}$ for NO~Lup, -9.0 to +2.0 $\rm{km} \, \rm{s}^{-1}$ for Sz~108B, and -5.4 to +1.0 $\rm{km} \, \rm{s}^{-1}$ for Lup~818s. The median velocity of the detected CO emission is consistent with that expected from the radial velocities of Lupus stars, $\rm{RV_{Lup}}=2.8\pm4.2\rm{kms^{-1}}$ \citep[see][]{Frasca17}. For NO~Lup, emission is measured over a broad range of velocities $\sim{19}$\,$\rm{km}\, \rm{s}^{-1}$. The spectrum may be consistent with the double-peaked structure typical of spectrally resolved, inclined rotating disks. Therefore, whilst the inclination of the NO~Lup disk was not well constrained from its continuum emission (see Table~\ref{tab:2DFit}), the CO spectrum suggests that this source may be inclined with respect to the plane of the sky. In both Sz~108B and Lup~818s, the spectral line shows only a single sharp peak; the lower negative peak of Sz~108B close to $-2\rm{km} \, \rm{s}^{-1}$ is likely due to cloud contamination absorbing the red-shifted side of the disk.  \\
\newline
To determine the uncertainty on the CO line fluxes, we found for all non-contaminated sources, for an aperture with a size of $(1.2\times1.0)\arcsec$ located at the image centre, the integrated emission uncertainty is $\rm{rms_{CO} = 3.0\,mJy}$ per 0.42\,km\,s$^{-1}$ channel  (based on the emission excluding channels within $\pm20$\,km\,s$^{-1}$ of the spectral line).
For the sources affected by cloud contamination the rms is higher once uv-baseline cuts have been applied, and for these we instead find the integrated emission uncertainty $\rm{rms_{CO} = 4.7\,mJy}$ per 0.42\,km\,s$^{-1}$ channel (i.e., the uncertainty increased by ${\sim}50\%$).
Although the channel width is $\Delta v = 0.42$\,km\,s$^{-1}$, the effective spectral bandwidth is ${\sim}2.667$ larger than this, since adjacent ALMA channels are not fully independent~\footnote{For a complete discussion, see \url{https://safe.nrao.edu/wiki/pub/Main/ALMAWindowFunctions/Note_on_Spectral_Response.pdf}}.
With these considerations, and by including an additional 10\% flux calibration error in quadrature, we find the integrated fluxes and uncertainties for NO~Lup, Sz~108B and Lup~818s as shown in Table~\ref{tab:class3ObjsGas}.
Further, for an assumed spectral line width of ${\sim}5$\,$\rm{km} \, \rm{s}^{-1}$, we place $3\sigma$ upper bounds on the CO flux of all 29 other class III survey sources as either $<0.19$ or $<0.12 \, \rm{Jy} \, \rm{km} \, \rm{s}^{-1}$ (i.e., depending on whether these sources were affected by cloud contamination or not respectively). \\
\newline
By further assuming that the CO gas is in Local Thermodynamic Equilibrium (LTE), has a low optical depth (i.e., $\tau{<}1$), and a temperature of $50\rm{K}$, we derived CO gas masses for NO~Lup, Sz~108B and Lup~818s, and upper limit masses on the 29 of our survey target sources. 
These calculations used equations 2 and 8 (for the J=3-2 transition) from \citet{Matra15}, for which all derived $M_{\rm{CO,LTE}}$ masses are shown in Table~\ref{tab:class3ObjsGas}. 
If the CO abundance (i.e., (x$_{\rm{CO}}=\rm{[CO]}$/$\rm{[H_2]}$) has a protoplanetary disk-like abundance \citep[i.e., $10^{-4}$, as per][]{Williams14}, we find the total gas-to-dust mass ratio for NO~Lup of $1.0\pm0.4$. For the class III sources with dust measurements but no CO counterpart, the $3\sigma$ upper limit gas-to-dust ratios are <0.25, <0.16 and <0.60 for MU~Lup, J155526.2-333823 and Sz~107, respectively.
A caveat to the above calculations is that if the emission comes from a sufficiently compact region, it may be optically thick, which would imply that the values stated in Table~\ref{tab:class3ObjsGas} could all be larger. 
Class II protoplanetary disks are usually optically thick in CO, but this is not necessarily the case for the low mass class~III disks examined here, with dust masses comparable to debris disks (see later, Fig.~\ref{fig:diskMassAge}). 
Reinforcing this point, we measure class~III CO fluxes that are either upper limits, or in the case of NO~Lup, a factor of ${\sim}$5 times lower than the average F$_{\rm{CO}}$ of the Lupus class~II sources \citep[see][]{Ansdell16}, and similar in line luminosity to nearby, optically thin CO-bearing debris disks \citep[see][]{Kral17}. 
Therefore, while higher angular resolution observations are needed to better estimate disk sizes and optical depth, we present optically thin masses as a reasonable estimate of the CO contents in these low F$_{\rm{CO}}$ class~III sources, but note that the estimates for the gas-rich class~II sources (e.g., Sz~108B) are likely underestimated. 
A second caveat is that the assumption of LTE may not apply if H$_2$ is not abundant in these disks, e.g. if CO gas is secondary, that is, released from planetesimals instead of being primordial \citep[e.g.,][]{Dent14, Marino16, Moor17, Matra17}. If the CO gas is not in LTE, then CO gas mass limits are typically much higher as shown by \citet{Matra15, Matra18B}, but this does not mean that the total gas-to-dust ratio is necessarily higher due to the absence of H$_2$.
We discuss the implications of this further in $\S$\ref{sec:discNOLupSz107}.

\subsection{Summary of CO Analysis}
By inspecting the CO emission from the 30 sources in our survey, whilst we found in 16 images large scale emission from presumed CO in the Lupus cloud, we found that one of the class III stars with a continuum detection also has a CO counterpart, NO~Lup, that will be discussed further in \citep{Lovell20}. After accounting for the likely galactic CO emission from the Lupus clouds, we also detected CO towards both class II stars for which we also detected continuum emission, Sz~108B and Lup~818s. None of the other class III stars or background/other (type C) detections contained CO emission, and upper limit CO fluxes/masses were derived for all 29 class III sources. From these measurements we have constrained the CO emission coincident with all class III and II sources present in this survey.

\section{Discussion}
\label{sec:discussion}
\subsection{Summary of Circumstellar Material around Class III Stars}
In $\S$\ref{sec:photometry} we showed the full set of YSOs that comprise our Lupus sample, and how the [K]-[12] and [K]-[24] photometry compares for the class II and III sources. We showed that the class II and III sources could be distinguished photometrically consistent with the \citet{Lada84} classification, since the class II's all have either $\alpha_{\rm{K-24}}>-1.6$ or $\alpha_{\rm{K-12}}>-1.6$, which is not the case for the class III's (although some of these only have upper limits to their mid-IR emission). For the majority the former criterion using the 24\,$\mu$m photometry is sufficient to classify the sources (i.e., all except AKC2006-18). Furthermore, whilst the class II sources showed a large range in the colour-colour values, the class III sources were mostly confined to near-stellar colours. Nevertheless, detailed consideration of their SEDs showed that low level non-photospheric emission is present for 6/30 of the class III sources. Our analysis led us to different conclusions about the nature of the circumstellar material for different stars in the 30 class III YSOs.

\subsubsection{Likely Debris Disks: MU Lup and J155526.2-333823}
These both have mm-sized dust which is inferred to be at 10s of au, with one of these resolved as a ring at 80\,au. Neither demonstrates any CO J=3-2 emission. Upper limits on the CO gas masses imply a $3\sigma$ upper limit on the gas-to-dust ratios of <0.25 and <0.16 (for MU~Lup and J155526.2-333823 respectively) if these had abundances similar to protoplanetary disks. We therefore consider these to be inconsistent with gas dominated primordial disks. In addition, while these also fell inside the $\alpha_{\rm{IR}}{<}-1.6$ classification for class III YSOs at both $12\mu \rm{m}$ and $24\mu \rm{m}$, low level excess emission is present at 24\,$\mu$m for both and at 12\,$\mu$m for MU Lup. Neither shows evidence of ongoing accretion on to their stars. In conjunction, we find these two sources consistent with being young debris disks.

\begin{figure}
    \includegraphics[width=0.475\textwidth]{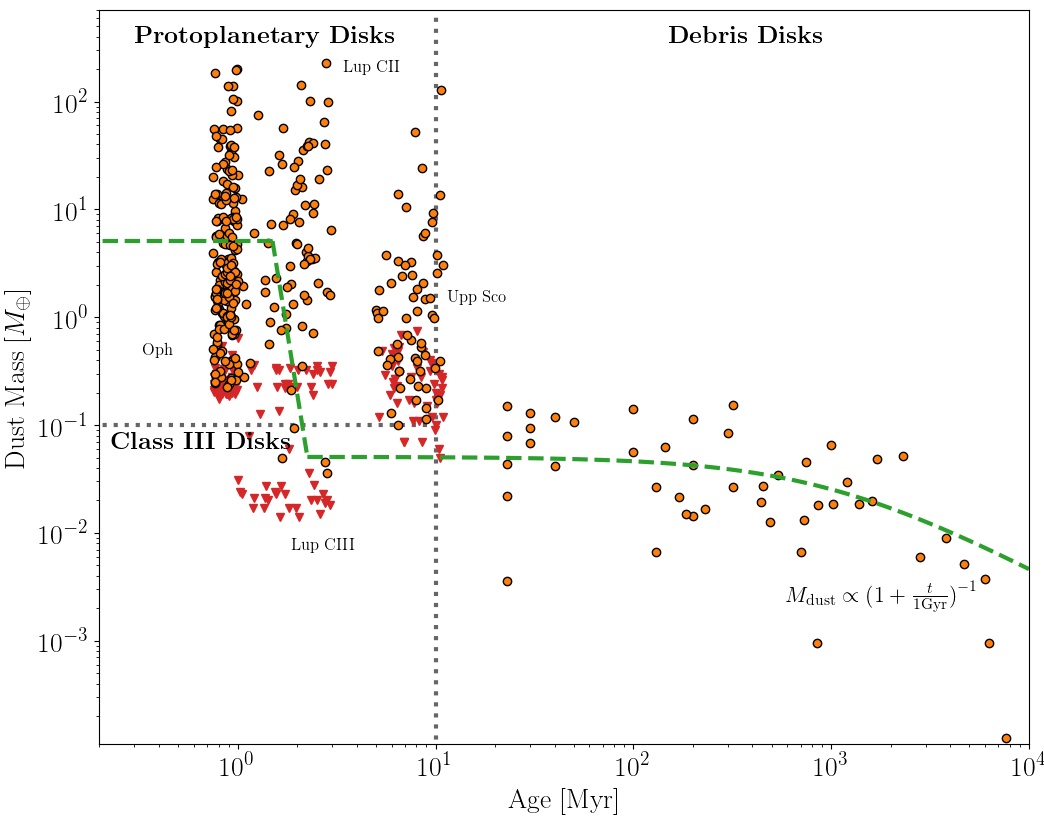}
    \caption{Disk mass as a function of age for known debris disks and protoplanetary disk populations, alongside the 30 class III Lupus sources outlined in this work. The vertical dotted line at 10\,Myr is only an approximate division between where protoplanetary and debris disks might be expected, and the horizontal line is the approximate division between sources in star forming regions that are considered class II and class III (noting that counter examples exist on either side of these lines). We plot as a green dashed line an illustrative path that a system may follow as it evolves from being a protoplanetary disk to a debris disk, via the class III stage explored in this work. }
    \label{fig:diskMassAge}
\end{figure}

\subsubsection{Uncertain if Debris Disks or Protoplanetary Disks: NO~Lup and Sz 107}
\label{sec:discNOLupSz107}
NO~Lup is found to have both dust and gas. The SED of the dust component can be fitted as a single temperature component which would put it at ${\sim} 3$\,au, but it is equally likely that there is an additional component at 10s of au (or a broad disk extending that far) that dominates the sub-mm continuum emission. 
Most surprising is the bright gas which has a high velocity width ($\sim 19\,\rm{km} \, \rm{s}^{-1}$) with a gas-to-dust mass ratio measured as $1.0\pm0.4$, based on the assumption of the CO gas being in LTE and having x$_{\rm{CO}}=0.0001$. Whilst such assumptions are reasonable for protoplanetary disks, they may not be for NO~Lup. For example, if the LTE assumption is invalid (i.e., in the case that the observed CO gas is secondary and instead replenished by planetesimals), so too may be the assumption that x$_{\rm{CO}}=0.0001$ (i.e., since primordial H$_2$ gas may have already dispersed, and in such a system x$_{\rm{CO}}$ could be much higher). Therefore, whilst non-LTE conditions can result in higher CO gas masses (as previously discussed in $\S$\ref{sec:COII_III}), the physics involved which violate such conditions (i.e., the dispersal of H$_2$ gas) can produce lower total gas masses. If this is the case for NO~Lup (and indeed the other 3 sub-mm detected sources), then this total gas-to-dust mass ratio could be much lower than calculated, and the disk would instead be dust dominated. To assess how much second generation gas we may expect around NO~Lup, we can compare it with the $\sim 10$\,Myr-old M-dwarf, TWA~7, where second-generation CO gas was recently detected \citep{Matra19B}. If planetesimals in the NO~Lup disk have the same volatile content as those around TWA~7, we can use the properties of its dust disk (setting the rate at which planetesimals are destroyed) to determine the level of second generation CO that may be produced. Accounting for the differences in stellar mass and luminosity, fractional luminosity, belt radius and width (assuming $r{\sim}30$\,au and a dr/r of 0.5 for NO~Lup), we would expect ${\sim}18$ times more CO gas mass around NO~Lup than TWA~7. Thus, by taking the same LTE, optically-thin excitation conditions for TWA~7, we would then predict ${\sim}2.4\times10^{-5}M_{\oplus}$ of second-generation CO to be present around NO~Lup. This is only a factor ${\sim}2$ lower than the CO gas mass measured (see Table~\ref{tab:class3ObjsGas}), and thus these estimates show broad consistency between the measured CO gas mass and a scenario in which this may be continuously replenished from exo-cometary ice (i.e., one in which the CO gas may no longer be primordial). This will be explored further in \citet{Lovell20}. \\ 
\newline
Sz~107 has mid-IR colours that puts it close to other class II transition disks, but its sub-mm dust mass is orders of magnitude lower (the dust masses of nearby [K]-[12] and [K]-[24] class II YSOs Sz~84 and Sz~111 are ${\sim}7.7$ and ${\sim}75 M_{\oplus}$, respectively). This suggests that whilst the mm-grains in this system have dispersed, modest levels of mid-IR emission persist (i.e., at similar levels to much more massive transition disks). Sz~107 therefore may provide a useful constraint on the rate at which this rapid mass dispersal takes place (i.e., since this is likely to have happened relatively recently), and also on the order, since it reinforces that the mm-sized grains in the outer disk seem to disappear before the mid-IR emission from the inner regions is finally cleared \citep[e.g.,][]{Wyatt15}. Like NO~Lup its dust emission could be fitted by a single temperature component within a few au, but there could be an additional component at 10s of au (or a broad disk). There is no evidence for gas, but the limits are not as strong due to confusion from cloud emission. Nevertheless, we could only place a $3\sigma$ upper limit on the total gas-to-dust ratio of <0.6 (i.e., whilst it is likely this system is dust dominated, it may still have a large reservoir of gas comparable in mass to its dust). Neither of these sources showed evidence of ongoing accretion on to their stars.\\
\newline
We find these two sources therefore cannot be as clearly defined, being debris disk-like in terms of their dust mass and accretion properties, but having [K]-[24] excesses that are only slightly lower than known class II disks or transition disks. Moreover, in the case of NO~Lup, the gas detection is consistent with a protoplanetary disk composition and a gas-to-dust ratio of unity, but is also with the gas being secondary. These two systems are therefore prime candidates for further deep ALMA imaging to constrain their physical sizes, CO gas masses and origins in further detail.

\subsubsection{Hot Dust from Asteroid Belts or Planet Formation: MV Lup and CD-31 12522}
Two stars have 24\,$\mu$m-only excesses, and neither showed evidence of ongoing accretion on to their stars. The SED fitted temperatures puts the dust in these systems within 1\,au, which could be from a close-in asteroid belt \citep{Wyatt08, Chen09, Morales09, Kennedy14}, or dust created during collisions between forming planets \citep[e.g., the shake-down of Kepler planetary systems][]{Kenyon04, Genda15, Wyatt16, Su20}. We therefore place these in a separate category to the other class III YSOs, since while the mid-IR excess could be an indicator that planetesimal formation has taken place (and/or that planet formation is ongoing), it is not clear that this has taken place beyond a few au (i.e., at debris disk radii) given the limits set by the non-detection of sub-mm dust.

\subsubsection{Photospheres: The Remaining Class III Stars}
The remaining 24 class III’s appear to be bare photospheres as they do not exhibit infrared excesses or CO gas. For 20/24 of these class IIIs there is no evidence of ongoing accretion on to their stars, (the remaining 4 of  which we were unable to confirm, see $\S$\ref{sec:Halpha}). In all 24 cases however, the SED analysis concluded that no significant excess emission is present and so we believe that these are appropriately classed, but lack evidence for planetesimal formation, either in the mid-IR or in the sub-mm.

\subsubsection{The Effect of Stellar Binarity?}
\label{sec:discBinary}
\citet{Zurlo20} observed 29/30 of the same class~III stars in our survey for binarity (J160714.0-385238 was too faint to observe), within a separation range of 20-1200\,au, and found that our sample contains 6 binaries. 
Three of these have wide (>100\,au) separations: Sz~108B at ${\sim}$683\,au, Sz~116 at ${\sim}$245\,au, and CD-35~10498 at ${\sim}$172\,au, and 3 of these have intermediate (20-100\,au) separations: J154306.3-392020 at ${\sim}$24\,au, CD-39~10292 at ${\sim}$47\,au and V1097~Sco at ${\sim}$21\,au. 
Stellar binaries are known to impact the detection rate of disks, and thus may have influenced those observed in our class~III survey, however for circumstellar emission with the extent probed by our observations, only those with intermediate separations are likely to significantly influence circumstellar disk properties \citep{Jensen96, Alexander12, Rosotti18, Manara19, Long19}. 
Indeed, \citet{Yelverton19} showed that the detection rate of older debris disks was lower for binary separations between 25-135\,au, thus if the circumstellar emission measured in our class~III survey is dust from the progenitors of the debris disks observed at 10s of au, then a similar relationship may exist. 
While no circumstellar emission was detected towards the 3 class~III stars with intermediate separation binaries, a detection rate of 0/3 is consistent with being the same as that of the class~III stars without confirmed intermediate separation binaries (6/26). 
Thus there is no evidence in our sample that stellar binarity influences the detection fraction of circumstellar emission, although we cannot rule out that the binary companions of J154306.3-392020, CD-39~10292 and V1097~Sco accelerated the dispersal of their disks, and we cannot comment on the influence of binaries with separations \textit{outside} of the 20-1200\,au range probed by \citet{Zurlo20}. 
There is also no evidence that intermediate separation binaries are over-represented in the class~III sample compared with that of class~II's, as might be expected if such binaries disperse their protoplanetary disks more quickly; e.g., \citet{Zurlo20} found intermediate separation binaries towards 4/76 of their Lupus class~II sample, which is consistent with the incidence rate of 3/29 towards class~III's. 
Larger sample sizes would help to ascertain the influence of binarity on disk detection rates.

\begin{figure}
    \includegraphics[width=0.475\textwidth]{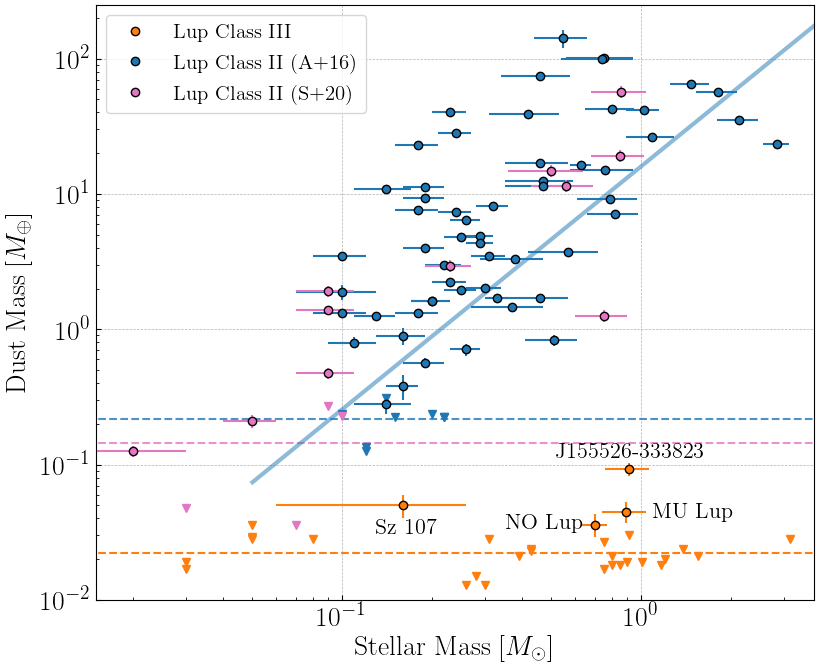}
    \caption{Disk masses in Lupus as a function of stellar mass for the class II and class III surveys. The dashed lines show the mean upper limit mass from the class II data in blue \citep{Ansdell16}, purple \citep{Sanchis20} and for the class III sources (this work) in orange. The blue solid line shows the relationship for class II YSOs between their stellar mass and dust mass from \citet{Ansdell16}.}
    \label{fig:diskMassStelMass}
\end{figure}

\subsection{The Lupus Disk Mass Distribution}
This survey of 30 class III stars was sensitive to disk masses down to $0.024M_\oplus$, which for context we include on a plot of disk mass vs age in Fig.~\ref{fig:diskMassAge}, alongside the survey populations of Upper Sco \citep{Barenfeld16}, Lupus (class II) \citep{Ansdell16} and Ophiuchus \citep{Cieza19, Williams19}, and for the debris disk population in \citet{Holland17}. 
We note here that in comparing the masses of disks in different populations it is most important that the value of $\kappa_\nu$ is used consistently. 
Our value of 3.5\,cm$^{2}$\,g$^{-1}$ (which does not account for the gas mass, since we are comparing their dust masses) is consistent with that used to estimate the dust masses for the Lupus class~II, Upper~Sco and Ophiuchus disks. 
However, the $\kappa_\nu$ used to estimate the debris disk masses in \citet{Holland17} and dust masses in \citet{Sanchis20} were both lower by ${\sim}60\%$, and thus we re-scaled these accordingly. 
For illustrative purposes we randomly assigned the ages of stars in Upper Sco, Lupus and Ophiuchus between 5-11\,Myr, 1-3\,Myr and 0.75-1.0\,Myr respectively (consistent with the ages of stars in these regions). This plot demonstrates that whilst the masses of protoplanetary disks are significantly higher than those of the much older debris disks, the dust masses measured for the Lupus class IIIs are consistent with those of the much older debris disk population. It also shows that there has been a significant improvement in sensitivity compared with previous surveys which has allowed us to detect the lowest mass disks at 1-3\,Myr. Whilst it may appear from the plot that an increase in dust mass is seen from those of the class III's we detected at ${\sim} 2$\,Myr and the brightest debris disks at ${\sim}$10-100\,Myr, similar to that discussed in \citet{Turrini19}, this may be due to a spectral type bias; i.e., the class III sample is dominated by K and M stars, whereas the debris disk sample is dominated by G, F and A type stars which we would expect to host more massive disks. The same caveat applies when comparing the detections on this plot between different regions; e.g., the large number of $0.1-0.3M_\oplus$ disk detections in Upper Sco could result from the different stellar mass, disk mass and disk radius distribution compared with Lupus \citep{Ansdell16, Hendler20}, but also from other features of the sample selection as noted later in this section. \\
\newline
Here we combine these class III results with the previous surveys of class II stars (as outlined in $\S$\ref{sec:sources}), which were sensitive to disk masses of ${\sim}0.2M_\oplus$, to consider the disk mass distribution in Lupus as a whole.
The previous class II surveys had showed that there is a dependence of disk mass on stellar mass, and Fig.~\ref{fig:diskMassStelMass} shows disk mass vs stellar mass for the combined sample \citep[excluding class II sources without stellar masses, see][for the complete list]{Ansdell16, Ansdell18, Sanchis20}. Note that for the class II star V1094~Sco, we re-derived the disk mass from the flux in \citet{Terwisga18} using equation 1 of \citet{Ansdell16} and the Gaia DR2 distance to this source. 
This recovers the aforementioned correlation for the class II population, with higher mass disks found only around higher mass stars, and lower mass disks only around lower mass stars (for which there was also a greater proportion of non-detections). Fitting to the detected disks in the class II population, \citet{Ansdell16} found a positive relationship between $\log M_{\rm dust}$ and $\log M_\star$ with a slope $1.8\pm0.4$ (which we re-plot on our Fig.~\ref{fig:diskMassStelMass}).\\
\newline
The detected class III disks all have a similar dust mass, and fall below the detection threshold of the class II survey.
Whilst the class II and III detections cover a range of stellar masses, when non-detections are included it may be clearer that neither this survey nor the \citet{Ansdell16} survey were particularly sensitive to disks around the lowest mass stars ${<}0.30M_\odot$. The \citet{Ansdell16} class II survey deliberately excluded Brown Dwarf YSOs (i.e., those with masses ${<}0.1M_\odot$), but we include those analysed by \citet{Sanchis20} which observed sources down to YSO stellar masses of ${\sim}0.02M_{\odot}$ (of which 4/9 with ${<}0.1M_\odot$ are found with upper limit disk masses).
For the sub-sample of 31 stars in Lupus that are ${>}0.7M_\odot$ (including both class II and III stars), there were no non-detections of class II stars, and so this sub-sample can be considered to have been searched down to a uniform depth of $0.024M_\oplus$.
Considering such stars it is clear from Fig.~\ref{fig:diskMassStelMass} that there is a large gap in disk mass between the lowest mass class II disk at ${\sim} 2M_\oplus$ and the highest mass class III disks at ${\sim} 0.09M_\oplus$.
The same is true when considering the 54 stars with masses $M_{\star}>0.3M_\odot$, but the gap appears to get smaller because of the lower disk masses of the class II’s.
For even lower stellar masses, there is a large fraction of class II’s with non-detections, and the correlation would suggest that many of these simply have disks just below the detection threshold of that survey.\\
\begin{figure}
    \includegraphics[width=0.475\textwidth]{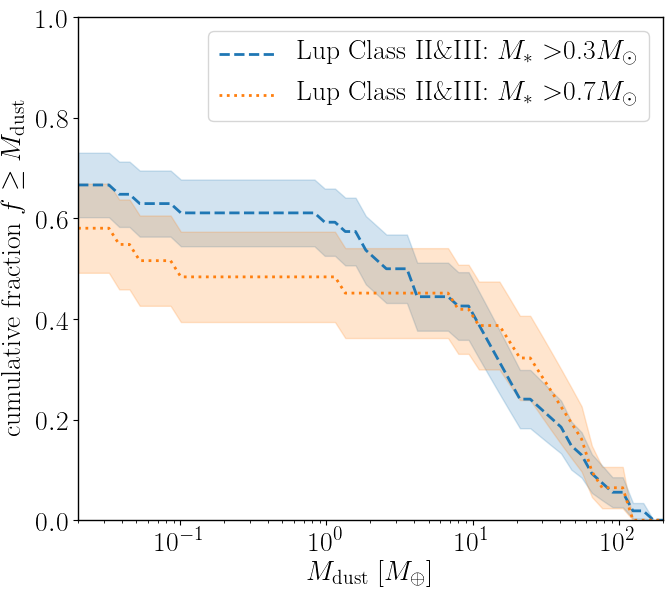}
    \caption{Cumulative fraction of disks with a mass greater than those in the binned range for the combined Lupus class II and III surveys, for all stars with either $M_{\star}>0.3M_{\odot}$ (blue dashed) or those with $M_{\star}>0.7M_{\odot}$ (orange dotted). The shaded regions represent $1\sigma$ confidence intervals.}
    \label{fig:diskMassDistComb}
\end{figure}
\newline
Rather than comparing the disk mass distribution of class II’s with class III’s, we first consider in Fig.~\ref{fig:diskMassDistComb} the disk mass distribution for the combined sample of class II’s and III’s, but cutting this in stellar mass.
This was calculated by comparing the number of stars with disks detected at a given mass with the number of stars for which such detections were possible (which is usually the full sample until the survey sensitivity is reached, though this is variable due to the different distances to the stars), with asymmetric error bars given by Poisson statistics \citep[][see also $\S$\ref{sec:disccomparison}]{Gehrels86}.
This further quantifies the gap mentioned above.
Also evident from Fig.~\ref{fig:diskMassDistComb} is that a larger fraction of higher mass stars have lost their protoplanetary disks at this age (50\% have $M_{\rm dust}>0.3M_\oplus$ for $M_{\star}>0.7M_\odot$) compared with lower mass stars (60\% for $M_{\star}>0.3M_\odot$). \\
\newline
The presence of a gap in disk masses (see Figs.~\ref{fig:diskMassStelMass} and~\ref{fig:diskMassDistComb}) suggests that the process responsible for dispersing proto-planetary disks acts quickly, perhaps being triggered once the disk mass has decreased below a certain threshold. This is reminiscent of what has been called "two-timescale behaviour" in the observational literature about disk dispersal \citep[see e.g., the seminal papers of][]{Strom89, Skrutskie90}, on the basis of infrared excesses. Because infrared excesses trace only the warm inner regions, our results imply that the disk dispersal process removes also the bulk of the disk's dust mass that is concentrated at 10s of au on roughly the same timescale, depleting this by at least an order of magnitude. Theoretically, the best explanation for the two-timescale behaviour is the process of photo-evaporation \citep[see e.g.,][]{Clarke01, Alexander14, Ercolano17}, i.e., a disk wind driven by the gas thermal energy. Photo-evaporation disperses the inner part (<1\,au) of the disk when the mass accretion rate onto the star becomes lower than the mass-loss rate of the wind. Because there is currently a large uncertainty about photo-evaporation mass-loss rates depending on which energy bands of the radiation from the central star provide the energy to drive mass-loss, it is interesting to estimate the mass-loss rates implied by our observations. A lower proto-planetary disk mass threshold of ${\sim}3M_{\rm{Jup}}$ (such as that found in Lupus for solar mass stars, assuming a dust-to-gas ratio of 100), for a typical viscous time-scale of 1\,Myr \citep{Rosotti17, Lodato17} would imply a photo-evaporative mass-loss rate of ${\sim}3\times 10^{-9}M_{\odot}$\,yr$^{-1}$. This value is in the range of what can be explained by current models \citep{Gorti09, Owen11, Picogna19} including far ultra-violet and X-ray photons, while being too high to be explained by earlier models based only on extreme ultra-violet radiation \citep[see e.g.,][]{Alexander06}. \\
\newline
As an alternative, over the last few years it has been proposed that disk evolution and dispersal is controlled by disk winds driven by the disk magnetic field, rather than a thermal wind as in photo-evaporation. Because these models are still in their infancy, depending on how the magnetic field flux evolves over the disk lifetime they can predict a wide variety of outcomes \citep[see e.g.,][]{Armitage13, Suzuki10, Bai16}, ranging from a slow "fading" of the disk to a behaviour similar to that predicted by photo-evaporation. Our observational results constrain that only the latter alternative is compatible, at least for ${>}0.3M_\odot$ stars with disks that disperse within ${\sim}2$\,Myr.\\
\newline
An important caveat is that our observations are not sensitive to gas mass, rather to dust mass which is dominated by mm-cm-sized grains. Photo-evaporation, or other types of disk winds, remove the gas, and while these can also entrain grains up to ${\sim}1\,\mu$m in size \citep{Takeuchi05} they cannot be directly responsible for the rapid removal of larger dust grains. While \citet{Alexander07} showed that the interplay with dust dynamics can remove the dust in the inner part of the disk (<1\,au), dust at larger distances still remains in the disc. The problem of where the mm-sized dust ends up was highlighted by \citet{Wyatt15} where it was argued that this must be removed in collisions, either grinding it into small dust or growing it into larger bodies. This feasibility of the former possibility was demonstrated by \citet{Owen19}, who showed that collisional destruction can grind the grains into micron-sized dust that is removed by radiation pressure. The latter possibility was also demonstrated by \citet{Carrera17} which showed that the dust can grow into planetesimals, either through collisions or via the streaming instability. What our observations show is that whichever of these the dispersal mechanisms is acting, its effect on the mm-sized dust must be both rapid and complete, in removing over 90\% of the dust mass. Although depletion may not be so substantial for lower mass stars (i.e., since the gap is not so pronounced), the transition disk of Sz~84 has a disk >100 times more massive than the disk of the class III star, Sz~107, despite these having approximately the same stellar mass (i.e., 0.18 and 0.16$M_\odot$ respectively), and mid-IR excess (see Fig.~\ref{fig:photometryC2C3}).
Deeper observations of the lower mass class II sources (i.e., to at least the same sensitivity as this work) would therefore allow us to quantify if this gap persists for the lowest mass YSOs between the class II and class III stages.

\subsection{Comparison with Class IIIs in other Star Forming Regions}
\label{sec:disccomparison}
\begin{figure}
    \includegraphics[width=0.475\textwidth]{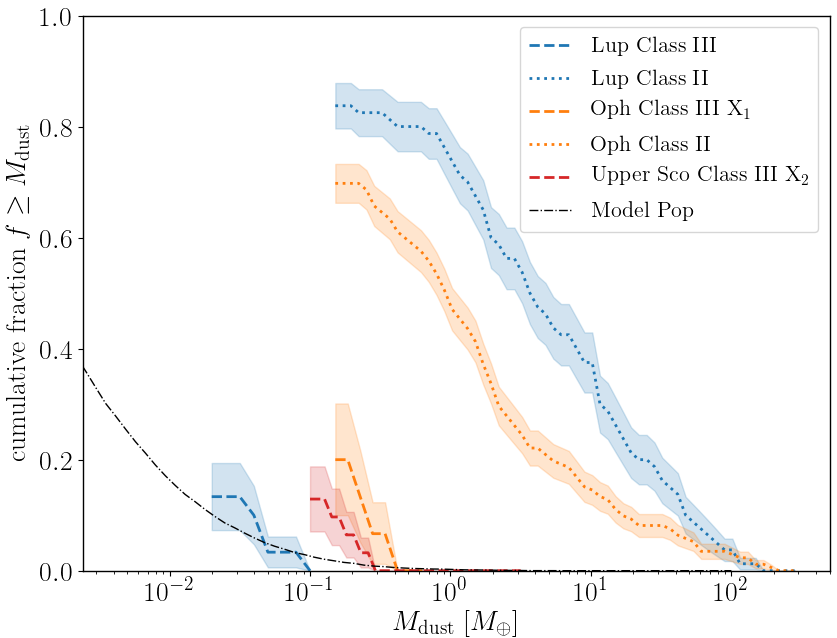}
    \caption{Cumulative fraction of disks with a mass greater than those in the binned range for Lupus sources (blue), where class III are dashed and class II are dotted, and similarly for Ophiuchus (in orange) and class III sources in Upper Sco (in red), noting that for both Ophiuchus and Upper Sco the class III sources were selected based on different constraints to our Lupus survey, and are hence labelled as Class III X$_1$ or Class III X$_2$ here. The shaded regions represent $1\sigma$ confidence intervals.}
    \label{fig:diskMassDist}
\end{figure}

The disk mass distribution is shown in Fig.~\ref{fig:diskMassDist} in the format shown commonly in other surveys of star forming regions, wherein the samples are split into their different classes. The complete sample of Lupus YSOs is included, excluding only sources without confirmed stellar masses. This was produced by binning all such sources in their disk masses, and evaluating the uncertainty on the fraction in each bin using binomial statistics \citep[i.e., for the total number of sources detected, and the total number of sources in each bin, see][]{Gehrels86}.
This reinforces that the class III’s in Lupus are orders of magnitude lower in mass than their class II counterparts.
This class III distribution will be compared in $\S$\ref{sec:popModel} with that expected for debris disks at this age, which shows excellent agreement. \\
\newline
First we use Fig.~\ref{fig:diskMassDist} to compare the Lupus class III disk mass distribution with that of the only other two surveys that observed class III stars in Ophiuchus and Upper Sco \citep[][respectively]{Williams19, Barenfeld16}. 
However, it is important to note that there are two reasons why these three class III samples may not be equivalent.
First, \citet{Ansdell16} demonstrated that there is a different stellar mass distribution for Lupus compared with Upper Sco YSOs from \citet{Barenfeld16}, which may be important given the trend of class II disk mass with stellar mass in Fig.~\ref{fig:diskMassStelMass}, and a similar trend for debris disk masses \citep{Greaves03}. 
Second, and perhaps most importantly, the class III sample selection criteria are different for the different surveys.
While all have the same requirement for class III stars not having large 12 and 24\,$\mu$m excesses (i.e., they are all samples of class III stars that lie in the green shaded region of Fig.~\ref{fig:photometryC2C3}), the Ophiuchus and Upper Sco surveys set additional constraints to exclude class III's in the bottom left corner for which there was no evidence for mid-IR excess emission.
The ODISEA survey in Ophiuchus excluded class III stars with [K]-[24]<1 (equivalent to excluding those in our survey in the cluster of near photospheric fluxes in the bottom corner of Fig.~\ref{fig:photometryC2C3}; see the horizontal green line on that figure) and the survey of Upper Sco excluded all class III stars with bare photospheres \citep[see the discussion in both][]{Luhman12, Barenfeld16}.
This complicates the comparison between these different associations, and to acknowledge the different sample selection criteria we add the suffix X$_1$ and X$_2$ to class III in reference to the Ophiuchus and Upper Sco samples in Fig.~\ref{fig:diskMassDist}. \\
\newline
We first compare Lupus with Ophiuchus. The class II’s in Lupus and Ophiuchus have comparable mass distributions, although the masses are slightly lower for Ophiuchus. This is different to expectations of disks decreasing in mass with age, given that Ophiuchus is slightly younger than Lupus at ${\sim} 1$\,Myr. However, this may simply reflect a stellar mass distribution with more lower mass stars given the correlation in Fig.~\ref{fig:diskMassStelMass}, or it may be due to differences in the local environment \citep[see further discussion in][]{Williams19}. The possibility of overlapping populations (and therefore ages) in the YSOs of Ophiuchus \citep{Luhman20}, and presumably Lupus, should also be considered. Nevertheless, it could also be the case that this mass distribution reflects an increase in dust production (e.g., from planetesimal collisions) at the slightly older age of Lupus in comparison to Ophiuchus \citep[e.g., as indicated by the toy model of][]{Gerbig19}.
The Ophiuchus class III X$_1$ survey was not as deep as that performed here, instead reaching a sensitivity comparable with the class II surveys of ${\sim} 0.2M_\oplus$.
Thus it appears surprising that the rate of detection of class III's is comparable for the two surveys, which translates into the disk mass distribution for Ophiuchus class III’s looking like it is higher than those of Lupus by around half to one order of magnitude (see Fig.~\ref{fig:diskMassDist}).
However, as noted above, sources in the Ophiuchus class III X$_1$ sample all have large [K]-[24] excesses, which we have shown implies a larger disk mass, given that mid-IR excess is likely present.
A deep survey of all Ophiuchus class III's would be required for a proper comparison.
We can however, consider what a Lupus class III X$_1$ survey would have found, noting that 4 of our class III's meet the class III X$_1$ K-24>1 colour threshold (excluding those that are confused or which have mid-IR upper limits). Two of these had sub-mm detections, both with disk masses almost an order of magnitude lower than those in \citet{Williams19}.
While the highest class III disk masses could be rarer in Lupus than in Ophiuchus, this is more likely a result of small number statistics.
For example, we should not be surprised by the fact that we did not detect disks more massive than ${\sim}0.1M_\oplus$ in our class III sample, since the model in \S~\ref{sec:popModel} predicts these should occur at a rate $\sim 3$\% (i.e., about 1 in 30).
Such high disk masses would be expected to occur more frequently in a class III X$_1$ sample. \\
\newline
In the case of Upper~Sco, we are also not comparing with a consistently selected sample nor observing depth. In the \citet{Barenfeld16} survey all sources were measured to a sensitivity of ${\sim} 0.1M_\oplus$, and only sources named "Debris/Ev. Trans" overlap with the same constraints we set for our class III sub-sample. As shown by \citet{Luhman12}, there is a large population of low [K]-[12] vs [K]-[24] sources (i.e., bare photospheres) which were not observed, whereas in our class III sample such bare photospheres were included (i.e., the detection fractions cannot be compared as directly as implied by Fig.~\ref{fig:diskMassDist}). Nevertheless, we found in our sample 6 stars not deemed to be bare photospheres \citep[i.e., 6 stars which would have met the][criteria and fall in what we are calling a class III X$_2$ sample]{Luhman12}. We detected sub-mm circumstellar dust in 4 of these, and all of these were less massive than the 5/31 Upper Sco detections. Similar to the case in Ophiuchus, the lack of high mass class III disks could be down to small number statistics; e.g., if the rates of $>0.1M_\oplus$ class III X$_2$ disks is the same in the two regions then we would expect to have detected $\sim 1$ (i.e., $6\times5/31$) in Lupus. However, it remains possible that the two disk mass distributions are different either due to evolution, since Upper~Sco is older \citep[i.e., 5-11\,Myr, see][]{Mathews12}, or because of the different stellar mass distributions (i.e., for which in comparison to our Lupus sample, there are a higher fraction of M-types and a lower fraction of K-types in Upper~Sco, along with a number of G-type stars which did not feature in our sample). To consider these possibilities, comparisons between samples must be done in a consistent manner, and to fully consider the complete disk mass distribution of class III stars, sources without excesses (including bare photospheres) should be included.

\subsection{Population Model}
\label{sec:popModel}
\begin{figure}
    \includegraphics[width=0.475\textwidth]{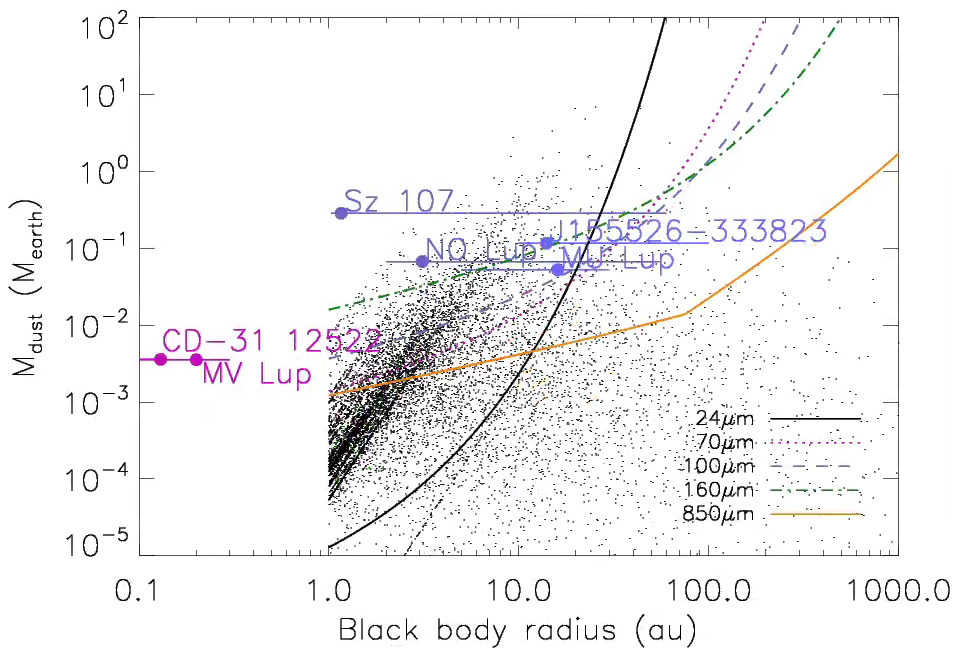}
    \caption{Top: Predicted population dust mass and radii for a sample of 10,000 disks based on the models of \citet{Wyatt07, Sibthorpe18}, and the expected detection thresholds at $24\mu \rm{m}$, $70\mu \rm{m}$ (i.e., with \textit{Spitzer}), $100\mu \rm{m}$, $160\mu \rm{m}$ (i.e., with \textit{Herschel}) and $850\mu \rm{m}$ (i.e., with \textit{ALMA}). Shown are the positions of the 4 sub-mm detections (lilac) and the 2 hot IR-excesses (pink). Note that the plotted radii are approximated by the black body SED fits (based on temperature and fractional luminosity which have a large range of values see Table~\ref{tab:SEDBB}) but that these span a wide range of values consistent with their detections/non-detections with either Spitzer (at 24$\mu$m) or ALMA (at 850$\mu$m).}
    \label{fig:MdiskRadiiRatePopModel}
\end{figure}

The population of debris disks seen around older main sequence stars has been well characterised by far-IR surveys of both nearby field stars of co-eval populations in moving groups. 
These show a decay in disk brightness with stellar age that can be understood within the context of a population model in which all stars are born with a planetesimal belt that depletes due to steady state collisional erosion on a different timescale according to its radius and initial mass \citep{Wyatt07}. 
We use such a population model, specifically that with parameters that fit observations of nearby Sun-like stars from \citet{Sibthorpe18}, to predict what this debris disk population would have looked like at the age of Lupus, for which we used the stellar masses (see Table~\ref{tab:class3Objs}), temperatures and luminosities (derived for the sources as modelled by our SEDs in $\S$\ref{sec:sed}), and $\kappa_\nu{\sim}3.5$\,cm$^{2}$\,g$^{-1}$ to be representative of the Lupus class III sample and estimated consistently with our previous mass calculations. 
Fig.~\ref{fig:MdiskRadiiRatePopModel} shows the model predictions for the dust masses and black body radii (for a distribution ranging between 1-1000\,au) of a population of 10,000 debris disks after just ${\sim}2\rm{Myr}$ of evolution, and in Fig.~\ref{fig:diskMassDist} the cumulative disk mass distribution is shown alongside the measured rate from this survey. \\
\newline
There are reasons why extrapolating this population model back to ${\sim}2$\,Myr could result in erroneous predictions. For example, the model is constrained by the 20\% of nearby stars for which their debris disks can be detected. This means that the disks of the remaining 80\% of the population, other than being too faint to detect at old ages, are unconstrained. The model has accounted for these by assuming they have small radii $\ll 5$\,au, since this means they have collisionally depleted below the detection threshold by Gyr ages, whereas this might not be the case for Lupus ages. Indeed, Fig.~\ref{fig:MdiskRadiiRatePopModel} shows that the model population includes a large fraction of small bright disks. However, where the population model is expected to make robust predictions is for bright disks with radii $>5$\,au, since it is those that evolve into the disks that remain detectable after Gyrs of evolution. \\
\newline
This population model was used to predict the detection rate for this survey, and finds at a mass sensitivity of ${\sim}0.024M_{\oplus}$ (i.e., one consistent with our survey) a rate of ${\sim}7\%$. Thus it was anticipated that, if class III stars host debris disks that are representative of the population seen around older stars, we should achieve detections of debris disks for $N_{\rm{det}}=2\pm \sqrt{2}$ of the sample, consistent with the observed detection rate of 4/30. The debris disk mass distribution predicted in Fig.~\ref{fig:diskMassDist} is also close to that observed for the class III sample. Furthermore, the model predicted that the population of non-detectable disks would have a mean disk flux of $0.0136$\,mJy. Using Eq.\ref{eq:dustmass} at the $141$\,pc average class III source distance, we find this flux to correspond to a mass of $0.0026M_{\oplus}$, consistent with (but slightly lower than) the measured stacked non-detection mass of $0.0048\pm0.0017M_{\oplus}$. This demonstrates that the model has both accurately predicted the detection rate of Lupus class III disks, and the mean disk mass of those below our detection threshold. \\
\newline
While for two of class III detections there remains some uncertainty about whether primordial gas or dust is present (i.e., NO~Lup and Sz~107), the two remaining detections (i.e., of MU~Lup and J155526-333823) are compatible with the predicted debris disk population. The good agreement of the model predictions with the disks observed around class III stars has a number of implications. First, it would seem that whatever processes have caused the 30 class III stars at this age to have lost their protoplanetary disks faster than the sample that are class II, those processes have not adversely affected the formation of planetesimals at 10s of au. This suggests that planetesimal formation, or more specifically the formation of the parent bodies of the debris disks seen around older stars, occurs earlier than 2\,Myr, also in agreement with studies of young (class II) Brown Dwarfs in Ophiuchus and Lupus \citep[e.g.,][respectively]{Testi16, Sanchis20}. \\
\newline
Having inferred that many (if not all) the detected class III disks are examples of young debris disks, given these are already bright this further suggests that debris disks are already stirred at this age (i.e., collisions are grinding planetesimals into dust, rather than collisions leading to growth). 
We confirm this for the 4 sub-mm detected disks by calculating the collisional lifetime of mm-cm sized grains \citep[i.e., the timescale over which bodies of this size collide in a collisional cascade, see][equation 15]{Wyatt08}, which for dr/r=0.5, e=0.1, $\rm{Q_{D}^{\star}}=10^{3}$\,J\,kg$^{-1}$, the stellar masses in Table~\ref{tab:class3Objs} and the disk radii in Table~\ref{tab:2DFit} (which for MU~Lup, NO~Lup and Sz~107 are upper limits), we find collision lifetimes in the ranges ${\sim}1-10$\,kyr, ${\sim}3-30$\,kyr, ${\sim}3-30$\,kyr and ${\sim}20-200$\,kyr for MU~Lup, J155526-333823, NO~Lup and Sz~107 respectively (i.e., all much less than the lower-bound age of Lupus of 2\,Myr, although only ${\sim}10\%$ in the case of cm-sized grain collisional lifetimes for the disk around Sz~107). Accounting for this range of ages, this means that even for recent protoplanetary disk dispersal (i.e., within 100-200\,kyr), the collisional lifetime of mm-cm grains in at least 3 of these systems is much shorter than the system age (i.e., the dust must be replenished via the break-up of larger bodies). \\
\newline
Further, we can estimate the lower limit on the maximum planetesimal size required for this dust replenishment to occur based on these collisional ages, and find a range of sizes between $10-300$\,m (i.e., 10s to 100s of metre-sized bodies are required to replenish the mm and cm sized grains observed in the sub-mm). This is in good agreement with the debris disk population model, and measurements of aged debris disks, which both require the presence of ${\sim}$\,km-sized planetesimals.
Although there are models in which debris disks are stirred by the growth of large Pluto-sized bodies within them \citep[so-called self-stirring models,][]{Kenyon10}, at tens of au on timescales of 1-3\,Myr these would be unable to stir debris disks (i.e., in order to produce the observed dust). For example, in the resolved 80\,au belt of J155526.2-333823, this would take >1\,Gyr, far in excess of the age of Lupus.
Self-stirring of 1-3\,Myr old disks at tens of au by Pluto-sized bodies is therefore incompatible with our observations.If these belts are not self-stirred, an alternative scenario may be one in which the stirring is forced by secular interactions between the planetesimals and unseen planets. Equation 15 of \citet{Mustill09} requires any such perturbing planets to be close to the belt and massive (i.e., given the timescale of 2\,Myr). For example, in the case of J155526.2-333823, a planet 10\,au inside the belt with a relatively low eccentricity of e=0.1 would need to have more mass than 2M$_J$ in order to stir the disk within the the age of Lupus. On the other hand, disks may simply be born stirred \citep[as discussed by][]{BoothR16}, or stirred during the process of protoplanetary disk dispersal.

\section{Conclusions}
\label{sec:conclusions}
We report the ALMA observations of 30 class III stars in the $1-3$\,Myr old Lupus Star Forming Region, having measured these to the deepest sensitivity thus far, to probe the disk masses present in each system. We detected significant continuum emission from 4 class III sources, 2 nearby class II sources (within the observation field of view), and marginally detect emission from the 26 class III sources without individual detections after stacking. We also report the first CO gas detection around a class III star, NO~Lup. \\
\newline
We attribute this continuum emission to circumstellar dust present in these Young Stellar Objects, and determine the range of dust masses from these 4 detections as $0.036-0.093M_{\oplus}$. Through stacking our sub-sample of non-detections, we tentatively report dust mass present at the level ${\sim}0.0048M_{\oplus}$. By an order of magnitude, these disks are less massive than those previously detected in other star forming regions. We found one of the class III detections to be spatially resolved, for which we constrained the size of the disk to be $80\pm15\rm{au}$, and for the three unresolved class III detections, we set upper limits on their radii (all being $<60\rm{au}$). \\
\newline
The detection of 4 class III disks is in accordance with steady-state debris disk model predictions, which suggest that the planetesimal belts observed around older systems (i.e., the ${\sim}10$s of au debris disks) are already in place by ${\sim}2\rm{Myr}$. By comparing the frequency of detections as a function of disk mass, for the first time, we probe the disk mass distribution of YSOs down to ${\sim}0.024M_{\oplus}$ levels, and show that the evolution of dust mass between the class II and III stages is both rapid and complete, removing over 99\% of dust. We show how this changes when considering different cuts in stellar mass, demonstrating that this may not be as substantial for lower mass stars. By also comparing this disk mass distribution with the debris disk evolution model, we show very good agreement between the predicted detection rate across the complete range of masses in our class III sample. \\
\newline
Considering the complete sample of class II and III Lupus YSOs outlined here (comprising 30/126 class III and 96/126 class II), we show that the processes which remove the protoplanetary disks around these stars have not affected the formation of planetesimals at tens of au (i.e., the planetesimals that form the parent bodies of debris disks). This further implies that such young disks are already stirred, such that they are bright enough to be detected.

\section*{Acknowledgements}
We thank the anonymous reviewer for their comments which improved the quality of this work. We also thank Alice Zurlo for useful discussions in relation to stellar multiplicity in Lupus. 
JBL is supported by an STFC postgraduate studentship.
GMK is supported by the Royal Society as a Royal Society University Research Fellow.
GR acknowledges support from the Netherlands Organisation for Scientific Research (NWO, program number 016.Veni.192.233).
M.T. has been supported by the UK Science and Technology research Council (STFC), and by the European Union's Horizon 2020 research and innovation programme under the Marie Sklodowska-Curie grant agreement No. 823823 (RISE DUSTBUSTERS project).
MK gratefully acknowledges funding by the University of Tartu ASTRA project 2014-2020.4.01.16-0029 KOMEET. 
This project has received funding from the European Union's Horizon 2020 research and innovation programme under the Marie Sklodowska-Curie grant agreement No 823823 (DUSTBUSTERS).
This work was partly supported by the Deutsche Forschungs-Gemeinschaft (DFG, German Research Foundation) - Ref no. FOR 2634/1 TE 1024/1-1.
This work was partly supported by the Italian Ministero dell'Istruzione, Università e Ricerca (MIUR) through the grantProgettiPremiali 2012 iALMA (CUP C52I13000140001) and by the DFG cluster of excellence Origin and Structure of the Universe (www.universe-cluster.de).

\section*{Data Availability}
This work makes use of the following ALMA data: $ADS/$ $JAO.ALMA$ $\#2018.1.00437.S$. 
ALMA is a partnership of ESO (representing its member states), NSF (USA) and NINS (Japan), together with NRC (Canada), MOST and ASIAA (Taiwan), and KASI (Republic of Korea), in cooperation with the Republic of Chile. The Joint ALMA Observatory is operated by ESO, AUI/NRAO and NAOJ.


\bibliographystyle{mnras}
\bibliography{mainBody} 


\appendix
\section{SEDs of 24 Lupus class III stars without excesses}
\label{sec:AppendixA}
\begin{figure*}
    \includegraphics[width=0.88\textwidth]{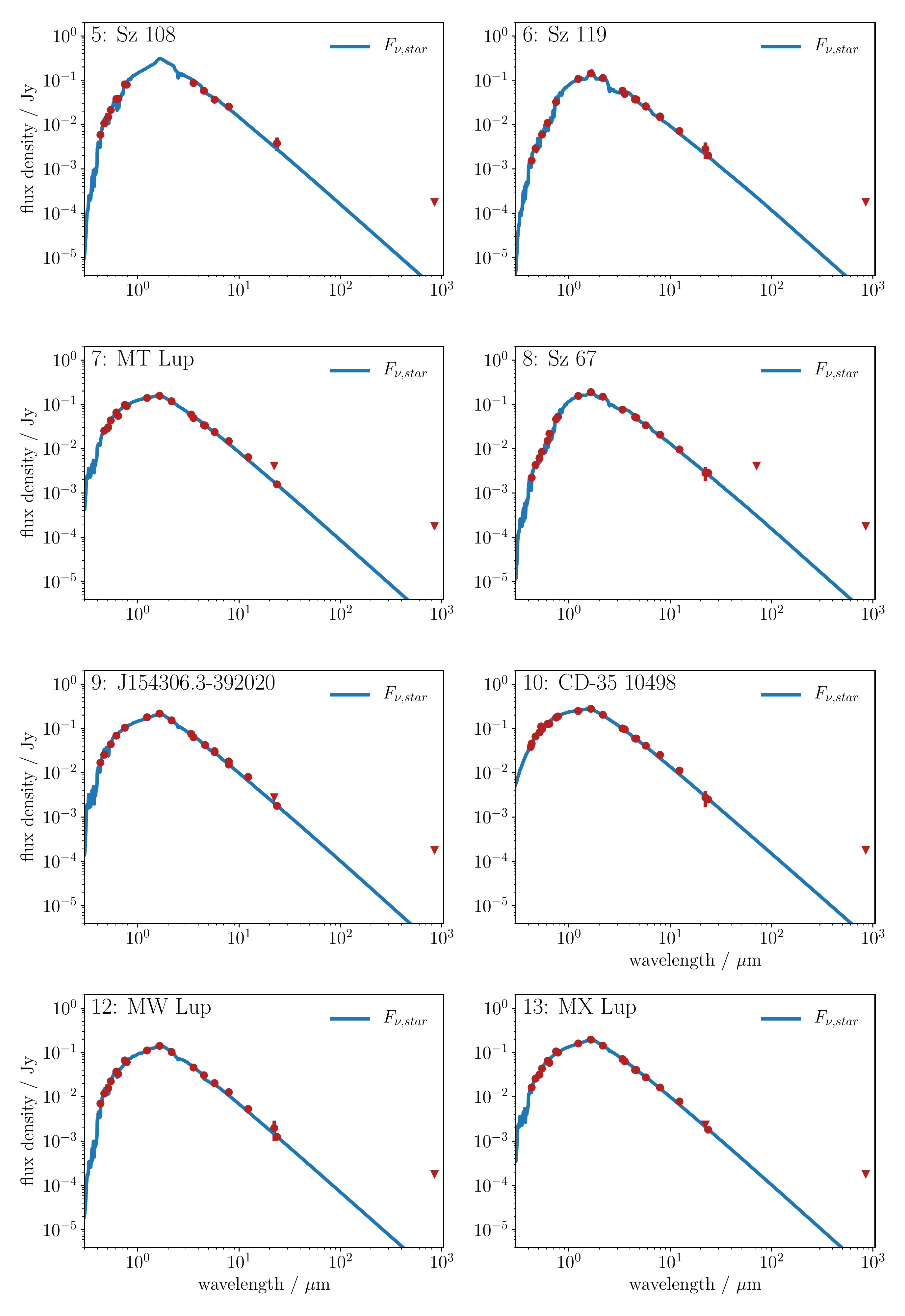}
    \caption{SEDs for sources 5, 6, 7, 8, 9, 10, 12 and 13.}
    \label{fig:SED5+}
\end{figure*}

\begin{figure*}
    \includegraphics[width=0.88\textwidth]{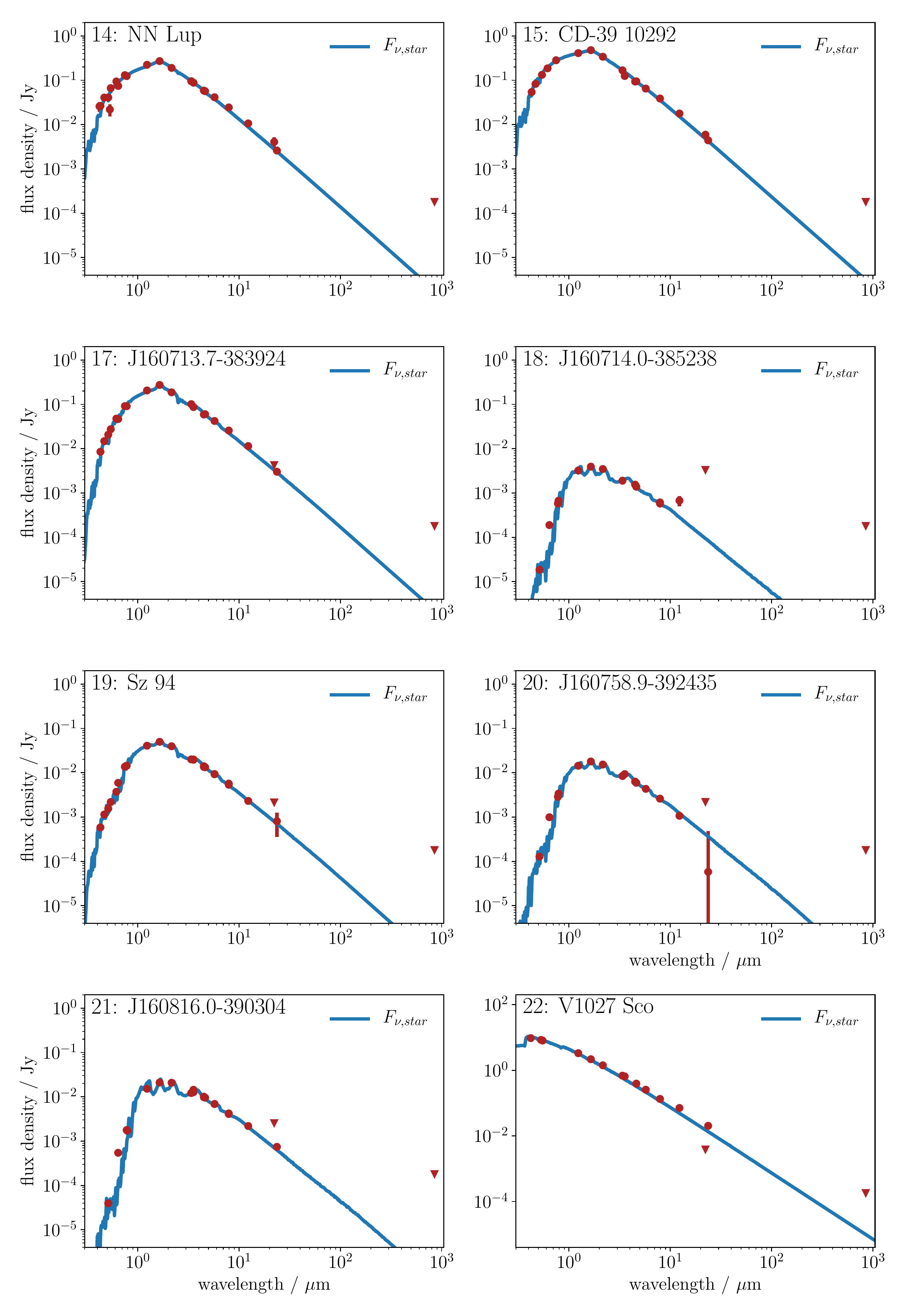}
    \caption{SEDs for sources 14, 15, 17, 18, 19, 20, 21 and 22.}
    \label{fig:SED14+}
\end{figure*}

\begin{figure*}
    \includegraphics[width=0.88\textwidth]{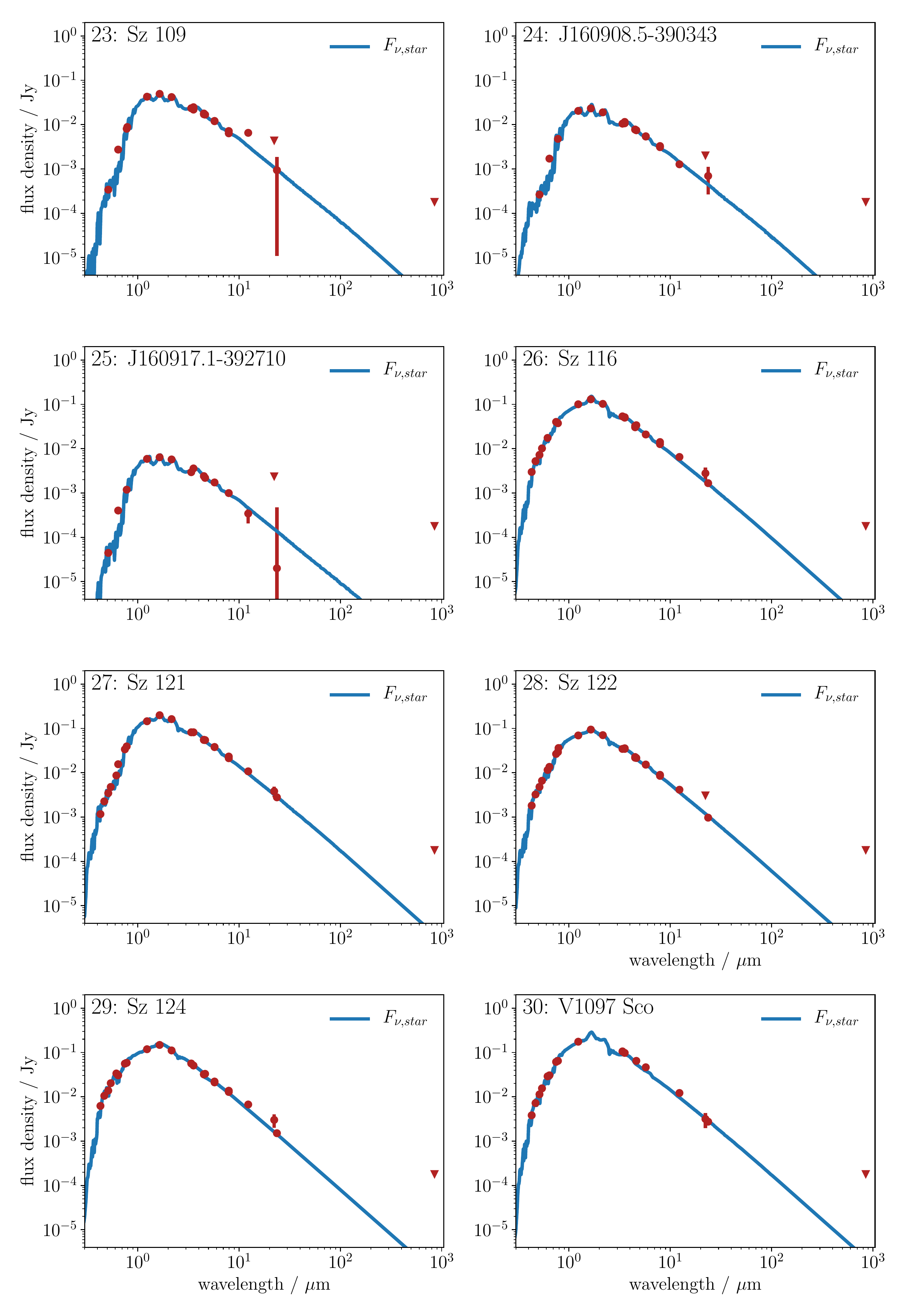}
    \caption{SEDs for sources 23, 24, 25, 26, 27, 28, 29 and 30.}
    \label{fig:SED24+}
\end{figure*}

All SEDs are provided in Figs.~\ref{fig:SED5+}, \ref{fig:SED14+} and~\ref{fig:SED24+} for the 24 class III stars in this survey without evidence for excess emission above the stellar photosphere.

\section{Two component black body modelling constraints}
\label{sec:AppendixAA}
\begin{figure*}
  \begin{center}
    \vspace{0.0in}
    \begin{tabular}{cc}
      \hspace{0.0in} \includegraphics[width=1.06\columnwidth]{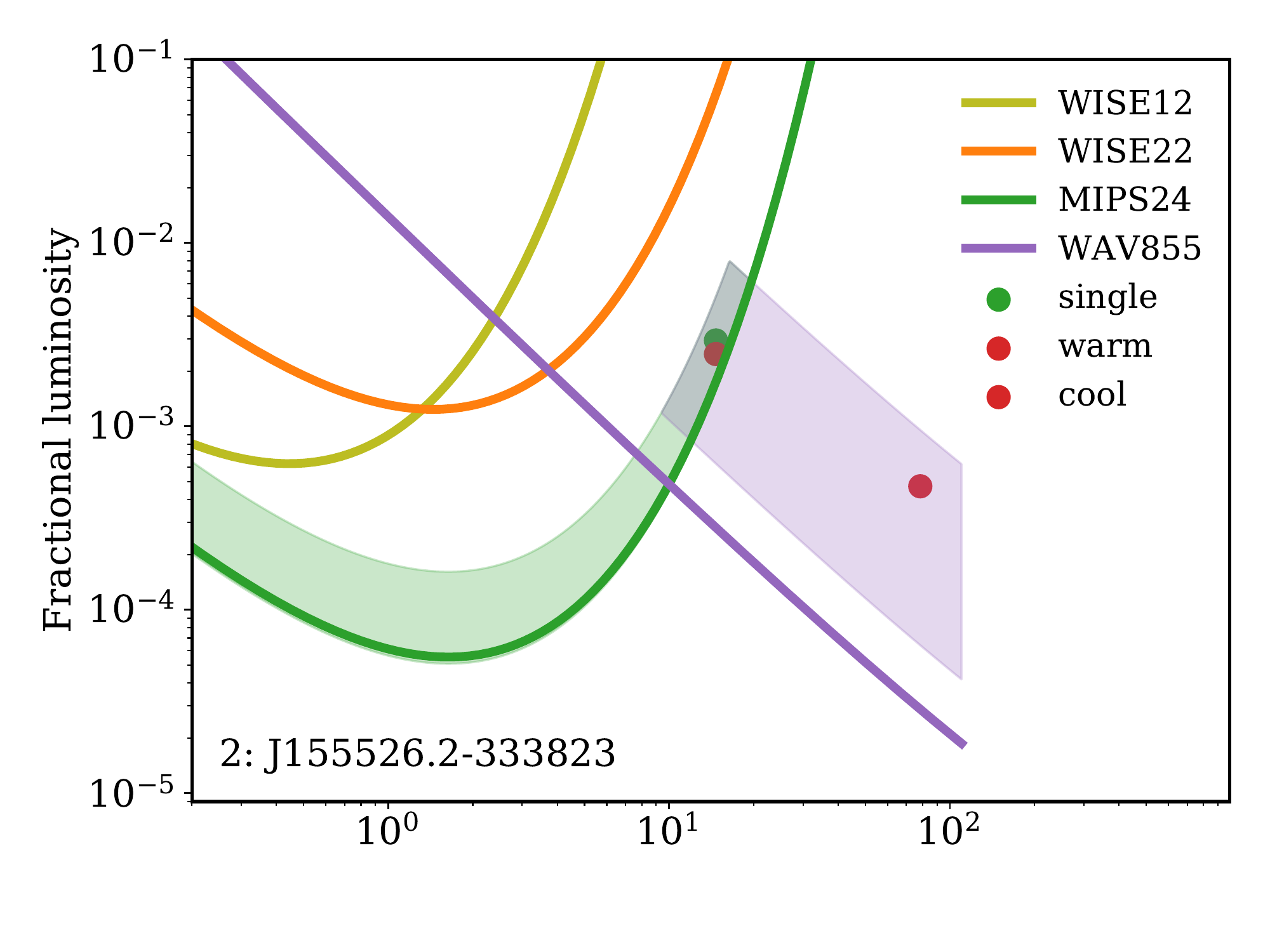} &
      \hspace{-0.27in} \includegraphics[width=1.06\columnwidth]{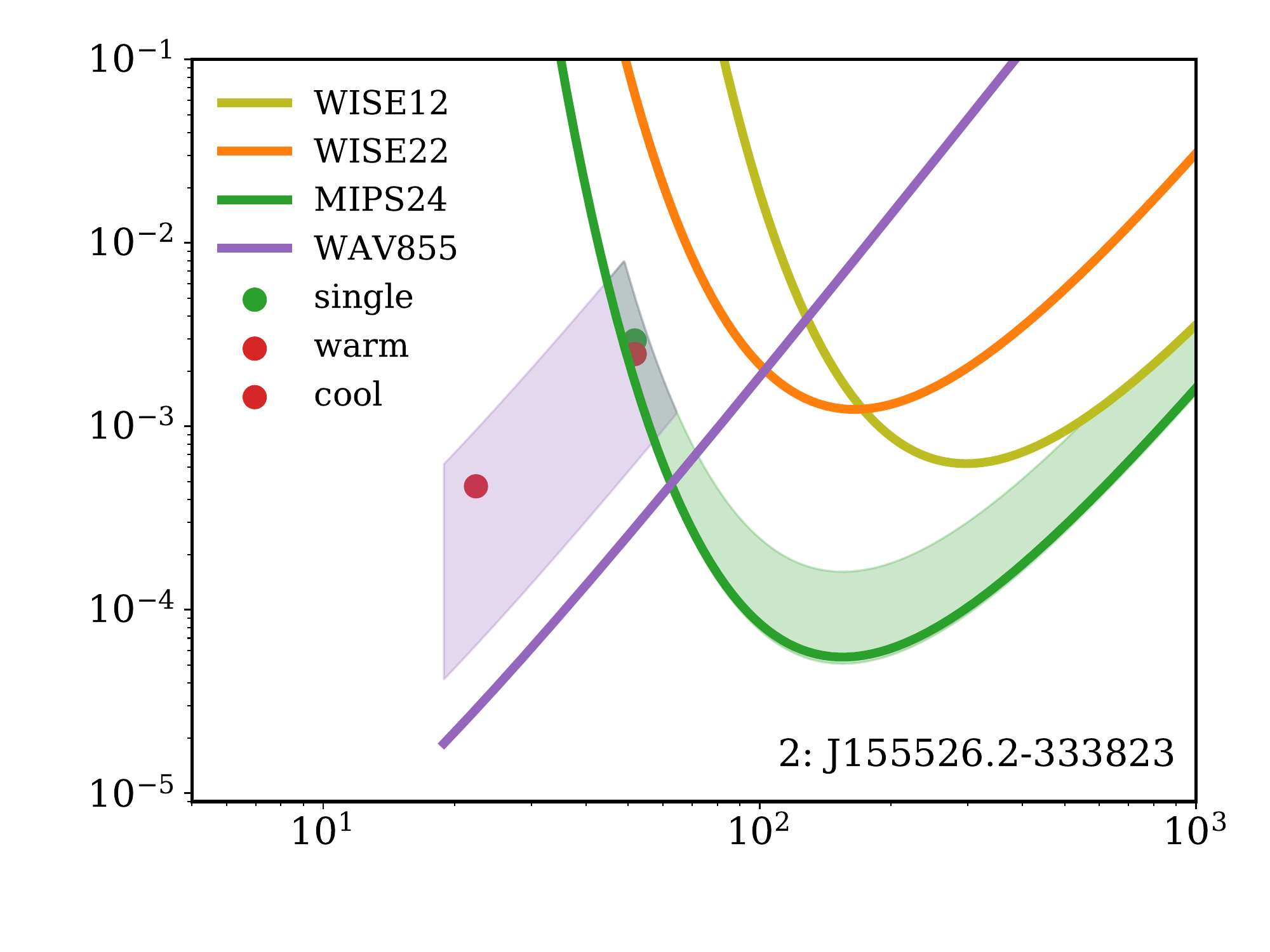} \\[-0.255in]
      \hspace{0.0in} \includegraphics[width=1.06\columnwidth]{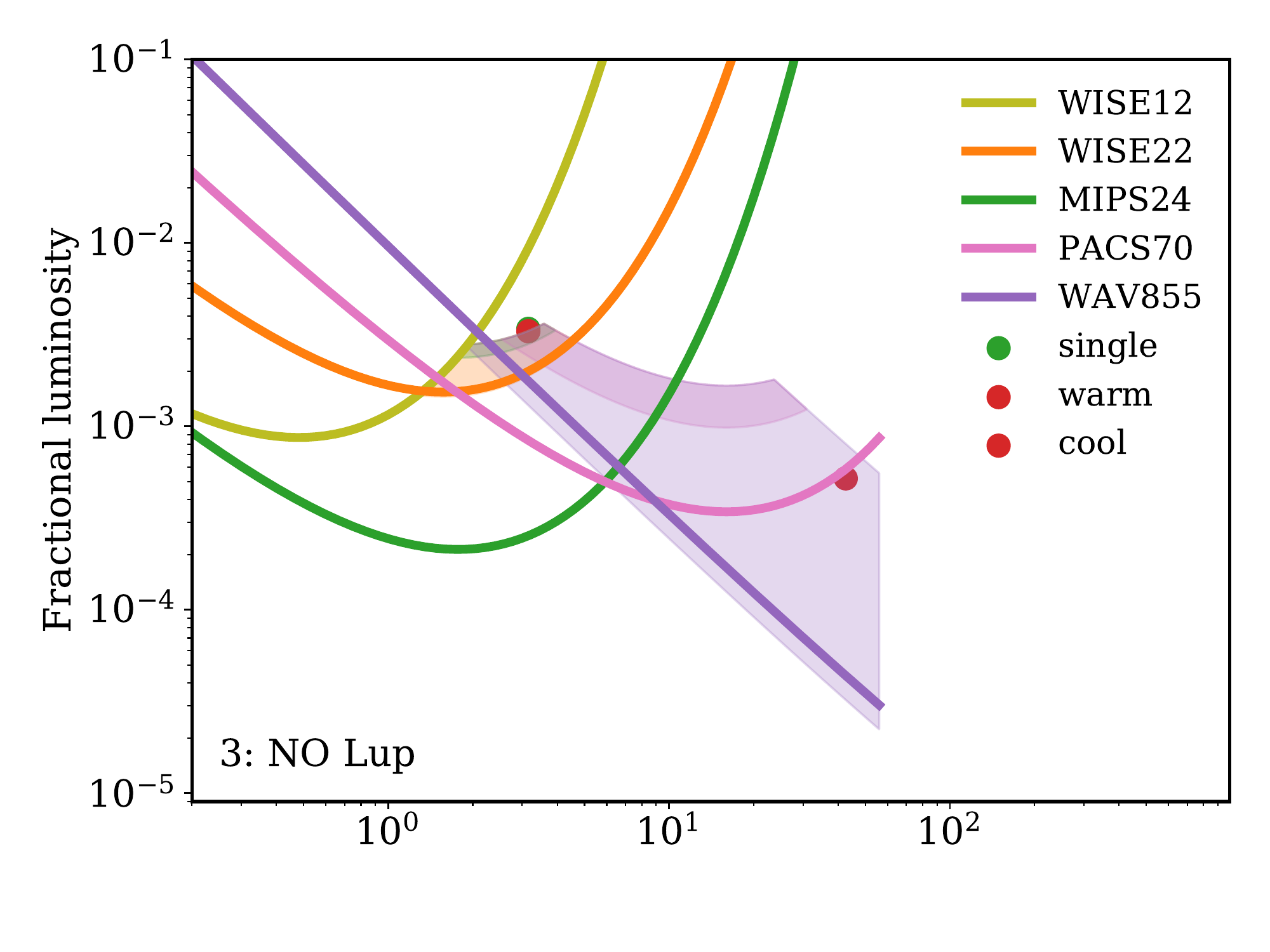} &
      \hspace{-0.27in} \includegraphics[width=1.06\columnwidth]{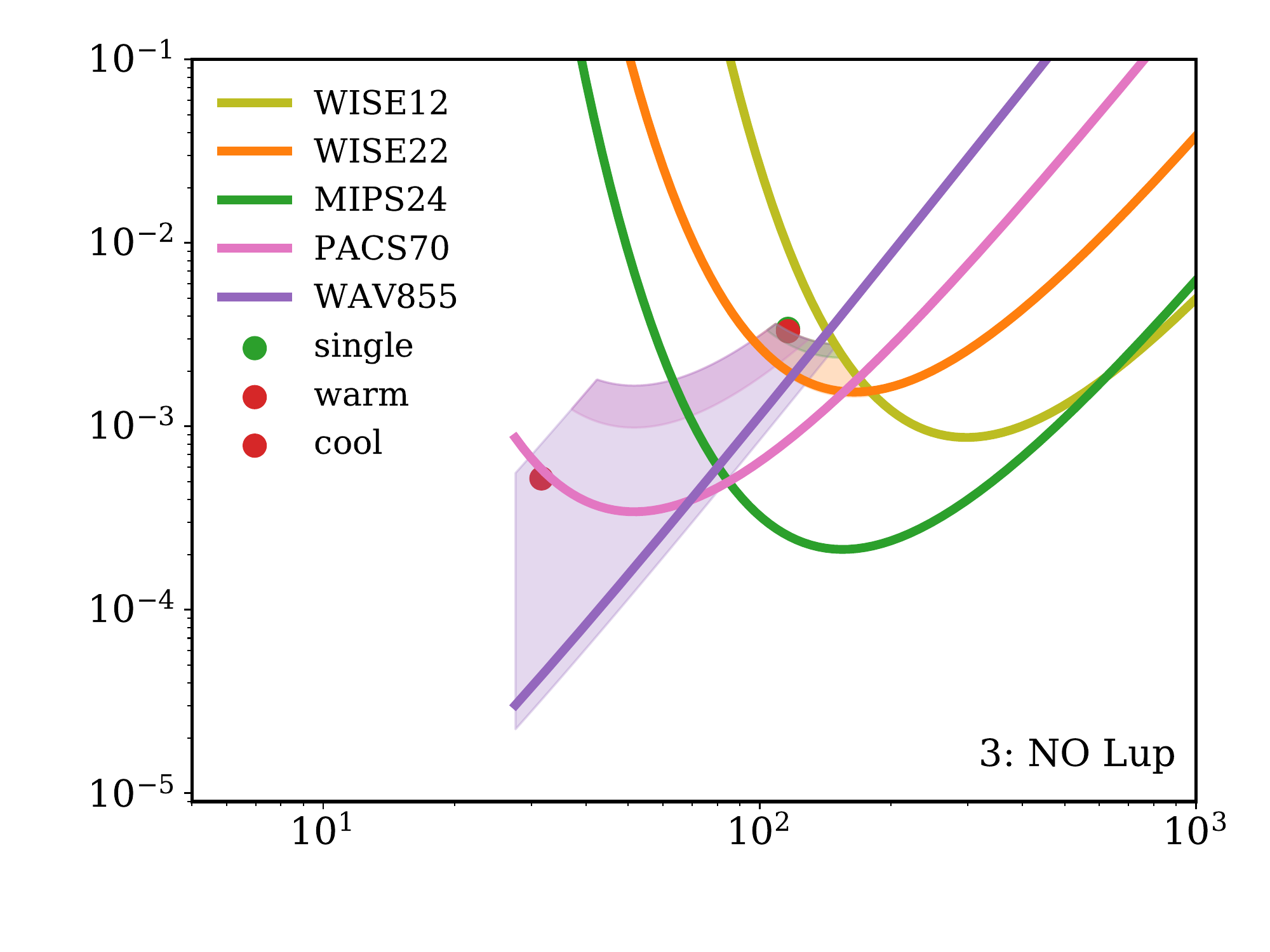} \\[-0.255in]
      \hspace{0.0in} \includegraphics[width=1.06\columnwidth]{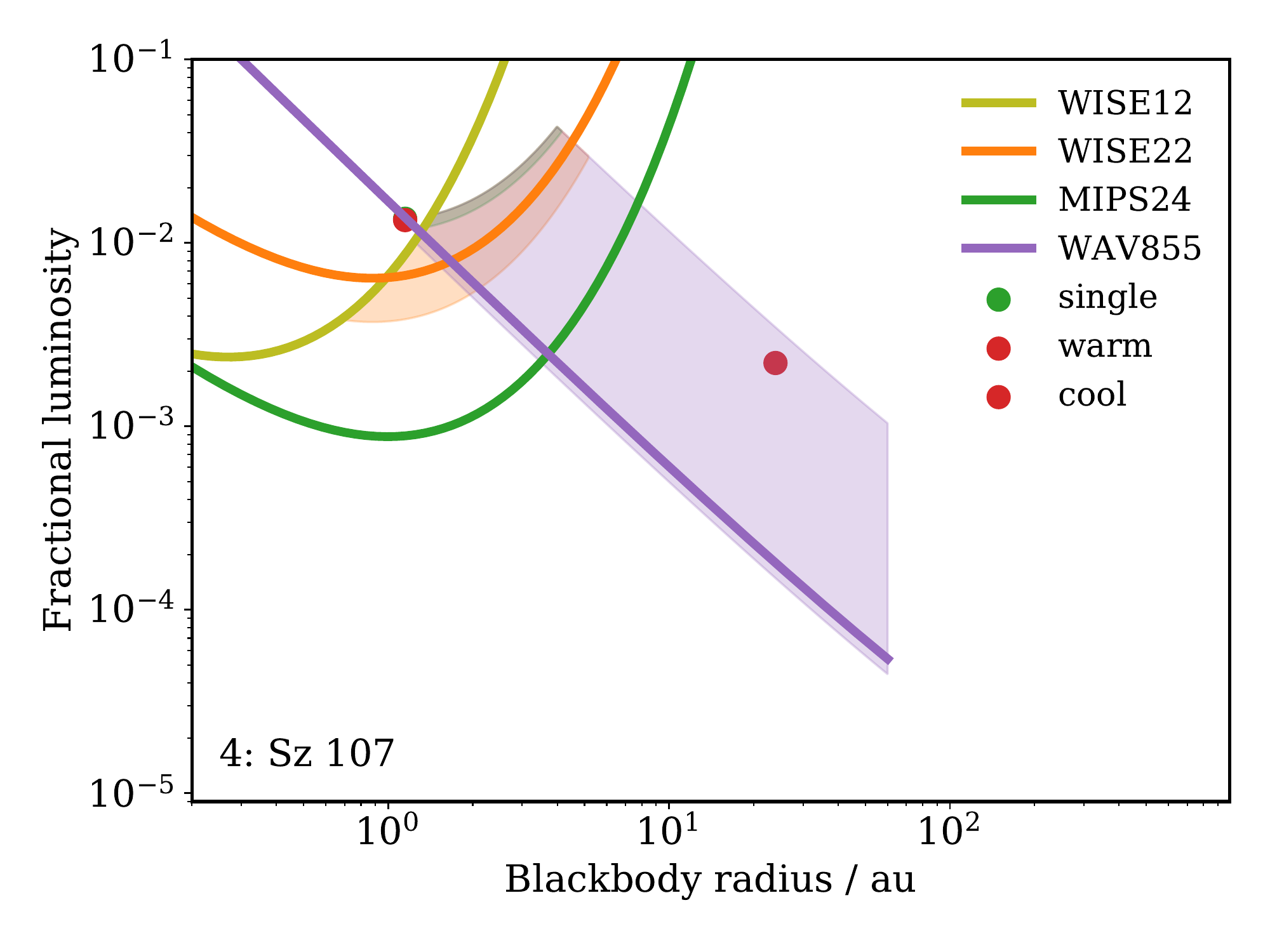} &
      \hspace{-0.27in} \includegraphics[width=1.06\columnwidth]{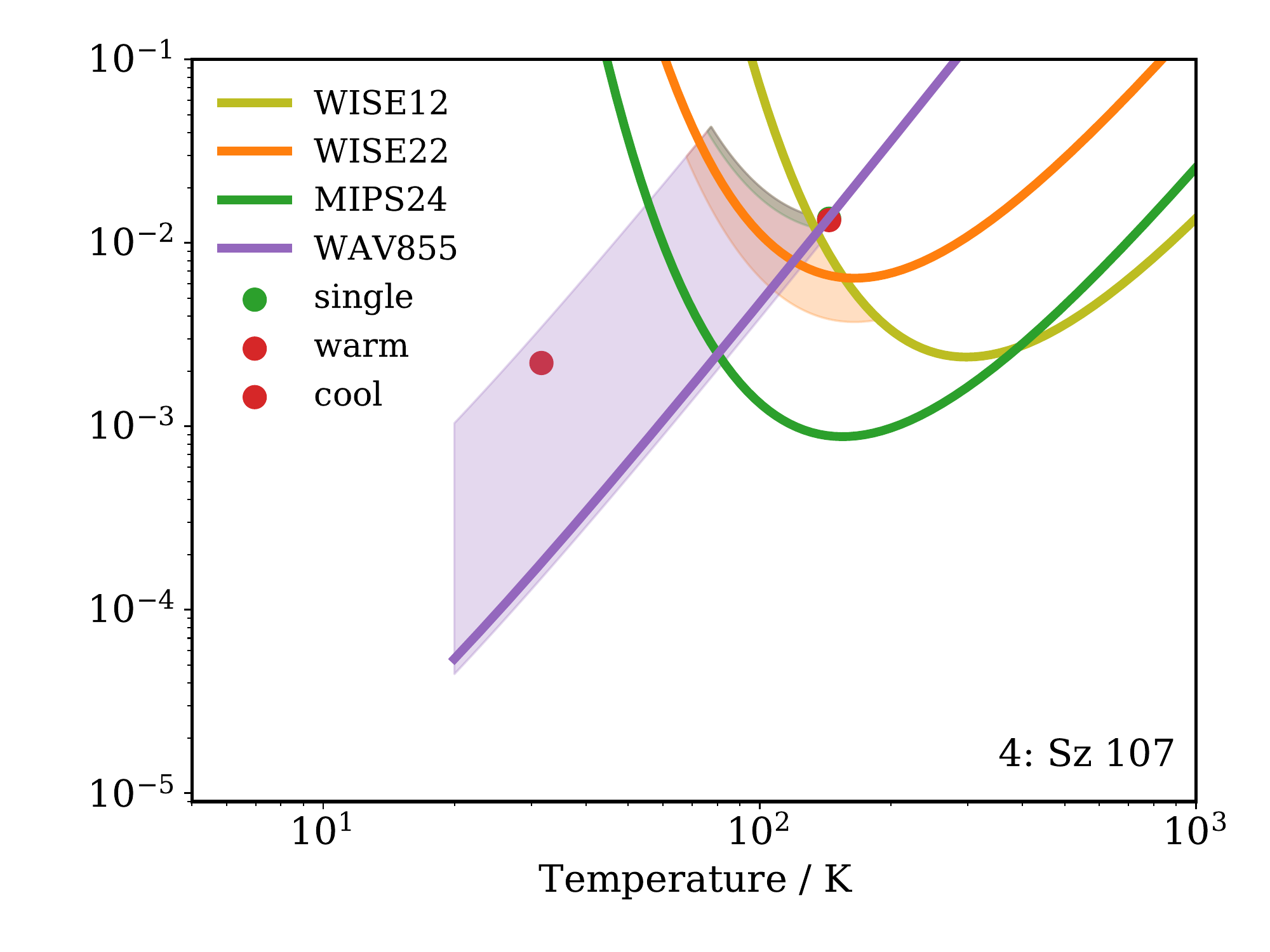}
    \end{tabular}
    \caption{Left column: Fractional luminosity versus blackbody radius for J155526.2-333823 (top), NO~Lup (middle) and Sz~107 (bottom). Right column: Fractional luminosity versus Temperature for J155526.2-333823 (top), NO~Lup (middle) and Sz~107 (bottom). All of these plots demonstrate the parameter space within which a 2-component model is consistent with the photometric data, and the constraints set by the measurements with different surveys (WISE, Spitzer, Herschel and ALMA). These map to the constraints detailed in the lower section of Table~\ref{tab:SEDBB}.}
   \label{fig:SEDSall6_appaa}
  \end{center}
\end{figure*}

Here we include fractional luminosity versus radius and temperature plots for the 3 sources determined to be consistent with having 2-temperature components of emission from their SEDs, all shown in Fig.~\ref{fig:SEDSall6_appaa}. We note that in all plots the lines represent the detection limits for each of the instruments, the shaded regions represent the region within which significant excesses may lie (for a given instrument). The green point represents the best fit single component blackbody fit, whereas the two red points represent permissible 2-component model values. The total fractional luminosity of any combined 2-component model is constrained by the measured source fluxes, and whilst there is overlap between permitted values for the cool and warm components, their point of intersection will lie at the position of the single blackbody fit. In the case of the ALMA measurements, the vertical region has been expanded to account for the wide range in allowed $\beta$ values (i.e., to fit the sub-mm region of the SED slopes). We provide a brief discussion of each source here. \\
\newline
For J155526.2-333823, significant detections of the source were made with both ALMA (WAV855) and Spitzer (MIPS 24). Disk components detected at just these wavelengths lie within the shaded regions in purple and green, respectively. The upper panel of Fig.~\ref{fig:SEDSall6_appaa} shows that a broad range of values are permitted by both shaded regions, although this is more tightly constrained for any single blackbody fit, which is bounded by the overlapping shaded region. From these plots we therefore constrain warm component as having upper and lower limit temperatures and radii from the x-axis minima and maxima of the purple shaded region as 20-60\,K and 10-100\,au, and a fractional luminosity based on the y-axis minima and maxima of the purple shaded region (5$\times10^{-4}$-0.01). The warm component has no lower limiting radius in this plot (and consequently no upper limit temperature) and these correspond to <15\,au and >50\,K, whilst the fractional luminosity limits can be seen to be identical to the cool component. \\
\newline
For both NO~Lup and Sz~107, (in the middle and lower panels of Fig.~\ref{fig:SEDSall6_appaa}) whilst not plotted here, their IRS spectra \citep[as measured by][]{Lebouteiller11} tightly constrain the allowed range of warm component emission in the dark shaded region, where the WISE22 (W4), MIPS24, (in the case of NO~Lup) PACS70, and ALMA detectable regions overlap.
This strong constraint means that the warm component of the 2 component model is the same as the single component.
For the cooler components of both of these two sources, a much broader parameter space is allowable. For NO~Lup, the cool component temperature and radii are limited between 25-120\,K and 2-56\,au (the upper limit radii constrained by our continuum modelling), with fractional luminosity limits between 0.00002-0.004. In the case of Sz~107 the cool component temperature and radii limits are instead 20-150\,K and 1-60\,au (the latter also limited by the continuum emission constraint) and the fractional luminosity ranges between 0.00005-0.4.

\section{Channel Maps of Complete Sample}
\label{sec:AppendixB}
\begin{figure*}
    \includegraphics[width=0.8\textwidth]{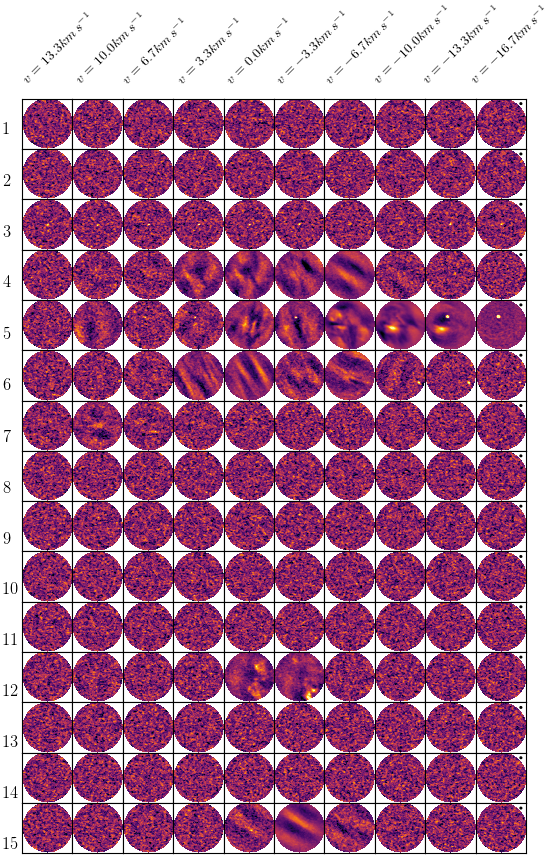}
    \caption{ALMA Channel Maps for target source IDs 1-15. Large scale emission was seen in sources 4, 5, 6, 7, 12 and 15 (prior to removing all baselines ${<}100\rm{k}\lambda$) assumed to be cloud emission from Lupus. In all images, North is up, East is left, and the beam is shown in the upper right of the right-most panel for each source. CO emission can be seen clearly in source 3 and 5, and towards the South East of source 6 for velocities $v{<}-6.7\rm{km}\,\rm{s}^{-1}$.}
    \label{fig:class3Spurious1}
\end{figure*}

\begin{figure*}
    \includegraphics[width=0.8\textwidth]{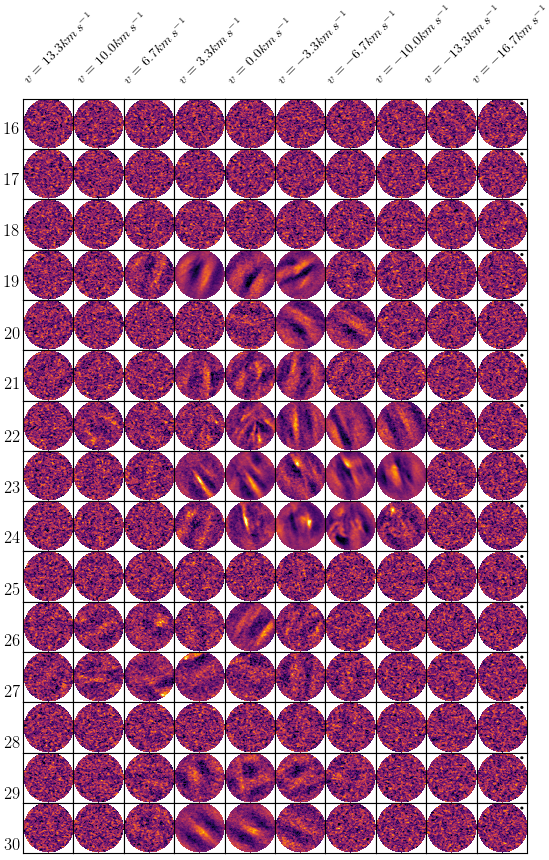}
    \caption{ALMA Channel Maps for target source IDs 16-30. Large scale emission was seen in sources 19, 20, 21, 22, 23, 24, 26, 27, 29 and 30 (prior to removing all baselines ${<}100\rm{k}\lambda$) assumed to be cloud emission from Lupus. In all images, North is up, East is left, and the beam is shown in the upper right of the right-most panel for each source.}
    \label{fig:class3Spurious2}
\end{figure*}

Channel maps are provided in Figs.~\ref{fig:class3Spurious1} and \ref{fig:class3Spurious2} for all 30 sources over a range of radial velocities to demonstrate the effect that Lupus CO cloud emission had on our data sets, with 16/30 of these suffering from this.

\bsp	
\label{lastpage}
\end{document}